\documentclass[pra,twocolumn,preprintnumbers,amsmath,amssymb,floatfix,longbibliography]{revtex4}
\usepackage{graphicx}
\usepackage{graphics}
\usepackage{psfrag}
\usepackage{hyperref}
\usepackage{multirow}
\usepackage[sort&compress]{natbib}
\usepackage{amssymb}
\usepackage{amsmath}
\usepackage{amsthm}
\usepackage{bbold}
\usepackage{bbm}
\usepackage{enumerate}
\usepackage{mathdots} 

\newcommand{\vare}{\varepsilon}
\newcommand{\sgn}{\mbox{sgn}}
\newcommand{\rmi}{{\rm i}}

\usepackage{xcolor}

\begin{document}

\hypersetup{pdftitle={title}}

\title{Infrared fixed points of higher-spin fermions in topological semimetals}

\author{Igor Boettcher}
\email{iboettch@umd.edu}
\affiliation{Joint Quantum Institute, University of Maryland, College Park, MD 20742, USA}

\begin{abstract}
We determine the fate of interacting fermions described by the Hamiltonian $H=\textbf{p}\cdot \textbf{J}$ in three-dimensional topological semimetals with linear band crossing, where $\textbf{p}$ is momentum and $\textbf{J}$ are the spin-$j$ matrices for half-integer pseudospin $j\geq 3/2$. While weak short-range interactions are irrelevant at the crossing point due to the vanishing density of states, weak long-range Coulomb interactions lead to a renormalization of the band structure. Using a self-consistent perturbative renormalization group approach, we show that band crossings of the type $\textbf{p}\cdot \textbf{J}$ are unstable for $j\leq 7/2$. Instead, through an intriguing interplay between cubic crystal symmetry, band topology, and interaction effects, the system is attracted to a variety of infrared fixed points. We also unravel several other properties of higher-spin fermions for general $j$, such as the relation between fermion self-energy and free energy, or the vanishing of the renormalized charge. An $\text{O}(3)$ symmetric fixed point composed of equal chirality Weyl fermions is stable for $j\leq 7/2$ and very likely so for all $j$. We then explore the rich fixed point structure for $j=5/2$ in detail. We find additional attractive fixed points with enhanced $\text{O}(3)$ symmetry that host both emergent Weyl or massless Dirac fermions, and identify a puzzling, infrared stable, anisotropic fixed point without enhanced symmetry in close analogy to the known case of $j=3/2$.
\end{abstract}

\maketitle

Topological semimetals are electronic materials that feature topologically protected band crossing points in their band structure \cite{GeimReview,RevModPhys.82.3045,WKrempa,RevModPhys.90.015001,GaoReview}. Upon tuning the chemical potential to any of these exceptional points, the low-energy physics is often captured by highly symmetric single-particle Hamiltonians known from particle physics \cite{BookWeinberg,PhysRev.60.61}. In particular, with the recent groundbreaking discovery of several compounds hosting fermions with large topological charge and nonzero chirality such as CoSi, AlPt, and PdBiSe, \cite{Bradlynaaf5037,PhysRevLett.119.206402,chang2018topological,PhysRevLett.122.076402,Rao2019ObservationOU,Sanchez,SchroeterNature,PhysRevB.99.241104,cano2019multifold}, studying the properties of higher-spin fermions in solids emerges as a new frontier of condensed matter physics \cite{PhysRevB.90.081112,PhysRevB.93.045113,PhysRevB.93.241113,PhysRevB.94.195205,PhysRevB.97.134521,PhysRevX.8.041026,PhysRevLett.121.157602,PhysRevLett.124.127602,PhysRevB.101.184503,lin2020dual}. In these materials, the topological features of the bulk electronic band structure are experimentally accessible and distinguishable through surface effects such as large Fermi arcs, or can be probed via angle-resolved photoemission spectroscopy (ARPES).

What do we consider a higher-spin fermion in this context? In a first attempt of a definition, we say a three-dimensional topological semimetal hosts a fermion with half-integer spin $j\geq 1/2$, if the $k \cdot p$ Hamiltonian close to a $(2j+1)$-fold band crossing point is given by
\begin{align}
 \label{int1} H_{\text{SO}(3)} (\textbf{p}) = 2 p_i J_i,
\end{align}
with $\textbf{p}$ the momentum measured from the crossing point and spin-$j$ matrices $J_i$ satisfying $[J_k,J_l]=\rmi \vare_{klm} J_m$, the factor of 2 introduced for later convenience. We implicitly sum over repeated indices $i=1,2,3=x,y,z$. The corresponding eigenenergies are labelled by $m=-j,\dots,j$ and read $E_m(\textbf{p}) = 2m|\textbf{p}|$. We restrict ourselves to linear band crossings here as they are closer to the relativistic dispersion found in particle physics. However, the exciting properties of spin-3/2 fermions at a quadratic band touching point \cite{abrikosov,abrben,luttinger} also attracted a lot of attention in recent years and have been investigated in Refs \cite{moon,PhysRevLett.113.106401,PhysRevX.4.041027,PhysRevB.92.045117,PhysRevB.92.035137,PhysRevB.93.205138,PhysRevB.93.165109,PhysRevLett.116.137001,PhysRevLett.116.177001,PhysRevB.95.085120,PhysRevB.95.075149,PhysRevLett.118.127001,PhysRevB.95.144503,PhysRevB.96.094526,PhysRevB.96.144514,BingNature,PhysRevB.96.214514,PhysRevLett.120.057002,PhysRevX.8.011029,Kimeaao4513,PhysRevB.97.205402,PhysRevB.97.125121,PhysRevB.98.104514,PhysRevX.8.041039,PhysRevB.99.054505,PhysRevB.99.125146,2018arXiv181104046S,PhysRevB.100.075104,PhysRevB.100.165115,PhysRevResearch.2.013230}.

The Hamiltonian in Eq. (\ref{int1}) is invariant under continuous rotations of momentum and spin taken from the group $\text{SO}(3)$. In real materials, the crystal structure breaks this continuous symmetry and we assume in the following that the remaining discrete symmetry is captured by the cubic rotational group $\text{O} \subset \text{SO}(3)$. As a result, cubic-only symmetric terms are allowed on the right-hand side of Eq. (\ref{int1}), their relative size being determined by the band structure of the material at hand. We therefore widen our definition of a higher-spin fermion to include any $(2j+1)$-fold linear band crossing point described by a Hamiltonian of the form $H(\textbf{p}) = p_i \mathcal{V}_i$ where $\mathcal{V}_i$ transforms as a vector under the cubic rotational group, i.e. according to the $T_1$ representation.

The band structure close to the band crossing point receives self-energy corrections due to the long-range part of the Coulomb interaction between electrons. The surprising finding of Isobe and Fu \cite{PhysRevB.93.241113} is that the rotation invariant Hamiltonian in Eq. (\ref{int1}) is unstable towards the inclusion of Coulomb interactions for $j=3/2$. Instead, depending on the parameters of the band structure, the system is attracted to one of two infrared renormalization group (RG) fixed points. One of them features enhanced $\text{O}(3)$ symmetry and is described by the "relativistic" Hamiltonian
\begin{align}
 \label{int2} H_{\text{O}(3)}(\textbf{p}) = p_i V_i,
\end{align} 
where the matrices $V_i$ satisfy the Clifford algebra,
\begin{align}
 \label{int3} \{V_k,V_l\} = 2\delta_{kl},
\end{align}
and the $(j+1/2)$-fold degenerate eigenvalues read $E_\pm(\textbf{p})=\pm |\textbf{p}|$. The second fixed point is only cubic symmetric and located at a seemingly arbitrary point in parameter space. The presence of such a stable infrared fixed point without any visibly enhanced symmetry is quite unusual, because many electronic systems feature emergent rotation or even Lorentz invariance at quantum critical points.

To understand the nature of the different fixed points, let us consider the cases $j=1/2$ and $j=3/2$ as illustrative examples. For $j=1/2$, the spin matrices are given in terms of the Pauli matrices, $2J_i = V_i = \sigma_i$, and so satisfy a Clifford algebra themselves. The Hamiltonian in Eq. (\ref{int1}) coincides with the Weyl Hamiltonian in this case. For $j=3/2$, on the other hand, the situation is less obvious. In this case, the most general cubic-symmetric higher-spin Hamiltonian reads
\begin{align}
 \label{int4} H_{3/2}(\textbf{p}) = p_i(u_1 J_i + u_2 J_i^3),
\end{align}
with $u_{1,2}$ two material parameters. Now introduce the matrices
\begin{align}
 \label{int5} V_i &= -\frac{7}{3}J_i+\frac{4}{3}J_i^3,\\
 \label{int6} U_i &= \frac{13}{6}J_i-\frac{2}{3}J_i^3
\end{align}
to obtain the Isobe--Fu Hamiltonian \cite{PhysRevB.93.241113}
\begin{align}
 \label{int7} H_{3/2}(\textbf{p}) = p_i(V_i+\alpha U_i).
\end{align}
We normalize momentum such that the prefactor of $p_iV_i$ is unity, so that $\alpha$ is the only free parameter. The matrices $V_i$ realize the Clifford algebra from Eq. (\ref{int3}), thus we obtain the relativistic Hamiltonian for $\alpha=0$. Importantly, a basis change brings the latter into the form $H_{\text{O}(3)}(\textbf{p}) \sim p_i(\mathbb{1}_2 \otimes \sigma_i^*)$, which shows that the relativistic system consists of two Weyl fermions of equal chirality. In contrast, a Dirac Hamiltonian comprises two Weyl fermions of opposite chirality in the massless limit. For $\alpha=2$, on the other hand, we obtain $H_{\text{SO}(3)}(\textbf{p})$ from Eq. (\ref{int1}). Finally, the stable cubic-only symmetric fixed point is located at $\alpha=2.296$. The three fixed points for $j=3/2$ are summarized in Tab. \ref{TabFP}.

In this work, we show that the behavior found for $j=3/2$ continues for larger $j$: The $\text{SO}(3)$-symmetric fixed point is unstable for $j=5/2$ and $7/2$, while the $\text{O}(3)$-symmetric one is stable, and we argue that this likely extends to $j>7/2$. Furthermore, for $j=5/2$ we identify a stable cubic-only symmetric infrared fixed point at a seemingly arbitrary point in parameter space. To arrive at these conclusions, we first generalize the matrices $V_i$ and $U_i$ to $j>3/2$, and discuss symmetries and topology of higher-spin Hamiltonians. We then derive general properties of the fermion and photon self-energy due to long-range interactions for arbitrary $j$, investigate the stability of the $\text{SO}(3)$ and $\text{O}(3)$ symmetric fixed points for $j\leq 7/2$, and eventually analyze the case of $j=5/2$ in considerate detail.

\begin{figure}[t!]
\centering
\begin{minipage}{0.48\textwidth}
\includegraphics[width=4cm]{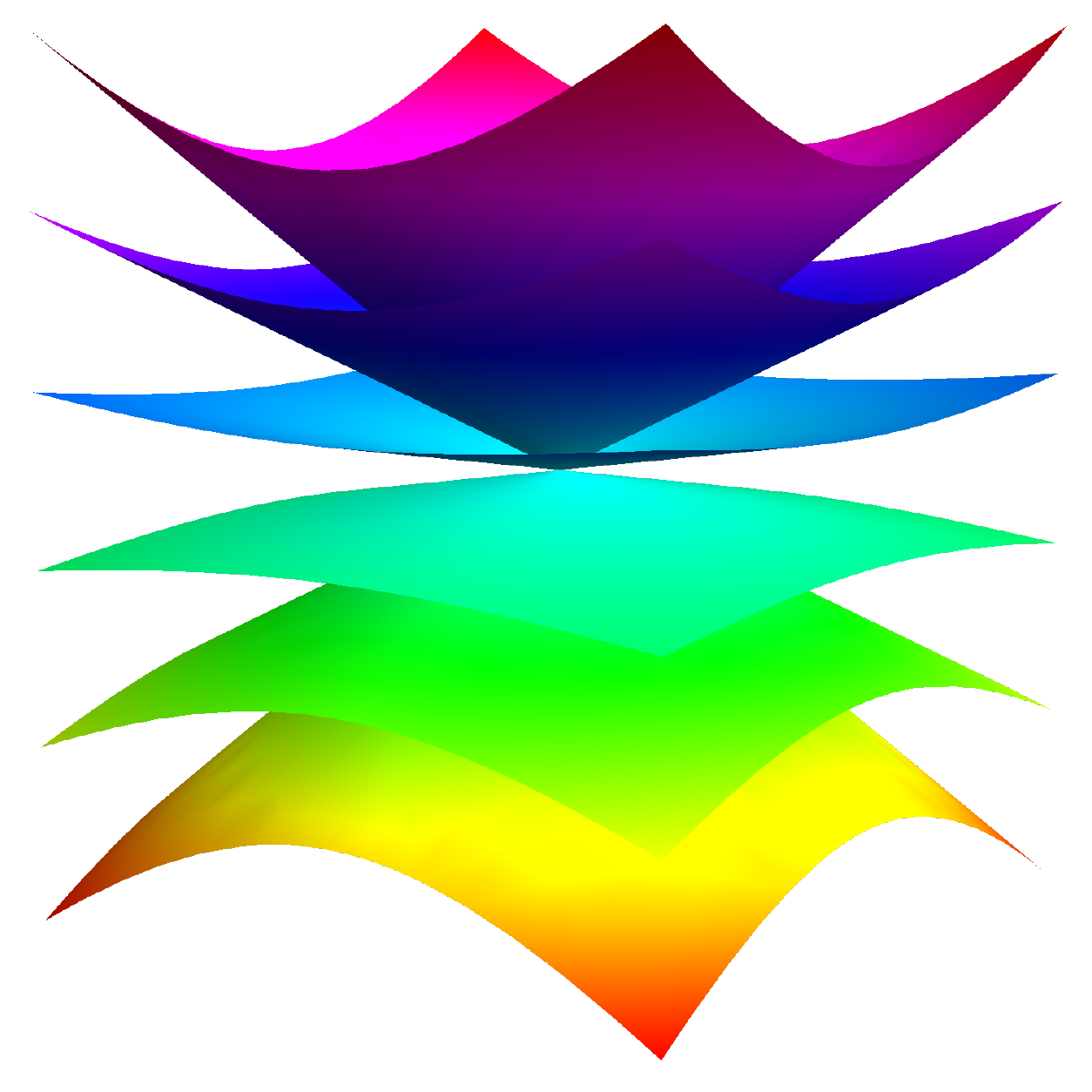}
\includegraphics[width=4cm]{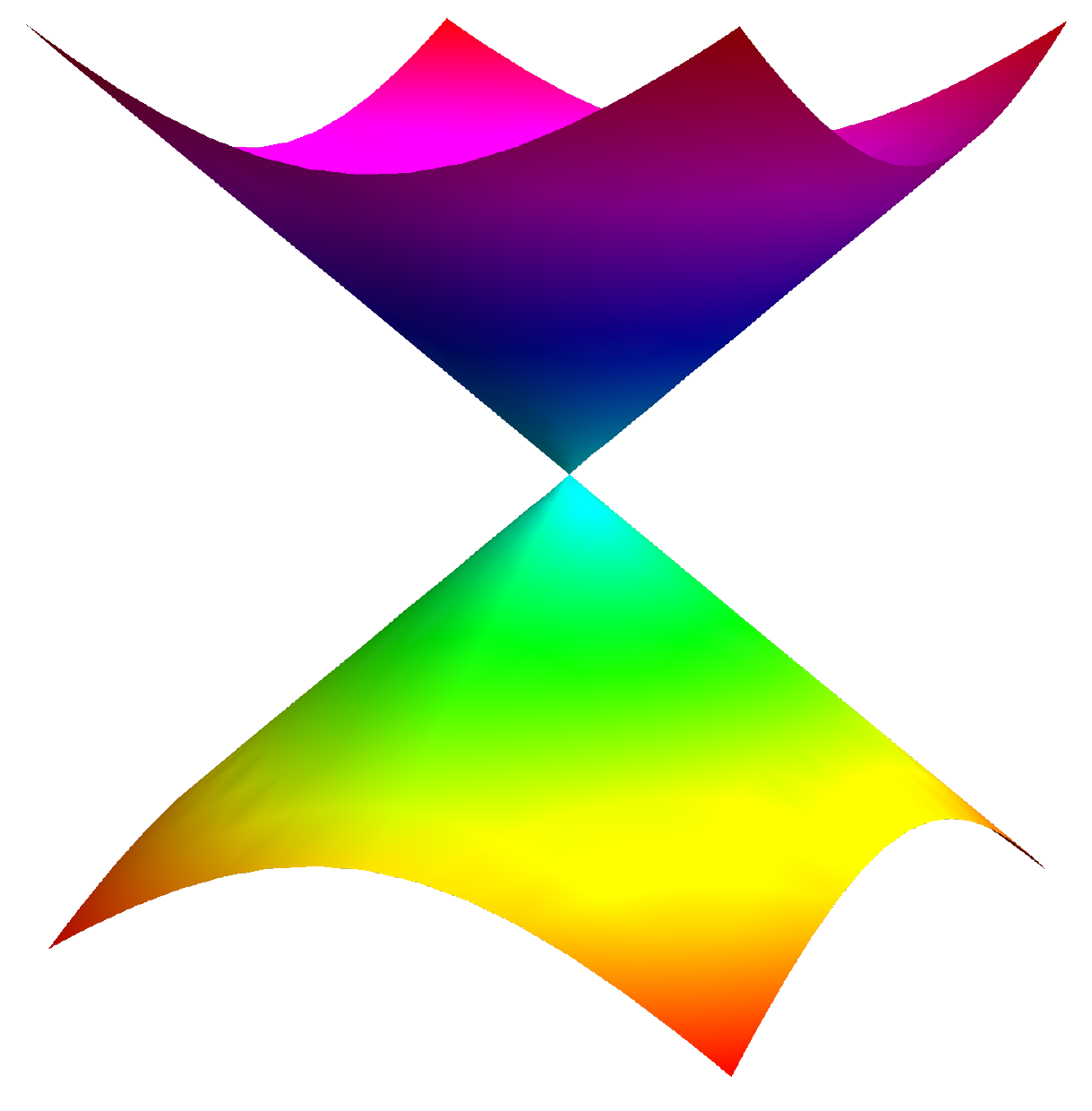}
\includegraphics[width=4cm]{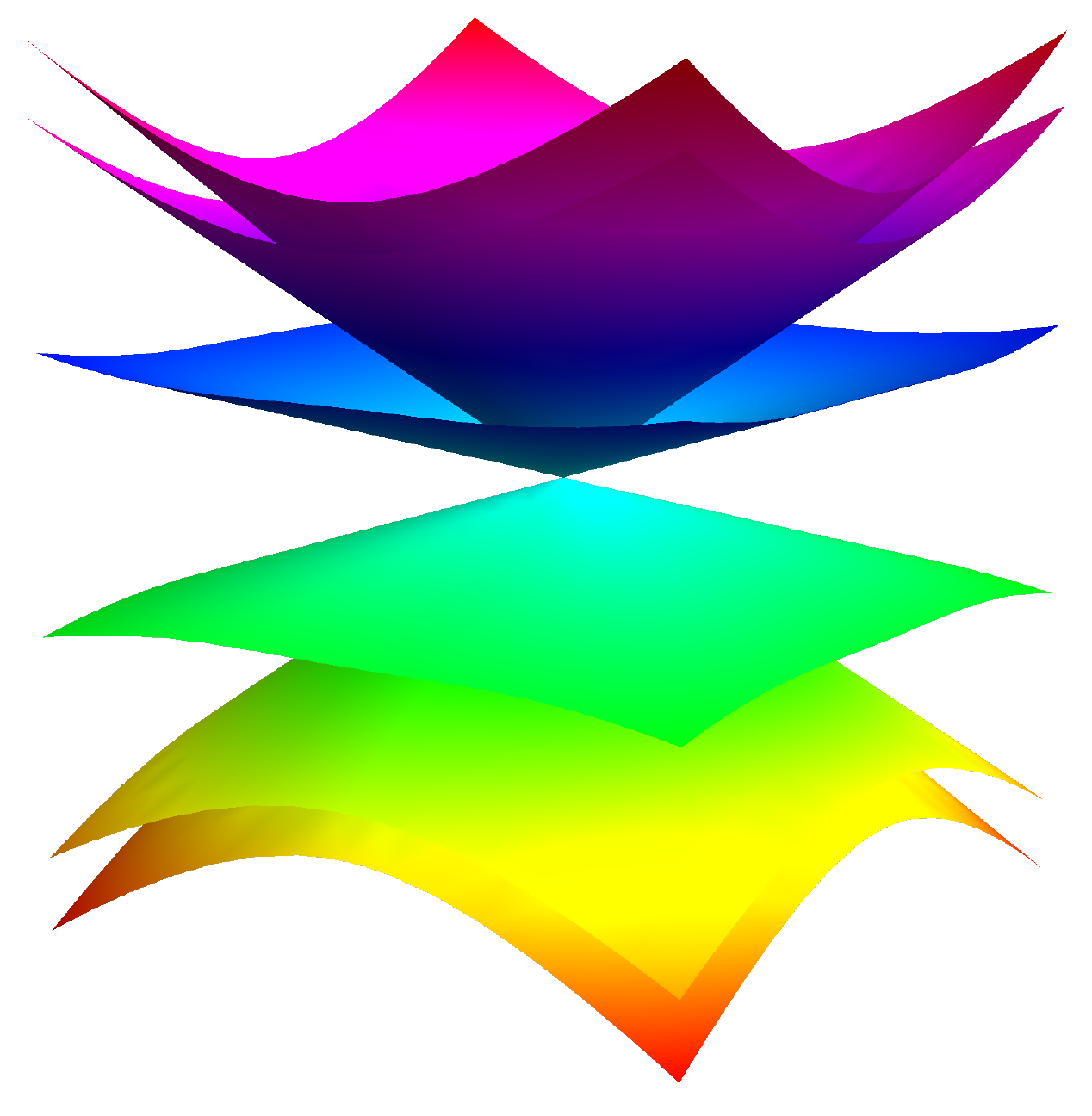}
\includegraphics[width=4cm]{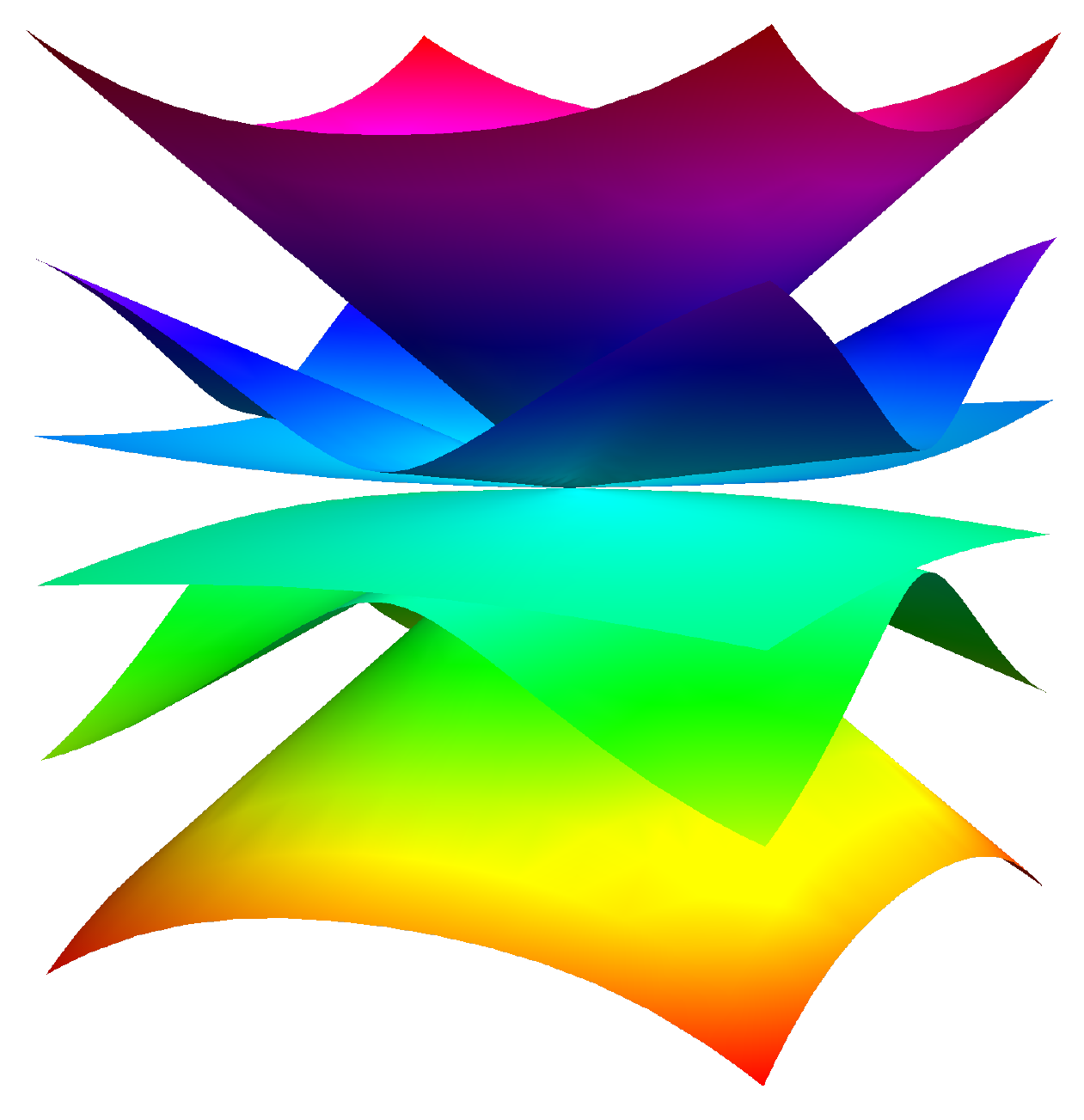}
\caption{Sixfold linear band crossing hosting spin-5/2 fermions in topological semimetals. We plot the energy bands for selected points of the parameters  $(\alpha,\beta,\gamma)$ entering the most general Hamiltonian in Eq. (\ref{high16}). \emph{Top left.} $\text{SO}(3)$-symmetric fixed point with single-particle Hamiltonian $H_{\text{SO}(3)} \propto p_i J_i$. This point is unstable towards inclusion of long-range interactions. \emph{Top right.} For $j=5/2$, three stable fixed points show an emergent relativistic $\text{O}(3)$ symmetry, comprising collections of Weyl and massless Dirac particles, see Tab. \ref{TabFP}. The six bands at these points are triply degenerate. \emph{Bottom left.} The system features a stable infrared fixed point without enhanced symmetry at $(\alpha,\beta,\gamma)=(1.172,-0.530,0)$. The existence of such "cubic" fixed points appears to be characteristic for higher-spin fermions. \emph{Bottom right.} For certain fine-tuned values of $(\alpha,\beta,\gamma)$, individual bands touch and lead to a change in the band topology. We show the dispersion for $(\alpha,\beta,\gamma)=(1,1,0)$, close to such a topological transition.}
\label{FigDispersion}
\end{minipage}
\end{figure}

\section{Higher-spin fermions}

\subsection{Lagrangian}\label{SecLag}

We consider a system of electrons with half-integer pseudospin $j$ at a linear band crossing point described the effective low-energy Lagrangian
\begin{align}
 \label{high1} L = \psi^\dagger\Bigl(\partial_\tau + H(-\rmi \nabla) +\rmi a\Bigr)\psi +\frac{1}{2\bar{e}^2} (\nabla a)^2,
\end{align}
with $\psi=(\psi_{j},\dots,\psi_{-j})^T$ a Grassmann field for the electrons and $\tau$ imaginary time. Long-range interactions are modeled by exchange of a real scalar photon $a$. The electric charge $\bar{e}$ appears in the kinetic part of the photon, but may also be put in front of the term $\psi^\dagger \rmi a \psi$ through a field redefinition $a \to \bar{e} a$. We assume the chemical potential to be at the crossing point. The model is defined with respect to an ultraviolet momentum cutoff $\Lambda$ so that the linear approximation to the Hamiltonian is valid for all momenta $p\leq \Lambda$. We define $e^2=\bar{e}^2/(2\pi^2)$ for later convenience.

The higher-spin nature of the linear band crossing point enters the Lagrangian through the $(2j+1)\times (2j+1)$ Hamiltonian matrix $H(\textbf{p})$.  The most general  cubic-symmetric Hamiltonian is a linear combination of $\mathcal{N}$ admissible terms $p_iK_i^{(1)},\dots, p_iK_i^{(\mathcal{N})}$ with certain matrices $K_i^{(n)}$. Identifying the value of $\mathcal{N}$ in dependence of $j$ is a crucial step in our analysis. For each $j$, the number $\mathcal{N}$ counts the number of irreducible vectors under the cubic group and is thus fixed by group theory. We discuss the procedure for finding $\mathcal{N}$ and the construction of the matrices $K_i^{(n)}$ for arbitrary $j$ in Appendix \ref{AppMat}. For instance, for $j=3/2$, the two irreducible vectors under the cubic group are 
\begin{align}
 \label{high9}  K_i^{(1)} &= \frac{2}{\sqrt{5}}J_i,\\
 \label{high10} K_i^{(2)} &= \frac{2\sqrt{5}}{3} \Bigl(J_i^3-\frac{41}{20}J_i\Bigr).
\end{align}
For $j=5/2$, assuming time-reversal symmetry, the four irreducible cubic vectors read
\begin{align}
 \nonumber K_i^{(1)} &= 2\sqrt{\frac{3}{35}}J_i,\\
 \nonumber K_i^{(2)} &=\frac{1}{3}\sqrt{\frac{5}{6}}\Bigl(J_i^3-\frac{101}{20}J_i\Bigr),\\
 \nonumber K_i^{(3)} &=\frac{1}{10}\sqrt{\frac{21}{2}}\Bigl(J_i^5 -\frac{145}{18}J_i^3 +\frac{11567}{1008}J_i\Bigr),\\
 \label{high14} K_i^{(4)} &= 3\sqrt{\frac{3}{10}}\Bigl(\{J_i,\hat{\mathcal{I}}\}+\frac{7}{18}J_i^5-\frac{95}{36}J_i^3+\frac{253}{96}J_i\Bigr),
\end{align}
where we define
\begin{align}
 \hat{\mathcal{I}} &=  \hat{\mathcal{I}}_{5/2} =\frac{1}{9}\Bigl(\sum_{k<l}(J_k^2J_l^2+J_l^2J_k^2) -\frac{283}{8}\mathbb{1}\Bigr).
\end{align}
The six matrices that can be constructed for $j=7/2$ are displayed in Eqs. (\ref{NewSpin1})-(\ref{NewSpin7}). We display a few more relevant values of $\mathcal{N}$ in Table. \ref{TabN}.

\renewcommand{\arraystretch}{1.8}
\begin{table}[t]
\begin{tabular}{|c|c|c|c|c|c|}
\hline
 $j$ & \ $3/2$ \ & \ $5/2$ \ & \ $7/2$ \ & \ $9/2$ \ & \ $11/2$ \ \\
\hline\hline
 $\mathcal{N}$ with $\mathcal{T}$-symmetry & 2 & 4 & 6 & 9 & 12 \\
\hline
 \ $\mathcal{N}$ without $\mathcal{T}$-symmetry \ & 2 & 5 & 8 & 13 & 18 \\
\hline
\end{tabular}
\caption{The number of terms $\mathcal{N}$ that transform as vectors under the cubic rotational group in the most general Hamiltonian $H(\textbf{p})$ is fixed by group theory. In this work, we restrict to time-reversal ($\mathcal{T}$) symmetric Hamiltonians, which feature a smaller number of admissible terms. The time-reversal operator $\mathcal{T}$ is defined in Eq. (\ref{sym1}).}
\label{TabN}
\end{table}
\renewcommand{\arraystretch}{1}

Although the Hamiltonian $H(\textbf{p})$ can be written in terms of the matrices $K_i^{(n)}$, practical calculations often simplify when using a different set of $\mathcal{N}$ orthonormal matrices that better implement the symmetries of the system. For $j=3/2$, these are the matrices $V_i$ and $U_i$ in Eqs. (\ref{int5}) and (\ref{int6}), with one real parameter $\alpha$. For $j=5/2$, we write the spin-5/2 Hamiltonian in terms of three real parameters $(\alpha,\beta,\gamma)$ as
\begin{align}
 \label{high16} H_{5/2}(\textbf{p}) = p_i(V_i+\alpha A_i +\beta B_i + \gamma C_i)
\end{align}
with orthonormal basis matrices
\begin{align}
 \nonumber V_i &= \sqrt{\frac{3}{35}}K_i^{(1)} + \frac{4}{3}\sqrt{\frac{2}{15}}K_i^{(2)}+\frac{8}{3}\sqrt{\frac{2}{21}}K_i^{(3)},\\
 \nonumber A_i &= \frac{4}{3}\sqrt{\frac{2}{35}}K_i^{(1)}+\frac{14}{9\sqrt{5}}K_i^{(2)}-\frac{25}{18\sqrt{7}}K_i^{(3)}-\frac{\sqrt{5}}{6}K_i^{(4)},\\
 \nonumber B_i &= \sqrt{\frac{32}{105}}K_i^{(1)}-\frac{2}{\sqrt{15}}K_i^{(2)}+\frac{1}{\sqrt{84}}K_i^{(3)}-\sqrt{\frac{5}{12}}K_i^{(4)},\\
 \label{high17} C_i &= \frac{4}{3}\sqrt{\frac{2}{7}}K_i^{(1)}-\frac{1}{9}K_i^{(2)}-\frac{2}{9}\sqrt{\frac{5}{7}}K_i^{(3)}+\frac{2}{3}K_i^{(4)}.
\end{align}
Their advantageous symmetry properties are explained in the next section. The $\text{O}(3)$ symmetric fixed point is reached for $(\alpha,\beta,\gamma)=(0,0,0)$.

More generally, we show below that matrices $V_i$ satisfying Clifford algebra can always be constructed as a linear combination of the $K_i^{(n)}$. Consequently, after an appropriate rescaling of momentum, the Hamiltonian for $j\geq 3/2$ can be written as
\begin{align}
 \label{high3b} H(\textbf{p}) =  p_i\Biggl(V_i+\sum_{n=1}^{\mathcal{N}-1}\alpha_n  U_i^{(n)}\Biggr),
\end{align}
with orthonormal matrices chosen such that
\begin{align}
 \label{high3c} &\mbox{tr}(V_kV_l) = (2j+1)\delta_{kl},\ \mbox{tr}(V_kU_l^{(n)}) = 0\\
 \label{high3d} &\mbox{tr}(U_k^{(n)}U_k^{(n')}) = (2j+1)\delta_{kl}\delta_{nn'}.
\end{align}
The number of independent velocity coefficients $\vec{\alpha}=(\alpha_1,\dots,\alpha_{\mathcal{N}-1})$ is $\mathcal{N}-1$. The $\text{O}(3)$ symmetric fixed point corresponds to $\vec{\alpha}=0$.

Let us now construct the matrices $V_i$ for arbitrary $j$. They can be expressed in a representation independent fashion in terms of linear combinations of odd powers of $J_i$. The procedure is explained in Appendix \ref{AppMat}. This also implies that they are odd under $\mathcal{T}$. Using the standard representation for the spin matrices $J_i$, Eqs. (\ref{mat1})-(\ref{mat4}), we obtain the block-diagonal form
\begin{align}
 \label{high4} V_1 &= \mathbb{A}_N \otimes \sigma_1 = \begin{pmatrix} 0 & 0 & \sigma_1 \\ 0 & \iddots & 0 \\ \sigma_1 & 0 & 0 \end{pmatrix},\\
 \label{high5} V_2 &= \mathbb{A}_N \otimes \bar{\sigma}_2 = \begin{pmatrix} 0 & 0 & \bar{\sigma}_2 \\ 0 & \iddots & 0 \\ \bar{\sigma}_2 & 0 & 0 \end{pmatrix},\\
 \label{high6} V_3 &= \mathbb{1}_N \otimes \sigma_3 = \begin{pmatrix} \sigma_3 & 0 & 0 \\ 0 & \ddots & 0 \\ 0 & 0 & \sigma_3 \end{pmatrix},
\end{align}
with $\bar{\sigma}_2=(-1)^{j-1/2}\sigma_2$ being either $\sigma_2$ or $\sigma_2^*$, $N=j+\frac{1}{2}$ the number of Weyl points for $\vec{\alpha}=0$, $\mathbb{1}_N$ the $N$-dimensional unit matrix, and $\mathbb{A}_N=\sigma_1\otimes\dots\otimes \sigma_1$ the $N\times N$ matrix whose nonzero entries are units along the antidiagonal. Using this representation of the matrices $V_i$, one immediately verifies the validity of $\{V_k,V_l\}=2\delta_{kl}$. An interesting consequence of the above form is that 
\begin{align}
 \nonumber &V_i \text{ commutes with } J_i,\\
 \label{high7} &V_i \text{ anticommutes with } J_{k\neq i}.
\end{align}
This, in turn, implies that $V_i$ commutes with $K_i^{(n)}$ and anticommutes with $K_{k\neq i}^{(n)}$ for every $n$ in the $\mathcal{T}$-symmetric case, since the $K_i^{(n)}$ are composed of sums of odd powers of spin matrices. Note that $V_i$ itself is a linear combination of the $K_i^{(n)}$.

At last, let us comment on the role of interactions in Eq. (\ref{high1}). In three spatial dimensions, for linear band crossing, the electric charge $e$ is a marginal coupling in the RG sense. Put differently, our system is right at the critical dimension for the coupling $e$, so we can apply perturbation theory in three dimensions without the need for introducing additional dimensions. This is very advantageous since the algebra of the spin matrices crucially depends on the underlying dimension of space. The small parameter in our RG scheme is the charge $e$. As $e$ will be shown to diminish under RG, a self-consistent scheme merely requires that $e$ is small at some initial RG scale.

\subsection{Symmetries}\label{SecSym}

In this section, we discuss the discrete symmetries of the system. We first define the time-reversal operator, which implies a particle-hole symmetric spectrum. We identify additional discrete symmetries that relate different parameter regimes of the couplings $\vec{\alpha}$ in Eq. (\ref{high3b}). We illustrative these additional symmetries for $j=3/2$ and $j=5/2$.

The anti-unitary time-reversal operator is given by
\begin{align}
 \label{sym1} \mathcal{T} = V_2 \mathcal{K},
\end{align}
where $\mathcal{K}$ denotes complex conjugation. We have $\mathcal{T}^2=-\mathbb{1}$ and 
\begin{align}
 \label{sym2} \{\mathcal{T},H(\textbf{p})\}=0
\end{align}
for every fixed $\textbf{p}$. Equation (\ref{sym2}) defines time-reversal symmetry of the free electron system. It implies that for each eigenvalue $E(\textbf{p})$ of $H(\textbf{p})$, there exists an eigenvalue $-E(\textbf{p})$ for the time-reversed eigenstate. To ensure time-reversal symmetry, we only allow those cubic vectors $K_i^{(n)}$ that satisfy $\{\mathcal{T},K_i^{(n)}\}=0$. It is easy to see that these are precisely those cubic vectors that are made from products of an odd number of spin matrices by employing the property (\ref{high7}) and the fact that $J_{1,3}$ are real, while $J_2$ is imaginary, implying $\{\mathcal{T},J_i\}=0$. Since the photonic part of the Lagrangian is also time-reversal invariant, the system described by Eq. (\ref{high1}) features time-reversal symmetry.

To understand the additional discrete symmetries in the case of $j=3/2$, consider the unitary operator \cite{PhysRevLett.124.127602}
\begin{align}
 \label{sym3} \hat{\mathcal{W}} =  \frac{2}{\sqrt{3}}(J_x J_y J_z + J_z J_y J_x),
\end{align}
which squares to unity. One easily verifies that $V_i$ and $U_i$ in Eqs. (\ref{int5}) and (\ref{int6}) satisfy
\begin{align}
 \label{sym4} [V_i,\hat{\mathcal{W}}]= \{U_i,\hat{\mathcal{W}}\} = 0.
\end{align}
Hence, $\hat{\mathcal{W}}(V_i+\alpha U_i) \hat{\mathcal{W}}= (V_i-\alpha U_i)$ and the spin-3/2 system features a global $\mathbb{Z}_2$ symmetry with respect to $\alpha \to -\alpha$. Indeed, a sign-change in $\alpha$ can be undone by a field transformation $\psi \to \hat{\mathcal{W}}\psi$, and so all physical observables are symmetric with respect to $\alpha$ for $j=3/2$.

How does this symmetry generalize to $j=5/2$? For this spin, define the unitary operators
\begin{align}
 \label{sym5} \hat{\mathcal{I}} &= \frac{1}{9}\Bigl(\sum_{k<l}(J_k^2J_l^2+J_l^2J_k^2) -\frac{283}{8}\mathbb{1}\Bigr),\\
 \label{sym6} \hat{\mathcal{W}} &= \frac{1}{3\sqrt{3}}(J_xJ_yJ_z+J_zJ_yJ_x)+\frac{1}{2}(\mathbb{1}+\hat{\mathcal{I}}).
\end{align}
Both of them square to unity, they mutually commute, and $\hat{\mathcal{I}}$ commutes with $\mathcal{T}$. The operator $\mathcal{\hat{W}}$ does not have a distinct behavior under $\mathcal{T}$, but the traceless analogue
\begin{align}
 \label{sym7} \mathcal{W} = \frac{1}{\sqrt{18}}(J_xJ_yJ_z+J_zJ_yJ_x)
\end{align}
anticommutes with $\mathcal{T}$. Note that $\hat{\mathcal{I}}$ is invariant under cubic transformations. Among the many possible linear combinations of $K_i^{(1)},\dots,K_i^{(4)}$ for spin 5/2, we use these symmetry operators to construct the mutually orthogonal matrices $V_i, A_i, B_i, C_i$ in Eq. (\ref{high16}) in the following way. First, $V_i$ is defined to satisfy the Clifford algebra in Eq. (\ref{int3}), thus has the form given in Eqs. (\ref{high4})-(\ref{high6}). One verifies that $V_i$ commutes with both $\hat{\mathcal{I}}$ and $\hat{\mathcal{W}}$. The matrix $C_i$ is uniquely constructed such that 
\begin{align}
 \label{sym8} [\hat{\mathcal{I}},A_i]=[\hat{\mathcal{I}},B_i]= \{\hat{\mathcal{I}},C_i\}=0.
\end{align}
This implies that the system has a global $\mathbb{Z}_2$-symmetry with respect to $\gamma \to -\gamma$. Next, the matrices $A_i$ and $B_i$ are chosen such that
\begin{align}
 \label{sym9} \{\hat{\mathcal{W}},A_i \} = [\hat{\mathcal{W}},B_i] =0,
\end{align}
implying that for $\gamma=0$ the systems is invariant under $\alpha \to -\alpha$. These symmetries are particularly useful in the RG analysis of the vast parameter space spanned by the couplings $(\alpha,\beta,\gamma)$, see Sec. \ref{SecFix52}.

As the value of $j$ is increased, the higher-spin analogues of the symmetry operators $\hat{\mathcal{I}}$ and $\hat{\mathcal{W}}$ can be constructed from the irreducible tensors that transform under the $A_1$ and $A_2$ representation, see Tab. \ref{TabIrrep}. We leave this task for potential future work. In this context, note that for $j=3/2$ we have $ \sum_{k<l} (J_k^2J_l^2+J_l^2J_k^2) = \frac{51}{8}\mathbb{1}$, thus the operator $\hat{\mathcal{I}}$ is trivial in this case. This is a consequence of the Cayley--Hamilton theorem applied to the spin-$3/2$ matrices, see the discussion in Appendix \ref{AppMat}.

\subsection{Band structure and topology}\label{SecTopo}

Higher-spin fermions described by the Hamiltonian $H(\textbf{p})$ in (\ref{high3b}) feature a rich band structure and complex band topology. In this section, we first discuss the topology of the band crossing point at the $\text{O}(3)$ and $\text{SO}(3)$ symmetric fixed points for arbitrary $j$. We then discuss some more detailed aspects of $j=3/2$ and $j=5/2$. The energies as a function of the parameters $\vec{\alpha}$ fully determine the infrared fixed points through Eqs. (\ref{rg1})-(\ref{rg5}), as is explained in the next section. For a brief review of the formulas relevant for topology see Appendix \ref{AppTopo}.

Let us first discuss the topology of the continuous fixed points for general $j$. At the $\text{O}(3)$ symmetric fixed point with $\vec{\alpha}=0$, the Hamiltonian is given by (\ref{int2}), with the matrices $V_i$ from Eqs. (\ref{high4})-(\ref{high6}). After a suitable basis change, the Hamiltonian can be brought into a block-diagonal form and corresponds to $N=j+1/2$ copies of Weyl Hamiltonians. Due to the appearance of $\bar{\sigma}_2$ in $V_2$, the chirality/monopole charge of each Weyl particle is $(-1)^{j-1/2}$, and the resulting total monopole charge is $\mathcal{Q}=(-1)^{j-1/2}N$. This basis change is explained below at the example of $j=5/2$ and easily generalized to arbitrary spin. At the $\text{SO}(3)$ symmetric fixed point, the Hamiltonian is given by Eq. (\ref{int1}). The energy bands are labeled by quantum numbers $m=-j,\dots,j$ and energies $E_m(\textbf{p}) = 2m p$. The Chern number of the $m$th band is $2m$, and so the total monopole charge, defined as the sum of the Chern numbers of the positive energy bands, is given by $N^2$.

To discuss the band structure for spin $j$, we use the following convention. Due to time-reversal symmetry expressed by Eq. (\ref{sym2}), the energy spectrum is particle-hole symmetric. We can then focus on the $j+1/2$ positive energy bands, which we label $E_1(\textbf{p}),\dots,E_{j+1/2}(\textbf{p})$, with the $\vec{\alpha}$-dependence implicit. With this notation, the energy bands for $j=3/2$ read
\begin{align}
 \nonumber &E_{1,2}(\textbf{p}) = \mathcal{E}_\pm^{(3/2)}(\textbf{p},\alpha) \\
 \label{band1} &= \sqrt{(1+\alpha^2)p^2\pm \Biggl(\alpha^2 \Bigl[ 4 p^4+3(\alpha^2-4)\sum_{k<l}p_k^2p_l^2\Bigr]\Biggr)^{1/2}}.
\end{align}
We restrict to $\alpha \geq 0$ because of the $\mathbb{Z}_2$ symmetry discussed before. For $\alpha=0$ and $\alpha=2$, we recover the dispersion at the $\text{O}(3)$ and $\text{SO}(3)$ symmetric fixed points, respectively, given by $E_{1,2}=p$ and $E_{1,2}=3p,\ p$. For each band, we compute the Chern number $\mathcal{C}$ as described in Appendix \ref{AppTopo}. We define the total monopole charge $\mathcal{Q}$ as the sum of the Chern numbers of the positive bands. One finds \cite{PhysRevLett.124.127602} that the total monopole charge is $-2$ for $\alpha<1$, whereas it is $4$ for $\alpha>1$. At the particular point $\alpha=1$, the band structure undergoes a topological transition.

The topological transition at $\alpha=1$ for $j=3/2$ is characteristic for the behavior found extensively for larger $j$, and therefore deserves some additional comments. For a change of the topology of the band structure, reflected by a change in the Chern numbers of the individual bands, the energy bands $E_\lambda(\textbf{p})$ of $H(\textbf{p})$ must intersect at some point in momentum space. Typically these intersection points describe Weyl fermions. The Hamiltonians we consider here, however, are linear in momentum and thus scale invariant. This implies that there cannot be a single value of $\textbf{p}_0$ where bands intersect, but rather bands cross along \emph{lines} in momentum space originating from the origin. By including terms into the Hamiltonian that are quadratic in momentum, the line nodes shrink to point nodes, but the change in topology persists. Since the quadratic terms are irrelevant for the low-energy physics, they are of no further relevance for this work.

\begin{figure}[t!]
\centering
\includegraphics[width=8.5cm]{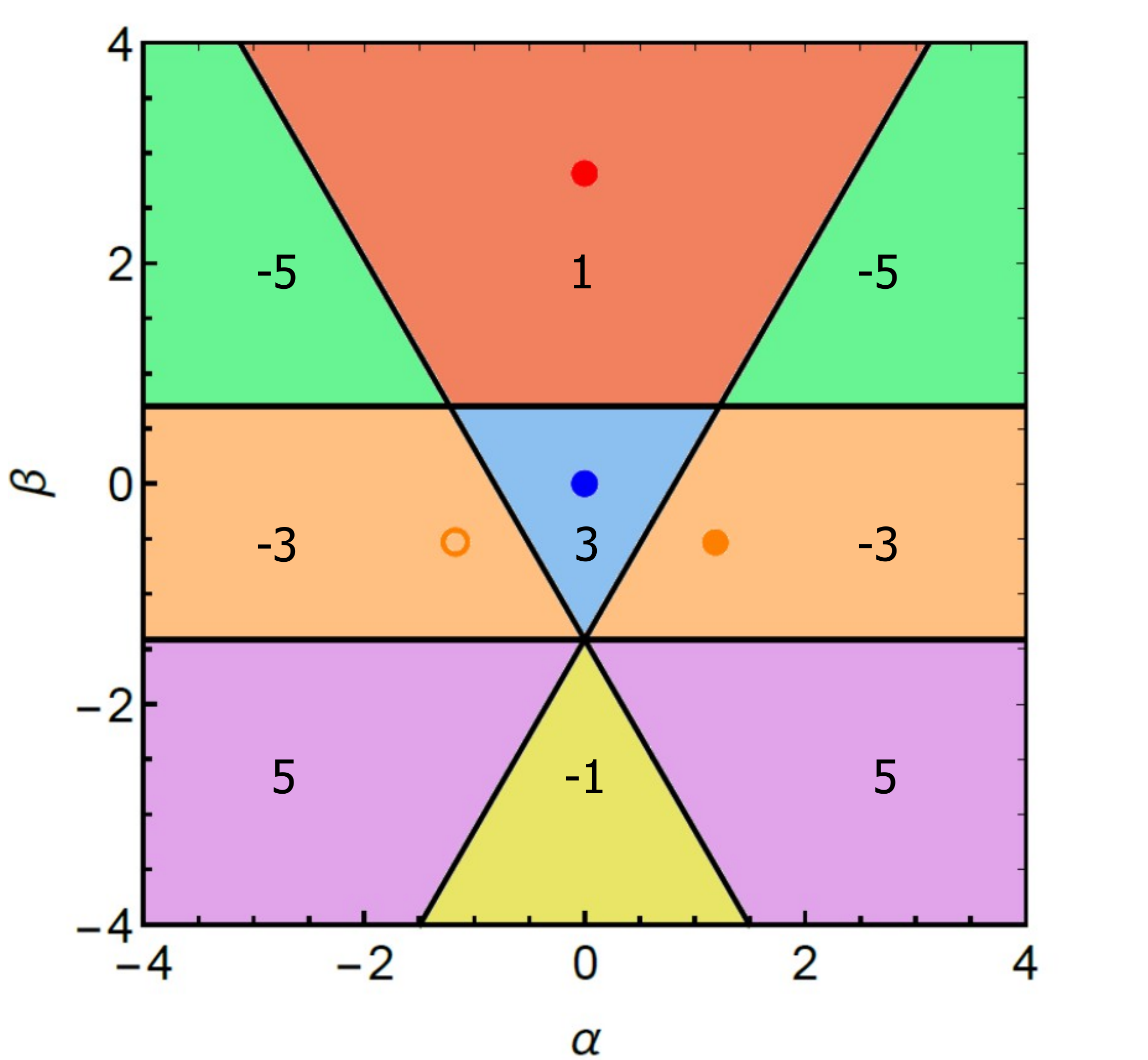}
\caption{Band topology of spin-5/2 fermions described by the Hamiltonian (\ref{high16}) in the $(\alpha,\beta)$ plane for $\gamma=0$. The distinct topological sectors are labeled by the total monopole charge $\mathcal{Q}$, defined as the sum of the Chern numbers of the positive energy bands. Note that all values are odd integers. Also note that for $\gamma=0$ the system is invariant under $\alpha \to -\alpha$. We show the stable $\text{O}(3)$ symmetric fixed point at $(\alpha,\beta)=(0,0)$ (blue dot), which comprises three Weyl particles of chirality $+1$ and total monopole charge $3$. The stable $\text{O}(3)$ symmetric fixed point at $(\alpha,\beta)=(0,2\sqrt{2})$ (red dot) features a Weyl fermion and a massless Dirac particle, and so has monopole charge $1$.}
\label{FigTopo}
\end{figure}

Due to cubic symmetry, the lines of topological transition are either along the $(1,0,0)$, $(1,1,0)$, or $(1,1,1)$ directions (and equivalent ones). In our example in Eq. (\ref{band1}), for $j=3/2$ and $\alpha=1$, the energy band $E_2(\textbf{p})$ vanishes along the $(1,0,0)$ direction, where it crosses with the negative energy band $-E_2(\textbf{p})$.  If either two positive or two negative energy bands touch, they may exchange Chern numbers, but they cannot change the total monopole charge. Hence $\mathcal{Q}$ can only change when a positive and a negative band intersect. For $j=5/2$, the total monopole charge in the $(\alpha,\beta)$ plane for $\gamma=0$ is shown in Fig. \ref{FigTopo}.

The spin 5/2 system with Hamiltonian (\ref{high16}) is invariant under $\gamma \to -\gamma$. For $\gamma=0$, we write $\alpha = \sqrt{2/3}\bar{\alpha},\ \beta = \lambda/\sqrt{2}$ and find
\begin{align}
 \label{band7} &E_1(\textbf{p}) = \pm (1-\lambda) p,\\
 \nonumber &E_{2,3}(\textbf{p}) =  \Biggl[\Bigl(\Bigl(1+\frac{\lambda}{2}\Bigr)^2+\bar{\alpha}^2\Bigr)p^2 \\
 \label{band8} &\pm \Biggl(\bar{\alpha}^2 \Bigl[(2+\lambda)^2p^4 +3\Bigl(\bar{\alpha}^2-(2+\lambda)^2\Bigr)\sum_{k<l}p_k^2p_l^2\Bigr]\Biggr)^{\frac{1}{2}}\Biggr]^{\frac{1}{2}}.
\end{align}
The band structure for $\gamma\neq 0$ cannot be expressed in closed form. However, the $\text{SO}(3)$ symmetric fixed point with energies $E_{1,2,3}=5p, 3p, p$ is located at
\begin{align}
 \alpha = \frac{1}{3}\kappa_{\rm c},\ \beta= \frac{1}{\sqrt{3}}\kappa_{\rm c},\ \gamma= \frac{\sqrt{5}}{3}\kappa_{\rm c},
\end{align}
with $\kappa_{\rm c} = 4\sqrt{2/3}$ defined in Eq. (\ref{rot2}). In the special case of $\beta=\gamma=0$ we have
\begin{align}
 \label{band3} E_1(\textbf{p}) &= p,\\
 \label{band4} E_{2,3}(\textbf{p}) &= \mathcal{E}_\pm^{(3/2)}(\textbf{p},\bar{\alpha}),
\end{align}
with the last line having the same form as in Eq. (\ref{band1}). On the other hand, for $\alpha=\gamma=0$ we find
\begin{align}
  \label{band5} E_1(\textbf{p}) &= |1-\lambda| p,\\
  \label{band6} E_2(\textbf{p}) &= E_3(\textbf{p}) = \Bigl|1+\frac{\lambda}{2}\Bigr|p.
\end{align}
The second and third band are degenerate in this limit.

The $j=5/2$ system features two fixed points with $\text{O}(3)$ symmetry that are distinct from the one at $\vec{\alpha}=0$. To see the difference, let us first discuss the conventional one with $\vec{\alpha}=0$. We shuffle the fermion components from $(\psi_{5/2},\psi_{3/2},\psi_{1/2},\psi_{-1/2},\psi_{-3/2},\psi_{-5/2})$ to $(\psi_{5/2},\psi_{-5/2},\psi_{3/2},\psi_{-3/2},\psi_{1/2},\psi_{-1/2})$ by means of the basis change matrix $\mathcal{S}_1$ in Eq. (\ref{band9}). In the new basis, the matrices $\tilde{V}_i=\mathcal{S}_1 V_i \mathcal{S}_1^{-1}$ read
\begin{align}
 \label{band10} \tilde{V}_1 = \begin{pmatrix} \sigma_1 & 0 & 0 \\ 0 & \sigma_1 & 0 \\ 0 & 0 & \sigma_1 \end{pmatrix},\ \tilde{V}_{2,3} = \begin{pmatrix} \sigma_{2,3} & 0 & 0 \\ 0 & -\sigma_{2,3} & 0 \\ 0 & 0 & \sigma_{2,3} \end{pmatrix}.
\end{align}
This block diagonal form makes the factorization into three Weyl Hamiltonians particularly transparent. Importantly, for a $2\times 2$ Weyl Hamiltonian of the form $H_{2\times 2} = \sum_{i} v_i p_i \sigma_i$, the chirality or monopole charge is given by $\sgn(v_1v_2v_3)$. Therefore, all three Weyl particles described by Eq. (\ref{band10}) carry equal chirality $1$ and the total monopole charge at this fixed point is $3$.

It is easy to find the remaining $\text{O}(3)$ symmetric fixed points by means of an ansatz $R_i = \frac{1}{c}(V_i +\alpha A_i + \beta B_i + \gamma C_i)$ and determination of the parameters $c$ and $(\alpha,\beta,\gamma)$ such that $\{R_k,R_l\}= 2\delta_{kl}$. We identify two solutions with $\vec{\alpha}\neq 0$, which are strikingly different in nature. First, for $(\alpha,\beta,\gamma)=(0,2\sqrt{2},0)$, the Hamiltonian can be written as
\begin{align}
 \label{band11} H_{\star}^{(1)}(\textbf{p}) = 3 p_i R_i^{(1)} 
\end{align}
with
\begin{align}
 \label{band12} R_i^{(1)} =\frac{1}{3}(V_i+2\sqrt{2}B_i).
\end{align}
Applying the basis change matrix $\mathcal{S}_2$ from Eq. (\ref{band9b}), the transformed matrices $\tilde{R}_i^{(1)}=\mathcal{S}_2 R_i^{(1)} \mathcal{S}_2^{-1}$ are
\begin{align}
 \label{band13} \tilde{R}_1^{(1)} = \begin{pmatrix} \sigma_1 & 0 & 0 \\ 0 & -\sigma_1 & 0 \\ 0 & 0 & \sigma_1 \end{pmatrix},\ \tilde{R}_{2,3}^{(1)} = \begin{pmatrix} \sigma_{2,3} & 0 & 0 \\ 0 & \sigma_{2,3} & 0 \\ 0 & 0 & \sigma_{2,3} \end{pmatrix}.
\end{align}
Note the difference in the number of minus signs in comparison to $\tilde{V}_i$. The Hamiltonian $H_{\star}^{(1)}(\textbf{p})$ describes two Weyl fermions of chirality $+1$ and one Weyl fermion of chirality $-1$. The two Weyl fermions of opposite chirality are equivalent to a four-component massless Dirac fermion. The third fixed point Hamiltonian with $\text{O}(3)$ symmetry reads
\begin{align}
 \label{band14} H_\star^{(2)}(\textbf{p}) = 3 p_i R_i^{(2)}
\end{align}
with
\begin{align}
 \label{band15} R_i^{(2)} = \frac{1}{3}\Bigl( V_i - 2\sqrt{\frac{2}{3}}A_i + \frac{4}{\sqrt{3}}C_i\Bigr).
\end{align}
The basis change with matrix $\mathcal{S}_3$ from Eq. (\ref{band9c}) yields the transformed matrices $\tilde{R}_i^{(2)}=\mathcal{S}_3 R_i^{(2)} \mathcal{S}_3^{-1}$ given by
\begin{align}
 \label{band16} \tilde{R}_{1,3}^{(2)} =\begin{pmatrix} \sigma_{1,3} & 0 & 0 \\ 0 & \sigma_{1,3} & 0 \\ 0 & 0 & \sigma_{1,3} \end{pmatrix},\ \tilde{R}_2^{(2)} = \begin{pmatrix} -\sigma_2 & 0 & 0 \\ 0 & -\sigma_2 & 0 \\ 0 & 0 & -\sigma_2 \end{pmatrix}.
\end{align}
We conclude that $H_\star^{(2)}(\textbf{p})$ describes three Weyl fermions of equal chirality $-1$ and total monopole charge $-3$.

\section{Renormalization group}

\subsection{Fermion self-energy and free energy}

\begin{figure}[t!]
\centering
\includegraphics[width=8.5cm]{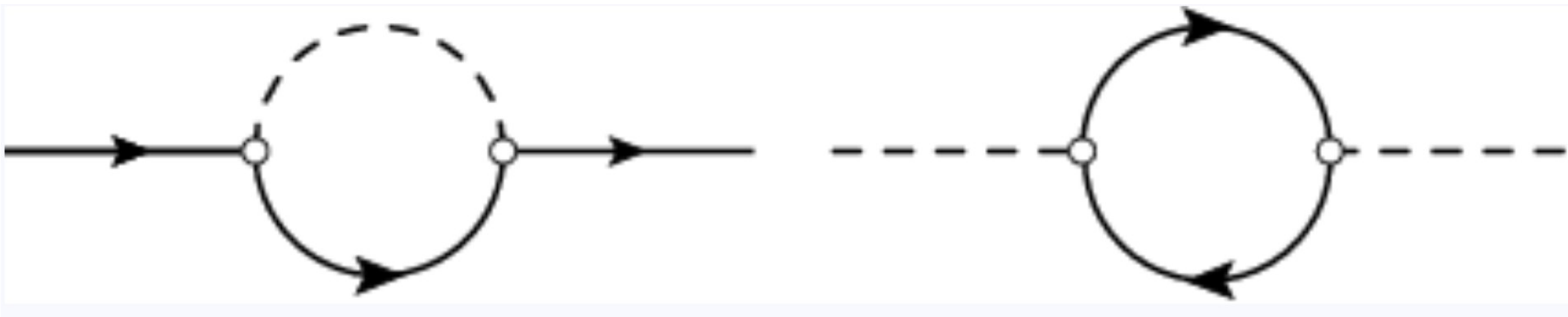}
\caption{One-loop diagrams contributing to the RG flow of the fermion self-energy (left) and photon self-energy (right). Here a continuous line depicts a fermion propagator and a dashed line depicts a photon propagator. The perturbative treatment is justified by a small value of the running charge $e$. We show that the beta function of $e^2$ is negative and, hence, its value decreases under RG. Consequently, as long as $e$ is small at some initial RG scale, it will always be small.}
\label{FigDiagrams}
\end{figure}

The RG flow of the system parameters $e$ and $\vec{\alpha}$ is induced by integrating out fluctuations in the momentum shell $\Lambda/b \leq p \leq \Lambda$ with $\Lambda$ the ultraviolet cutoff and $b>1$ \cite{herbutbook}, thereby effectively decreasing the ultraviolet cutoff and creating a running of the couplings $e(b)$ and $\vec{\alpha}(b)$. The flow of $e$ follows from the renormalization of the photon propagator, the fermion anomalous dimension $\eta$ and the flow of $\vec{\alpha}$ follow from the fermion self-energy. Both one-loop diagrams are depicted in Fig. \ref{FigDiagrams}. In this section, we focus on the fermion self-energy, assuming a small value of $e^2$. We confirm this assumptions in the next section.

For the present system, there exists an interesting connection between the fermion self-energy and the free energy at the one-loop level. The relation is the following: Write the eigenvalues of the single-particle Hamiltonian $H(\textbf{p})$ as $E_\lambda(\textbf{p}) = p \hat{E}_\lambda(\phi,\theta)$, with $\phi,\theta$ the usual angular variables in spherical coordinates. The eigenvalues depend on the  parameters $\vec{\alpha}$. Now define the dimensionless function
\begin{align}
 \label{rg1} f(\vec{\alpha}) = \sum_{\lambda=1}^{2j+1} \int_\Omega |\hat{E}_\lambda|,
\end{align}
with angular average
\begin{align}
 \label{rg2} \int_\Omega(\dots) = \frac{1}{4\pi}\int_0^{2\pi} \mbox{d}\phi\int_0^\pi\mbox{d}\theta\ \sin\theta\ (\dots).
\end{align}
It is easy to see that $f$ is simply (minus) the normal state free energy density at zero temperature given by
\begin{align}
 \label{rg3} F = -\frac{\Lambda^4}{32\pi^2} f.
\end{align} 
On the other hand, the RG flow equations for the velocity parameters $\vec{\alpha}$ are given by
\begin{align}
 \label{rg4} \dot{\alpha}_n = -\eta \alpha_n + \frac{e^2}{3(2j+1)} \frac{\partial f}{\partial \alpha_n} := h_n(\vec{\alpha}) e^2,
\end{align}
and the fermion anomalous reads
\begin{align}
 \label{rg5} \eta = \frac{e^2}{3(2j+1)} \Bigl( f(\vec{\alpha}) - \vec{\alpha}\cdot \frac{\partial f}{\partial \vec{\alpha}}\Bigr).
\end{align}
Here, a dot denotes a derivative with respect to $\log b$. For example, for $j=3/2$ we have
\begin{align}
 \label{rg5b} \dot{\alpha} = h(\alpha)e^2 = \frac{e^2}{12}\Bigl[ - f(\alpha)\alpha+(1+\alpha^2)f'(\alpha)\Bigr].
\end{align}
We derive these relations in Appendix \ref{AppSelf}.

Above equations have an interesting implication for the anomalous dimension at \emph{any} infrared fixed point of $\vec{\alpha}$. To see this, let $\vec{\alpha}_\star$ be a solution which satisfies $\dot{\alpha}_n=0$ for all $n$ and denote the corresponding anomalous dimension by $\eta_\star=\eta(\vec{\alpha}_\star)$. We have $h_n(\vec{\alpha}_\star)=0$ and 
\begin{align}
 \label{rg6}  0 =\vec{\alpha}_\star \cdot \vec{h}(\vec{\alpha}_\star) = -\eta_\star \vec{\alpha}_\star^2 + \frac{e^2}{3(2j+1)}\vec{\alpha}_\star \cdot  \frac{\partial f}{\partial \vec{\alpha}}(\vec{\alpha}_\star).
\end{align}
This implies the anomalous dimension at the fixed point to be
\begin{align}
 \label{rg7} \eta_\star= \frac{e^2}{3(2j+1)} \frac{f(\vec{\alpha}_\star)}{1+\vec{\alpha}^2_\star}.
\end{align}
Below, we use this relation to compute $\eta_\star$ for the relativistic and rotational invariant fixed points for arbitrary values of $j$. Unitarity requires $\eta>0$ and so we exclude regions of negative anomalous dimension as unphysical.

The stability of a given fixed point $\vec{\alpha}_\star$ is determined by the eigenvalues of the stability matrix
\begin{align}
 \label{rg8} M_{nn'} = \frac{\partial \dot{\alpha}_n}{\partial \alpha_{n'}}\Bigr|_{\vec{\alpha}=\vec{\alpha}_\star}= e^2 \frac{\partial h_n}{\partial \alpha_{n'}}(\vec{\alpha}_\star).
\end{align}
Stable fixed points feature only negative eigenvalues. Every positive eigenvalue corresponds to a repulsive direction in the parameter space spanned by $\{\alpha_n\}$. We denote the eigenvalues of $M$ by $\{\theta_n\}$. Note that Eqs. (\ref{rg4}) and (\ref{rg5}) imply that
\begin{align}
 \label{rg9} M_{nn'} = -\eta \delta_{nn'} + \frac{e^2}{3(2j+1)}\Bigl[ \alpha_n\vec{\alpha}\cdot \frac{\partial^2 f}{\partial \vec{\alpha}\partial \alpha_{n'}}+\frac{\partial^2 f}{\partial \alpha_n \partial \alpha_{n'}}\Bigr].
\end{align}
We derive this formula and a second method to compute $M$ in Appendix \ref{AppStab}.

\subsection{Charge renormalization}

Our perturbative RG analysis is built on the assumption that the charge $e>0$ remains small during the RG flow. In this section, we show that this is guaranteed for every $j$ and every choice of parameters $\vec{\alpha}$ as long as $e$ is small at some initial ultraviolet scale. For instance, the effective microscopic electric charge may be suppressed by a large dielectric constant. Furthermore, we explain an intriguing connection between the flow of the charge $e$ and the topology of the band structure.

Since the charge $e$ appears in the Lagrangian (\ref{high1}) as a prefactor of the photon kinetic term, charge renormalization is equivalent to the renormalization of the photon propagator, and therefore captured by the right diagram in Fig. \ref{FigDiagrams}. The flow equation for the charge can be written as
\begin{align}
 \label{rg10} \frac{\mbox{d}e^2}{\mbox{d}\log b} = -\eta e^2 - P(\vec{\alpha}) e^4,
\end{align}
where both $\eta$ and $P$ depend on the values of the couplings $\vec{\alpha}$. We now show that $P>0$. Together with a positive anomalous dimension this implies that $e\to 0$ as $b\to \infty$. This justifies the perturbative treatment in the above sense. However, it does not imply that the overall RG flow is trivial. Indeed, for every finite $b$, we have a nonvanishing charge $e>0$ that induces a renormalization of the band structure through fermion self-energy corrections, and so attracts the values of $\vec{\alpha}$ to an infrared fixed point or may result in a runaway flow. On the other hand, since $e_\star=0$ strictly at the fixed point, all critical exponent are trivial at the quantum critical points described here.

Denote again the eigenvalues and eigenvectors of $H(\textbf{p})$ in Eq. (\ref{high3b}) by $E_\lambda= p \hat{E}_\lambda$ and $|\lambda\rangle$, respectively, so that $H(\textbf{p})|\lambda\rangle = E_\lambda |\lambda\rangle$. We prove in Appendix \ref{AppPhot} that 
\begin{align}
 \label{rg11} P(\vec{\alpha}) =  2 \sum_{E_\lambda>0}\sum_{E_{\lambda'}<0}  \int_\Omega \frac{|\langle \lambda| h_3|\lambda'\rangle |^2}{(\hat{E}_{\lambda}-\hat{E}_{\lambda'})^3}
\end{align}
with $h_3= \partial H(\textbf{p})/\partial p_3 = V_3 + \sum_n \alpha_n U_3^{(n)}$. The sum extends over the positive and negative eigenvalues, respectively. This expression for $P$ is manifestly positive and is valid (perturbatively) for every time-reversal symmetric Hamiltonian linear in momentum. It fails, however, if the energy dispersion contains terms that are of higher power in momentum. Most famously, of course, the quadratic band touching Luttinger Hamiltonian inserted into the Lagrangian (\ref{high1}) features a stable Abrikosov fixed point close to four dimensions with $e_\star >0$ \cite{abrikosov,abrben}.

Some particular values of $P(\vec{\alpha})$ can be computed analytically. For the relativistic $\text{O}(3)$ symmetric fixed point at $\vec{\alpha}=0$ we have
\begin{align}
 \label{rg12} P(\vec{0}) = \frac{2j+1}{12}.
\end{align}
For the rotational $\text{SO}(3)$ symmetric fixed point we have $P_\star = \frac{2}{3}$ for $j=3/2$, and $P_\star = \frac{3}{2}$ for $j=5/2$.

\begin{figure}[t!]
\centering
\begin{minipage}{0.48\textwidth}
\includegraphics[width=\textwidth]{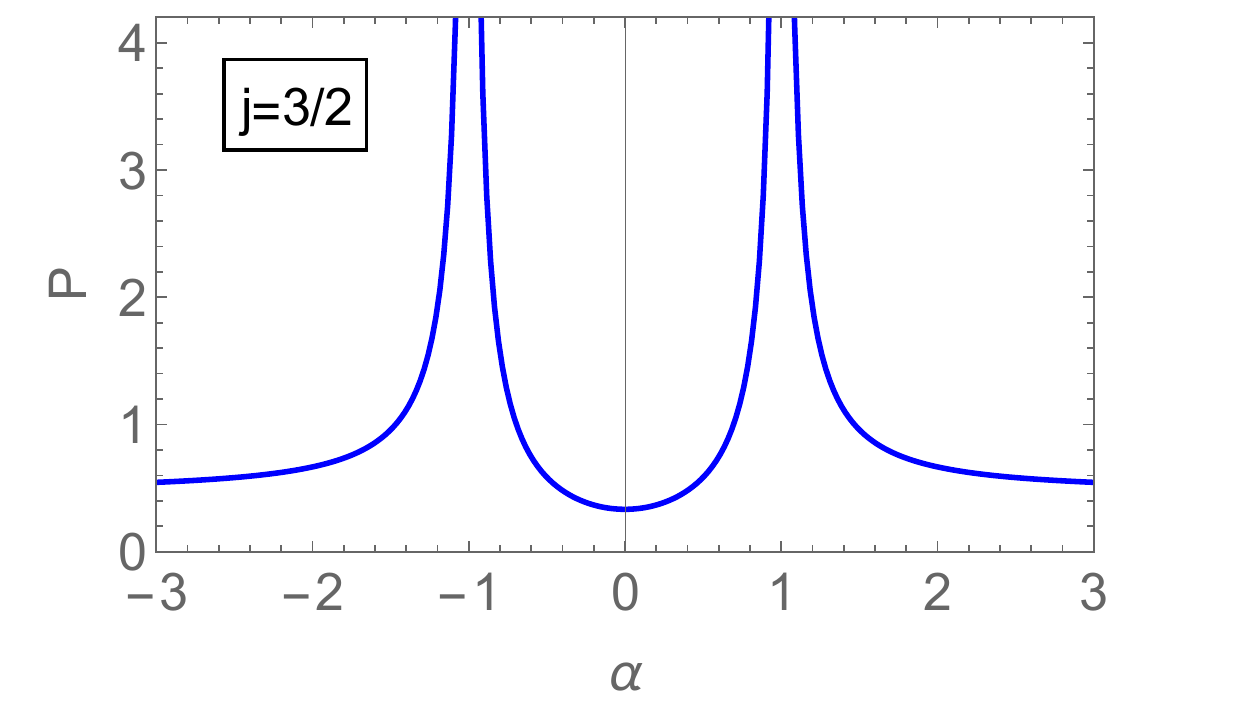}
\includegraphics[width=\textwidth]{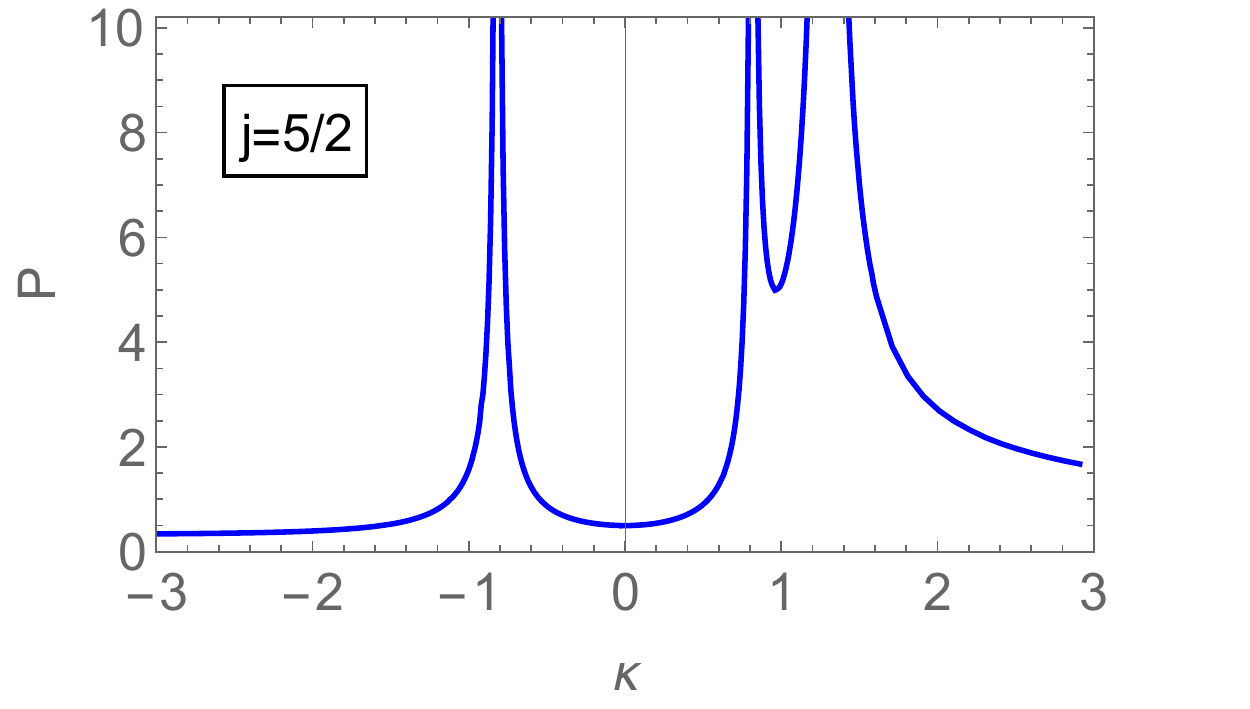}
\caption{Charge renormalization due to long-range interactions. The function $P(\vec{\alpha})$ is defined in Eq. (\ref{rg10}). A positive $P$ implies that the renormalized charge $e$ flows to zero in the infrared. The function features poles at those points in parameter space where the total monopole charge changes. For $j=3/2$ (upper panel), $P$ is symmetric in $\alpha$ and diverges at $\alpha=\pm1$. For $j=5/2$ (lower panel), we plot $P$ along the line $(\alpha,\beta,\gamma) = \frac{1}{3}(1,\sqrt{3},\sqrt{5})\kappa$, which connects the $\text{O}(3)$ symmetric fixed point at $\kappa=0$ with the $\text{SO}(3)$ symmetric fixed point at $\kappa_{\rm c}=4\sqrt{2/3}$. The restricted Hamiltonian along this line reads $H_{\rm 5/2}(\textbf{p})=p_i(V_i+\kappa U_i)$ with $U_i$ from Eq. (\ref{rot3}). The monopole charge changes for $\kappa=\pm \sqrt{2/3}$ and $\kappa=2^{7/6}/\sqrt{3}$.}
\label{FigP}
\end{minipage}
\end{figure}

Equation (\ref{rg11}) reveals an intriguing interplay between band topology and charge renormalization. For generic values of $\vec{\alpha}$, the energy bands $E_\lambda(\textbf{p})$ are distinct for all values of $\textbf{p}$ and the denominator of the integral is nonzero, implying $P<\infty$.  However, recall from the discussion in section \ref{SecTopo} that the parameter space $\vec{\alpha}$ is divided into topological sectors classified by the total monopole charge $\mathcal{Q}$. The latter changes when a positive and negative energy band intersect, which results in a divergence in the denominator of Eq. (\ref{rg11}). Consequently, $P\to +\infty$ and $e\to 0$ at these points of topological band transition and thus the flow of $\vec{\alpha}$ is effectively stopped. Indeed, the perturbative inclusion of long-range interactions cannot modify the band topology. The functions $\vec{h}=\dot{\vec{\alpha}}/e^2$ remain regular at the topological transitions. Two examples of $P$ for $j=3/2$ and $j=5/2$ are shown in Fig. \ref{FigP}.

\section{Infrared fixed points}

\subsection{O(3) symmetric fixed point for $\vec{\alpha}=0$}

\begin{figure}[t!]
\centering
\begin{minipage}{0.48\textwidth}
\includegraphics[width=1.05\textwidth]{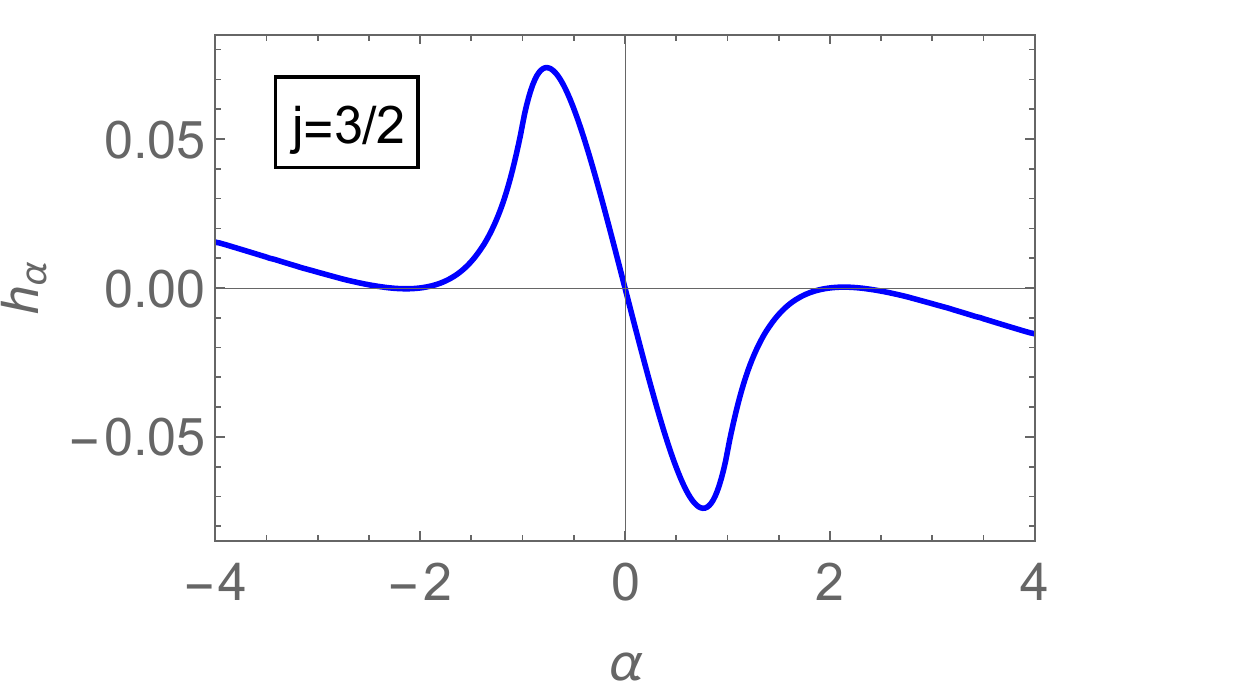}
\includegraphics[width=1.05\textwidth]{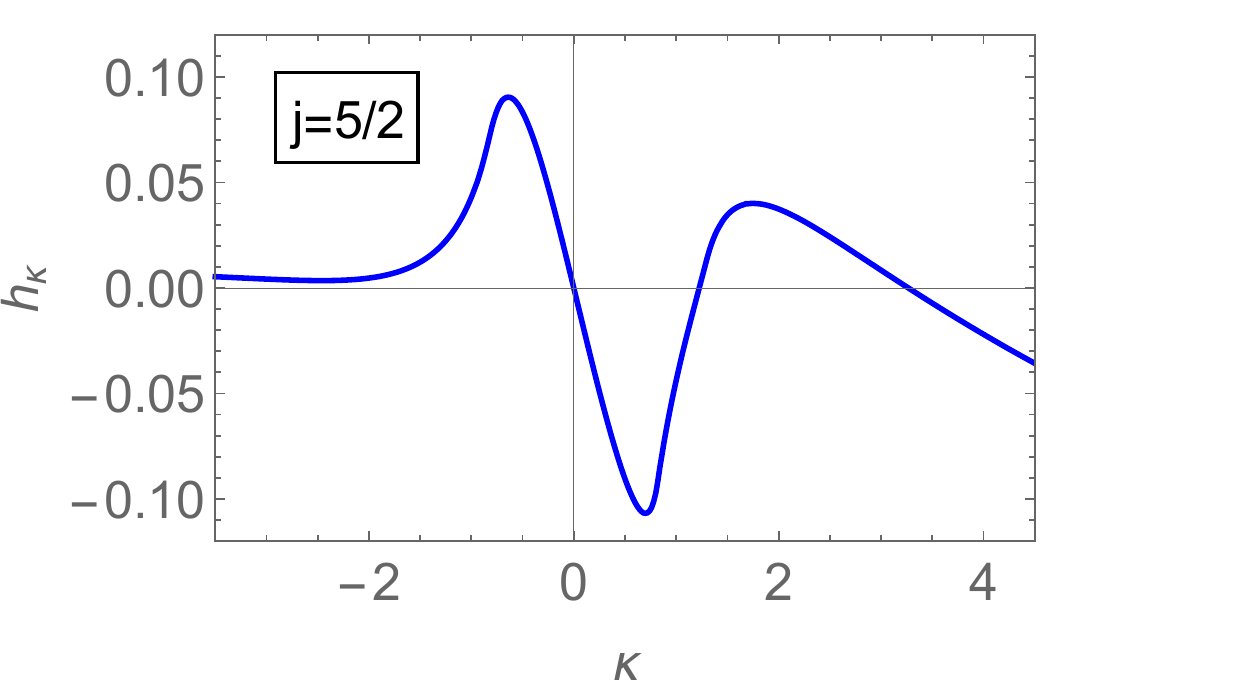}
\caption{Infrared fixed points from long-range interactions.  \emph{Upper panel.} For $j=3/2$, we plot the beta function of $\alpha$ defined by $\dot{\alpha}=h_\alpha(\alpha) e^2$. Besides the zero at $\alpha=0$, it features barely visible zeros at $|\alpha|=2,\ 2.296$. The fixed points for $\alpha=0$ and $|\alpha|= 2.296$ are stable, the $\text{SO}(3)$ symmetric fixed point for $|\alpha|=2$ is unstable. \emph{Lower panel.} For $j=5/2$, we plot the beta function $\dot{\kappa}=h_\kappa(\kappa) e^2$ along the line $(\alpha,\beta,\gamma) = (\frac{1}{3},\frac{1}{\sqrt{3}},\frac{\sqrt{5}}{3})\kappa$ as in Fig. \ref{FigP}. We find both continuous symmetry fixed points at $\kappa=0$ and $\kappa_{\rm c}=4\sqrt{2/3}$ to be stable along this direction. However, since the parameter space of velocity coefficients $(\alpha,\beta,\gamma)$ is three-dimensional for spin 5/2, the two orthogonal directions also need to be taken into account. We then find that the $\text{SO}(3)$ symmetric fixed point is actually unstable, whereas the relativistic one remains stable.}
\label{FigFermions}
\end{minipage}
\end{figure}

We first study the $\text{O}(3)$-symmetric relativistic fixed point with $\vec{\alpha}_\star=0$ and Hamiltonian $H_{\text{O}(3)}(\textbf{p})=p_iV_i$. This fixed point exists for every $j$ due to the enlarged continuous symmetry. Indeed, if we finetune the system such that $\vec{\alpha}=0$, then none of the $\text{O}(3)$-symmetric fluctuations can generate any $\vec{\alpha}\neq 0$. Emergent Lorentz invariance, reflected by the matrices $V_i$ satisfying the Clifford algebra relation (\ref{int3}), is a common phenomenon for low-energy electronic systems. Example beta functions with zeros at $\vec{\alpha}=0$ are shown in Fig. \ref{FigFermions}. We find that the relativistic fixed point is stable for all $j \leq 7/2$ and give reasons to believe that the stability extends to all $j$.

The anomalous dimension at the relativistic fixed point can be computed by utilizing Eqs. (\ref{rg1}) and (\ref{rg7}). The eigenvalues of $H_{\text{O}(3)}(\textbf{p}) = p_i V_i$ are $\pm p$, so that $|\hat{E}_\lambda|=1$ and
\begin{align}
 \label{rel1} f(\vec{0}) = \sum_{\lambda=1}^{2j+1} \int_\Omega 1 = 2j+1.
\end{align} 
Hence the anomalous dimension is given by
\begin{align}
 \label{rel2} \eta_\star = \frac{e^2}{3(2j+1)}f(\vec{0}) = \frac{e^2}{3}.
\end{align}
This result is independent of $j$, which can be explained by the fact that the system in this limit is comprised of $N=j+1/2$ independent Weyl fermions, each having anomalous dimension $e^2/3$.

In order to determine the stability of the fixed point, we compute the eigenvalues $\{\theta_n\}$ of the stability matrix. One can show that in this particular case, using the parametrization in Eq. (\ref{high3b}), the elements of the stability matrix are
\begin{align}
 \label{rel3} M_{nn'}= -\frac{e^2}{5}\delta_{nn'}-\frac{2e^2}{15(2j+1)}\mbox{tr}[U_3^{(n)} V_3 U_1^{(n')}V_1].
\end{align}
After the set of matrices $U_i^{(n)}$ has been determined for a given value of $j$ by a Gram--Schmidt procedure, it is straightforward to compute the stability of the $\text{O}(3)$ symmetric fixed point using this formula. The outcome is quite surprising: The stability eigenvalues $\{\theta_n^{(j)}\}$ for spin $j\leq 7/2$ are given by
\begin{align}
 \label{rel4} \{\theta_n^{(j=3/2)}\} &= \Bigl\{ - \frac{2}{15} \Bigr\}e^2,\\
 \label{rel5} \{\theta_n^{(j=5/2)}\} &= \Bigl\{ -\frac{2}{15},\ -\frac{2}{15},\ - \frac{1}{3} \Bigr\}e^2,\\
 \label{rel6} \{\theta_n^{(j=7/2)}\} &= \Bigl\{ -\frac{2}{15},\ -\frac{2}{15},\ -\frac{2}{15},\ - \frac{1}{3},\ - \frac{1}{3} \Bigr\}e^2.
\end{align} 
We first observe that all eigenvalues are negative and thus the fixed point is stable. Furthermore, the eigenvalues have the striking  pattern that at each order in $j$, the eigenvalues are either $-2e^2/15$ or $-e^2/3$. This leads us to conjecture that this behavior persists even for $j>7/2$, and so the $\text{O}(3)$ symmetric fixed point is always stable. We leave the proof of this conjecture for future work.

Let us comment on a remarkable feature of the spin-5/2 case, which may also extend to higher spin in a suitable form. Diagonalizing the stability matrix yields a preferred choice of basis constructed from linear combinations of the $U_i^{(n)}$. For $j=5/2$, the basis that diagonalizes $M$ is precisely the matrices $A_i, B_i, C_i$ in Eq. (\ref{high16}), constructed from their particular symmetry properties with respect to the symmetry operators $\hat{\mathcal{I}}$ and $\hat{\mathcal{W}}$. Further, any $\text{O}(2)$ rotation in the subspace spanned by $A_i$ and $C_i$ also diagonalizes $M$ in this case.

\subsection{SO(3) symmetric fixed point}

Next we study the rotational fixed point with $\text{SO}(3)$ symmetry and fixed point Hamiltonian $H_{\text{SO}(3)}(\textbf{p})=2p_iJ_i$. For the same reason as explained in the previous section for the $\text{O}(3)$ symmetric case, this fixed point always exists due to the enlarged continuous symmetry. However, its stability properties are drastically different, as we demonstrate here.

We begin the analysis by locating the rotational fixed point in the space of couplings spanned by $\vec{\alpha}$. Given the matrices $V_i$, we can always construct the matrix $U_i$ which is orthonormal according to $\mbox{tr}(U_kU_l)=(2j+1)\delta_{kl}$, $\mbox{tr}(V_kU_l)=0$ and satisfies
\begin{align}
 \label{rot1} H_{\text{SO}(3)}(\textbf{p}) = 2p_iJ_i = p_i (V_i +\kappa_{\rm c} U_i) 
\end{align}
for some critical coupling constant $\kappa_{\rm c}$. This equation is a direct generalization of the one-parameter case of spin 3/2 in Eq. (\ref{int7}). It turns out that the value of $\kappa_{\rm c}$ is fixed through spin algebra to be
\begin{align}
 \label{rot2} \kappa_{\rm c} = \sqrt{\frac{4}{3}j(j+1)-1},
\end{align}
see Appendix \ref{AppMat}, and consequently we simply define $U_i$ as
\begin{align}
 \label{rot3} U_i = \frac{1}{\kappa_{\rm c}}(2J_i-V_i).
\end{align}
Some particular values of $\kappa_{\rm c}$ for small $j$ are presented in Table \ref{TabRot}. By matching the Hamiltonian in Eq. (\ref{high3b}) with Eq. (\ref{rot1}), we can express $\kappa_{\rm c}$ in terms of the couplings $\vec{\alpha}$. In particular, $\vec{\alpha}^2_\star=\kappa_{\rm c}^2$ at the rotational fixed point.

Equation (\ref{rg7}), again, provides an elegant way to compute the anomalous dimension at the rotational fixed point for arbitrary $j$. Since the eigenvalues of $\hat{H}_{\text{SO}(3)}$ are given by $\hat{E}_m=2m$ with $m=-j,\dots,j$, we have
\begin{align}
 \label{rot4} f(\vec{\alpha}_\star) = 2\sum_{m=-j}^j \int_\Omega |m| = \frac{(2j+1)^2}{2}.
\end{align}
This implies that the anomalous dimension at the rotational fixed point is given by
\begin{align}
 \label{rot5} \eta_\star = \frac{2j+1}{8j(j+1)} e^2.
\end{align}
We list the values for small $j$ in Table \ref{TabRot}. For large values of $j$ we have $\eta_\star = e^2/(4j) +\mathcal{O}(j^{-2})$.

\renewcommand{\arraystretch}{1.8}
\begin{table}[t]
\begin{tabular}{|c||c|c|c|c|c|c|c|}
\hline
 \multicolumn{8}{|c|}{ $\text{SO}(3)$ symmetric fixed point} \\
\hline
 \ $j$ \ & \ $3/2$ \ & \ $5/2$ \ & \ $7/2$ \ & \ $9/2$ \ & \ $11/2$ \ & \ $13/2$ \ & \ $15/2$ \ \\
\hline\hline
 \ $\kappa_{\rm c}$ \ & \ $2$ \ & \ $4\sqrt{\frac{2}{3}}$ \ & \ $2\sqrt{5}$ \ & \ $4\sqrt{2}$ \ & \ $2\sqrt{\frac{35}{3}}$ \ & \ $8$ \ & \ $2\sqrt{21}$ \ \\
\hline
 \ $\eta_\star/e^2$  & \ $2/15$ \ & \ $3/35$ \ & \ $4/63$ \ & \ $5/99$ \ & \ $6/143$ \ & \ $7/195$ \ & \ $8/255$ \ \\
\hline
\end{tabular}
\caption{The rotational $\text{SO}(3)$ symmetric fixed point with Hamiltonian $2p_iJ_i$ can be parametrized as $p_i(V_i+\kappa_{\rm c} U_i)$. The critical coupling $\kappa_{\rm c}$ and the matrix $U_i$ are defined in Eqs. (\ref{rot2}) and (\ref{rot3}), respectively. The anomalous dimension $\eta_{\star}$ at the fixed point is given by Eq. (\ref{rot5}).}
\label{TabRot}
\end{table}
\renewcommand{\arraystretch}{1}

The rotational fixed point is unstable for $j\leq 7/2$. For $j=3/2$, the single eigenvalue of the stability matrix is
\begin{align}
 \label{rot6} \theta_1^{(j=3/2)} = \frac{4}{945}e^2.
\end{align}
The instability in this particular case was shown in Ref. \cite{PhysRevB.93.241113}. For $j=5/2$ and $j=7/2$ we find
\begin{align}
 \label{rot7} \{\theta_n^{(j=5/2)}\} &=   \Bigl\{-0.062,\ 0.061,\ 0.022\Bigr\} e^2,\\
 \label{rot8} \{\theta_n^{(j=7/2)}\} &= \Bigl\{-0.056,\ -0.040,\ 0.013,\ 0.026,\ 0.054\Bigr\} e^2.
\end{align}
In all cases considered, the rotational fixed point has at least one positive stability eigenvalue, and thus is unstable. In fact, the number of positive eigenvalues exceeds the number of negative ones for every $j$ considered. We thus conjecture that this trend extends to higher spin and that the rotational fixed point is likely always highly unstable.

\subsection{Fixed point structure for spin 5/2}\label{SecFix52}

In this section, we study the fixed point structure for $j=5/2$ with the Hamiltonian given by Eq. (\ref{high16}). Long-range interactions lead to a running of the three velocity parameters $(\alpha,\beta,\gamma)$ according to
\begin{align}
 \label{fix1} \dot{\alpha} & = h_1(\alpha,\beta,\gamma),\\
 \label{fix2} \dot{\beta} & = h_2(\alpha,\beta,\gamma),\\
 \label{fix3} \dot{\gamma} & = h_3(\alpha,\beta,\gamma).
\end{align}
The full expressions for the beta function  $(h_1,h_2,h_3)$ are presented in Eqs. (\ref{fix3c})-(\ref{fix3e}). The anomalous dimension $\eta(\alpha,\beta,\gamma)$ is given by Eq. (\ref{fix3b}). The rather vast three-dimensional parameter space, together with the sufficiently complicated expressions for the RG flow equations, implies that the problem of finding the infrared fixed points is essentially of numerical nature. However, several analytical or semi-analytical statements are possible. In fact, it turns out that only one fixed point (the cubic symmetric one) needs to be determined numerically, while the other ones are accessible analytically. We arrived at this conclusion by numerically scanning the volume $(\alpha,\beta,\gamma)\in [-3,3]^3$ for mutual zeros of the beta functions $(h_1,h_2,h_3)$ with three negative eigenvalues of the stability matrix. The corresponding infrared stable fixed points for $j=5/2$, together with the case of $j=3/2$ for comparison, are summarized in Tab. \ref{TabFP}. Our analysis cannot exclude cubic fixed points with $\vec{\alpha}^2>9$, but such large values would be untypical. In the following, we discuss some particular aspects of the RG flow and its fixed points.

\renewcommand{\arraystretch}{1.7}
\begin{table*}[t]
\begin{tabular}{|c|c|c|c|c|c|c|c|c|c|}
\hline \multicolumn{10}{|c|}{Infrared fixed points for $j=3/2$} \\
\hline\hline \  \ & \multicolumn{3}{|c|}{$\alpha\geq 0$} & $\eta_\star/e^2$  &$\mathcal{Q}$ & \multicolumn{3}{|c|}{$\theta_1/e^2$} & stability  \\
\hline\ $\text{O}(3)$: Weyl$^{-2}$ \ & \multicolumn{3}{|c|}{$0$} & $1/3$ & $-2$ & \multicolumn{3}{|c|}{$-2/15$} & stable \\
\hline \ $\text{SO}(3)$ \ & \multicolumn{3}{|c|}{$2$} & $2/15$ & $4$ & \multicolumn{3}{|c|}{$4/945$} & unstable\\
\hline \ Cubic \ & \multicolumn{3}{|c|}{$2.296$} & \ $0.119$ \ & $4$ & \multicolumn{3}{|c|}{$-0.003$} & stable\\
\hline\hline \multicolumn{10}{|c|}{Infrared fixed points for $j=5/2$} \\
\hline\hline \  \ & $\alpha$ & $\beta$ & $\gamma \geq 0$ & \ $\eta_\star/e^2$ \ & $\mathcal{Q}$ & $\theta_1/e^2$ & $\theta_2/e^2$ & $\theta_3/e^2$ & stability\\
\hline  $\text{O}(3)$: Weyl$^3$ & $0$ & $0$ & $0$ & $1/3$ & $3$ & $-1/3$ & $-2/15$ & $-2/15$ & stable\\
\hline \ $\text{O}(3)$: Dirac$+$Weyl$^{1}$ \ & $0$ & $2\sqrt{2}$ & $0$ & $1/9$ & $1$ & $-1/9$ & $-1/15$ & $-2/45$ & stable\\
\hline $\text{O}(3)$: Weyl$^{-3}$ & \ $-2\sqrt{2/3}$ \  & $0$ & $4/\sqrt{3}$ & $1/9$ & $-3$ & $-1/9$ & \ $-2/45$ \ & $-2/45$ & stable\\
\hline $\text{SO}(3)$ & $\frac{4}{3}\sqrt{2/3}$ & $\frac{4}{3}\sqrt{2}$ & \ $\frac{4}{3}\sqrt{10/3}$ \ & \ $3/35$ & $9$ & \ \ $-0.062$ \ & \ $0.061$ \ & \ $0.022$ \ & \ unstable \ \\
\hline \ Cubic \ & \ $1.172$ \ & \ $-0.530$ \  & $0$ & \ $0.190$ & \ $-3$ \  \ & $-0.19$ & $-0.10$ & \ $-0.005$ \ & stable\\
\hline
\end{tabular}
\caption{Infrared fixed points for higher-spin fermions with $j=3/2$ and $j=5/2$. We restrict to stable fixed points, where all eigenvalues $\{\theta_n\}$ of the stability matrix $M$ in Eq. (\ref{rg8}) are negative, but choose to include the unstable rotational fixed point with $\text{SO}(3)$ symmetry in the list. While this covers all infrared fixed points for $j=3/2$, we omit several repulsive ones for $j=5/2$. For the fixed points with relativistic $\text{O}(3)$ symmetry we use the following notation: Weyl$^\mathcal{Q}$ denotes a collection of Weyl fermions with total monopole charge $\mathcal{Q}$. Massless Dirac particles are labeled accordingly. Recall that a single Weyl fermion has monopole charge (or chirality) $\pm1$, whereas a massless Dirac fermion is comprised of two Weyl fermions with opposite chirality and thus has zero monopole charge.}
\label{TabFP}
\end{table*}
\renewcommand{\arraystretch}{1}

Let us start with some general remarks. We derived in Sec. \ref{SecSym} that the system is invariant under $\gamma \to -\gamma$ and so we can assume $\gamma \geq 0$. This also implies that the $(\alpha,\beta)$-plane for $\gamma=0$ is guaranteed to satisfy $h_3=0$, and, as a result, is likely to host fixed points where the lines of zeros of $h_1(\alpha,\beta,0)$ and $h_2(\alpha,\beta,0)$ intersect. This does, however, not imply that the plane spanned by $\gamma=0$ is stable in the third direction. Indeed, the derivative $\frac{\partial h_3}{\partial \gamma}(\alpha,\beta,\gamma\to 0^+)$ is generally nonzero and can lead to a growth of $\gamma$ if the RG flow is initialized at any value $\gamma\neq 0$. Furthermore, although the system for $\gamma=0$ is invariant under $\alpha \to -\alpha$, this only implies hat the locations of fixed points are symmetric in $\alpha$ in the $(\alpha,\beta)$ plane, but, again, this does not imply symmetry of the stability matrix at these fixed points. This explains why the cubic fixed point at $(\alpha,\beta,\gamma)=(1.172,-0.530,0)$ is stable, while the one at $(-1.172,-0.530,0)$ is not.

We verify that the three $\text{O}(3)$ symmetric fixed points associated to the fixed point Hamiltonians in Eqs. (\ref{int2}), (\ref{band11}), and (\ref{band14}) are stable. The system at these points, respectively, is equivalent to three Weyl fermions of positive chirality, a Weyl fermion of positive chirality and a Dirac fermion, and three Weyl fermions of negative chirality, see the discussion in Sec. \ref{SecTopo}. The anomalous dimension at these fixed points follows from Clifford algebra and the stability matrix can be computed analytically using Eqs. (\ref{stab8}).

\begin{figure}[t!]
\centering
\includegraphics[width=8.5cm]{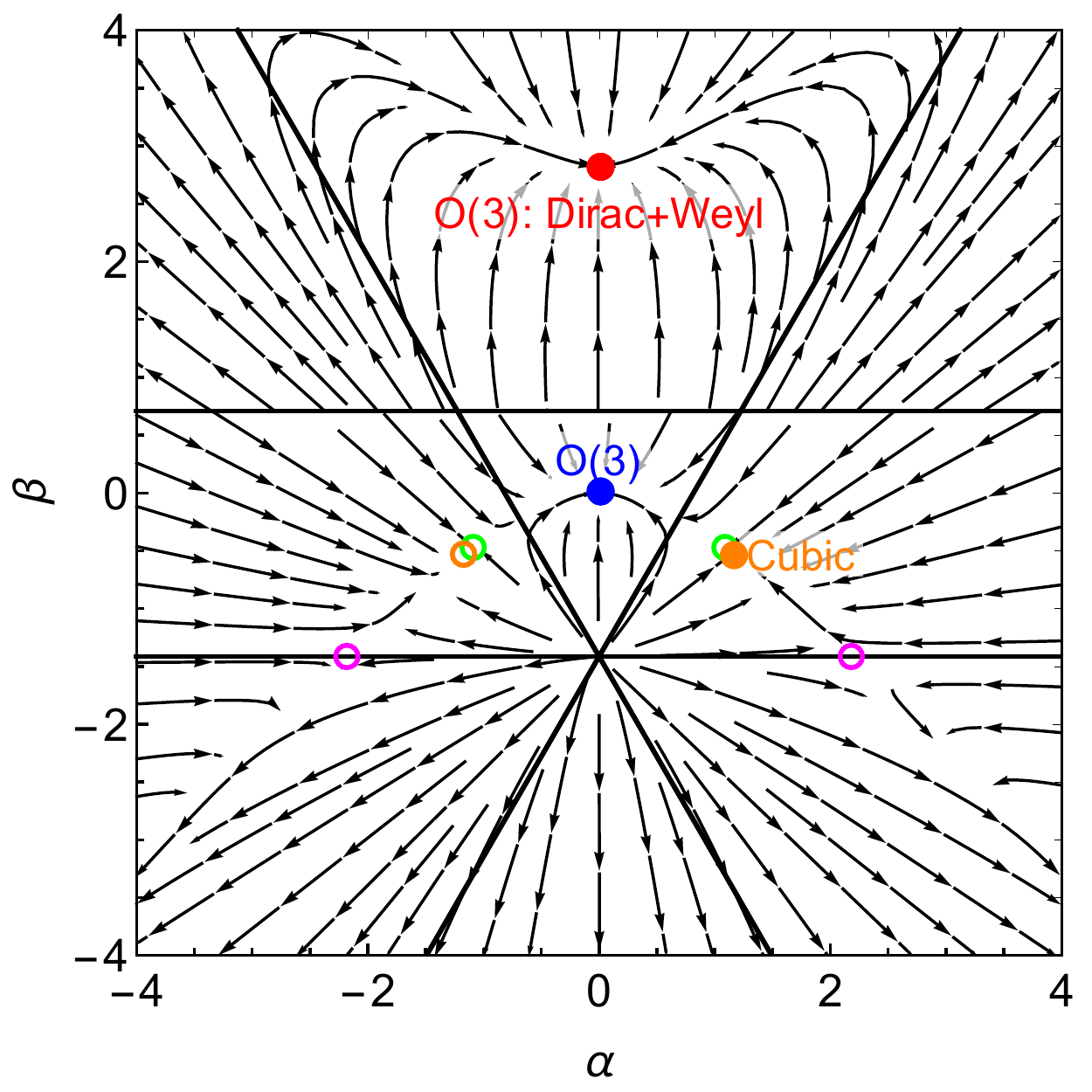}
\caption{RG flow for $j=5/2$ in the $(\alpha,\beta)$-plane for $\gamma=0$. Arrows point towards the infrared. We identify three infrared stable fixed points, indicated by the blue, red, and orange dots as in Fig. \ref{FigTopo}. The properties of the individual fixed points are summarized in Tab. \ref{TabFP}. The empty circles indicate unstable fixed points with at least one positive eigenvalue of the stability matrix. The straight lines border the distinct topological sectors as in Fig. \ref{FigTopo}. While the functions $(h_1,h_2)=(\dot{\alpha},\dot{\beta})/e^2$ remain finite along these lines, the RG flow is effectively stopped due to $e\to 0$ in our model of strictly linear band crossing. Among the two cubic-symmetric fixed points at $(\alpha,\beta)=(1.172,-0.530)$ (orange dot) and $(\alpha,\beta)=(-1.172,-0.530)$ (orange circle), only the first one is stable. This is not in conflict with the symmetry with respect to $\alpha\to -\alpha$ for $\gamma=0$, because the stability in the $\gamma$-direction depends on the properties for $\gamma\neq 0$.}
\label{FigFlow}
\end{figure}

The only additional stable fixed point besides the ones with enhanced $\text{O}(3)$ symmetry is a cubic symmetry fixed point at $(\alpha,\beta,\gamma)=(1.172,-0.530,0)$. Similar to the cubic fixed point at $\alpha=2.296$ for $j=3/2$, its location in the $(\alpha,\beta)$-plane appears to be bare of any distinctive features. Another common feature of both cubic fixed points is the presence of one unusually small negative eigenvalue of the stability matrix, rendering them almost marginal. In Fig. \ref{FigFlow} we plot the RG flow in the $(\alpha,\beta)$-plane for $\gamma=0$, i.e. we plot the derivatives of $(h_1,h_2)$. This plane contains three stable fixed points and several repulsive ones.

\section{Outlook}

In this work, we studied the band renormalization of higher-spin fermions in topological semimetals due to long-range interactions. Many of the constructions and properties we derived are valid for any half-integer $j\geq 3/2$, but for the sake of concreteness we often turned to $j=3/2$ and $j=5/2$ as illustrative examples.

One may wonder what can be learned from such an analysis that is relevant for experiments. Topological semimetals hosting "Rarita--Schwinger--Weyl" (RSW) fermions with effective pseudospin $j=3/2$, such as PdBiSe \cite{PhysRevB.99.241104}, have been synthesized recently and constitute an active frontier of quantum matter research. Importantly, the identification of candidate materials is guided by group theoretic studies of the consistency of $j=3/2$ fermions with any of the possible space groups \cite{Bradlynaaf5037}. Going further up in spin, the same analysis suggests \cite{Bradlynaaf5037} that sixfold band touchings are generically non-topological and so do not host the type of topological $j=5/2$ fermions discussed in this work. The question of interaction effects in this topologically trivial sector could be addressed with the same techniques as presented here. To realize the intriguing higher-spin physics discussed in this work, one may need to turn to other quantum simulation platforms, such as ultracold Fermi gases \cite{PhysRevB.84.165115,PhysRevLett.112.086401,PhysRevLett.113.033002}, although simulating long-range forces poses a serious challenge for the latter. We conclude that, currently, the case of $j=3/2$ is experimentally by far the most interesting and pressing one.

The band renormalization effects discussed in this work become experimentally relevant when the material parameters are such that the effective model of a linear band crossing point is faithful over a sufficiently large energy range. In view of the renormalization group flow, the ultraviolet cutoff of the model is given by the energy window where the dispersion is approximately linear, whereas the infrared cutoff is given by the temperature of the sample. If these two cutoffs are separated by at least one or two orders of magnitude, we expect band renormalization effects to be visible. From the experimental ARPES data on the RSW crossing point in PdBiSe reported in Ref. \cite{PhysRevB.99.241104}, we estimate a width of 0.25 eV (3000 K), while the measurements were taken at 20 K. The energy scales are thus sufficiently wide apart.

The cubic infrared fixed points may be observed through anisotropy of both ARPES data or transport. For instance, in order to distinguish the two $\mathcal{Q}=4$ fixed points for $j=3/2$ with ARPES, i.e. the unstable isotropic one with $\alpha=2$ from the stable anisotropic one with $\alpha=2.296$, a sufficiently good angular resolution is required. For $\alpha=2.296$, the energy difference of $E_2$ in Eq. (\ref{band1}) for momenta $\textbf{p}$ and $\textbf{p}'$ in the $xy$-plane is maximal for a relative angle of $45^\circ$, with a (rather small) relative energy difference of $17\%$ measured from the crossing point. Given the above band width from Ref. \cite{PhysRevB.99.241104}, this translates to an energy difference in the meV-range, while the experiment achieved an energy resolution of 100 meV and angular resolution of $0.07^\circ$. The tiny and thus currently unobservable effect can be explained by the value of $\alpha=2.296$ being close to the isotropic limit $\alpha=2$. For higher spin, the cubic and isotropic fixed points can be distinguished more easily, also because of the different total monopole charges $\mathcal{Q}$, see Table \ref{TabFP}.

Compared to the properties of Weyl fermions with spin 1/2, the case of spin 3/2 appears rather exotic and mysterious. However, when going to even larger values of $j>3/2$, as we did in this work, some features of interacting higher-spin fermions become clear that might otherwise have been hidden. Such are, for instance, the generic existence of matrices $V_i$ and $U_i$ that generalize the Isobe--Fu Hamiltonian to $j\geq 3/2$, the instability of the $\text{SO}(3)$ symmetric fixed point, the relation between the fermion self-energy and free energy in Eqs.  (\ref{rg4}) and (\ref{rg5}), the relation between charge renormalization and changes in the total monopole charge in Eq. (\ref{rg11}), or the curious pattern of the stability exponents at the $\text{O}(3)$ symmetric fixed point for $\vec{\alpha}=0$ in Eqs. (\ref{rel4})-(\ref{rel6}). Furthermore, compared to the unwieldy case of $j=5/2$ with three free parameters $(\alpha,\beta,\gamma)$, fermions with $j=3/2$ suddenly appear rather harmless. We therefore believe that the present analysis gives an original perspective on the physics of spin 3/2 fermions via the extension to higher spin.

In view of understanding general properties of higher spin fermions, it is encouraging that, despite the growth in complexity of the single-particle Hamiltonian for larger spin, the number of stable infrared fixed points (at least for $j=5/2$) increases only moderately. In fact, we only identified one cubic symmetric fixed point for $j=5/2$, which is in analogy to $j=3/2$. The remaining additional stable fixed points show an enhanced $\text{O}(3)$ symmetry and their properties mostly follow from Clifford algebra.

The reader may have noticed that the total monopole charges $\mathcal{Q}$ for a given $j$ are either all even or all odd. The reason for that is the following. The monopole charge of the Hamiltonians $H_{\text{O}(3)}(\textbf{p})=p_iV_i$ and $H_{\text{SO}(3)}(\textbf{p})=2p_iJ_i$ are $(-1)^{j-1/2}N$ and $N^2$, respectively, with $N=j+1/2$. If $N$ is even/odd, then $N^2$ is even/odd, consistent with our observation. Furthermore, the value of $\mathcal{Q}$ can only change when a positive and a negative energy band intersect, whereby the Chern number of the bands changes by an integer. Due to particle-hole symmetry, however, the intersecting bands have opposite Chern numbers and so the change in the monopole charge is a multiple of $+2$ or $-2$, which eventually confirms our observation.

The behavior of the topological invariants implies that fixed points which host massless Dirac particles are likely to appear for higher spins $j\geq 5/2$. The case of spin 5/2 with Hamiltonian $H_\star^{(1)}$ in Eq. (\ref{band11}) has been discussed in detail in this work. Interestingly, however, no fixed point with Dirac particle appears for $j=3/2$, although $\mathcal{Q}=0$ could, in principle, be reached from $\mathcal{Q}=-2,\ 4$. To understand this, note that for $j=3/2$ we can construct a Dirac Hamiltonian $H_{\rm D}(\textbf{p}) = p_i \Gamma_i$ with the matrices $\Gamma_1=\frac{1}{\sqrt{3}}\{J_y,J_z\}$, $\Gamma_2=\frac{1}{\sqrt{3}}\{J_x,J_z\}$, $\Gamma_3=\frac{1}{\sqrt{3}}\{J_x,J_y\}$. These matrices, indeed, satisfy the Clifford algebra property $\{\Gamma_k,\Gamma_l\}=2\delta_{kl}$. However, they transform under the $T_2$ representation of the cubic group (see Appendix \ref{AppMat}) and thus cannot be generated from the RG flow considered here. The absence of a massless Dirac fermion for $j=3/2$ can therefore be explained by an obstruction from symmetry. It would be interesting to see how this behavior extends to higher spin and whether it can be used to constrain the number of possible fixed points with enhanced $\text{O}(3)$ symmetry.

\acknowledgements I thank Igor Herbut, Fangli Liu, and Seth Whitsitt for insightful comments and inspiring discussions. I acknowledge funding by the DoE BES Materials and Chemical Sciences Research for Quantum Information Science program (award No. DE-SC0019449), NSF PFCQC program, DoE ASCR Accelerated Research in Quantum Computing program (award No. DE-SC0020312), DoE ASCR Quantum Testbed Pathfinder program (award No. DE-SC0019040), AFOSR, ARO MURI, ARL CDQI, AFOSR MURI, and NSF PFC at JQI.

\begin{appendix}

\section{Spin algebra}\label{AppMat}

\subsection{Matrices $J_i$, $V_i$ and $U_i$}

Let $j$ be a half-integer. We define the spin matrices $J_i$ satisfying $[J_k,J_l]=\rmi \vare_{klm}J_m$ through their usual representation given by
\begin{align}
 \label{mat1} (J_+)_{mm'} &= \sqrt{j(j+1)-mm'}\delta_{m,m'+1},\\
 \label{mat2} (J_-)_{mm'} &= \sqrt{j(j+1)-mm'}\delta_{m+1,m'},\\
 \label{mat3} (J_z)_{mm'} &=  m \delta_{m,m'}
 \end{align}
and
\begin{align}
 \label{mat4} J_x &= \frac{1}{2}(J_++J_-),\ J_y=\frac{1}{2\rmi}(J_+-J_-).
\end{align}
These are $(2j+1)\times (2j+1)$ matrices. We use the convention $m=j,\dots,-j$ so that the first entry of the diagonal matrix $J_z = \text{diag}(j,\dots,-j)$ is positive.

We now construct two matrices $V_i$ and $U_i$ with the following properties:
\begin{itemize}
 \item[(i)] They transform as $T_1$ under the cubic rotational group $\text{O}$. 
 \item[(ii)] The matrices $V_i$ satisfy $\{V_k,V_l\} = 2\delta_{kl}$.
\item[(iii)] They are orthogonal and normalized according to $\mbox{tr}(V_k V_l) = \mbox{tr}(U_kU_l) = (2j+1)\delta_{kl},\ \mbox{tr}(V_k U_l) =0$.
\item[(iv)] There exists a number $\kappa_{\rm c}>0$ such that $V_i +\kappa_{\rm c} U_i = 2J_i$.
\end{itemize}
For $j=1/2$ we have $V_i = 2J_i=\sigma_i$ and $U_i=0$, with $\sigma_i$ the Pauli matrices. We exclude this trivial case in the following and assume $j\geq 3/2$. At the end of this section we prove the property (\ref{high7}).

First we explicitly construct $V_i$ in the representation of Eqs. (\ref{high4})-(\ref{high6}). While it is clear that $V_i$ can be written as a superposition of the $K_i^{(n)}$ from the next section, it turns out that it is sufficient to write it as a linear combination of odd powers of $J_i$. Thus we make the ansatz
\begin{align}
 V_i = \sum_{\text{odd } \mu=1}^{2j} v_{\mu} J_i^{\mu} = ( v_1J_i + v_3 J_i^3+\dots+v_{2j}J_i^{2j}),
\end{align}
with coefficients $v_1,\dots,v_{2j}$ to be determined. Considering the diagonal matrix
\begin{align}
V_3 \stackrel{!}{=} \mathbb{1}_N \otimes \sigma_3 = \text{diag}(1,-1,\dots,1,-1),
\end{align}
we arrive at the condition 
\begin{align}
 (V_3)_{mm} = \sum_{\text{odd } \mu=1}^{2j} v_{\mu} m^{\mu} \stackrel{!}{=} (-1)^{m+j+1}
\end{align}
for the diagonal components of $V_3$. To determine all coefficients $v_\mu$, it is sufficient to only consider $m=\frac{1}{2},\dots,j$. We can phrase this as a problem of matrix inversion. Define the square matrix $\mathcal{A}$ with entries
\begin{align}
 \label{mat26} \mathcal{A}_{m\mu} = (-1)^{m+j+1}m^{\mu}.
\end{align}
for $m=\frac{1}{2},\dots, j$ and $\mu=1,\dots, 2j$ odd. We then have to invert $\sum_\mu \mathcal{A}_{m\mu} v_\mu=1$, which is solved by
\begin{align}
 \label{mat27} \begin{pmatrix} v_1 \\ v_3 \\ \vdots \\ v_{2j} \end{pmatrix} = \mathcal{A}^{-1} \begin{pmatrix} 1 \\ 1 \\ \vdots \\ 1 \end{pmatrix},
\end{align}
i.e. $v_\kappa = \sum_m (\mathcal{A}^{-1})_{\kappa m}$. Examples for small $j$ are:
\begin{align}
 \nonumber j=\frac{3}{2}:\ V_i ={}& -\frac{7}{3}J_i+\frac{4}{3}J_i^3,\\
 \nonumber j=\frac{5}{2}:\ V_i ={}&\frac{149}{60}J_i-2J_i^3+\frac{4}{15}J_i^5,\\
 \nonumber j=\frac{7}{2}:\ V_i ={}&-\frac{2161}{840}J_i+\frac{217}{90}J_i^3-\frac{22}{45}J_i^5+\frac{8}{315} J_i^7,\\
 \nonumber j=\frac{9}{2}:\ V_i ={}& \frac{53089}{20160}J_i-\frac{30571}{11340}J_i^3+\frac{179}{270}J_i^5\\
 \label{mat28} &-\frac{52}{945}J_i^7+\frac{4}{2835}J_i^9.
\end{align}

Next, we determine the matrices $U_i$ and the parameter $\kappa_{\rm c}$ for arbitrary $j$. For this purpose, we express $V_i$ in terms of the basis $K_i^{(n)}$ according to
\begin{align}
 \label{mat14} V_i = \sum_{n=1}^{\mathcal{N}} a_{n} K_i^{(n)}
\end{align} 
with real coefficients $a_{n}$. (The coefficients do not depend on $i$ due to cubic symmetry. The value of $\mathcal{N}\geq 2$ is not important in the following.) Normalization of $V_i$ implies
\begin{align}
 \label{mat15} \sum_{n=1}^{\mathcal{N}} a_{n}^2=1.
\end{align}
In order to satisfy (iv), $U_i$ needs to be of the form
\begin{align}
 \label{mat16} U_i = \frac{1}{\kappa_{\rm c}}\Bigl( b K_i^{(1)} - \sum_{n=2}^{\mathcal{N}} a_n K_i^{(n)}\Bigr),
\end{align}
with $\kappa_{\rm c}$ to be determined and, due to Eq. (\ref{mat12}),
\begin{align}
 \label{mat17} (a_1 + b)\sqrt{\frac{3}{j(j+1)}} \stackrel{!}{=} 2.
\end{align}
Normalization of $U_i$ yields
\begin{align}
 \label{mat18} 1 \stackrel{!}{=} \frac{\mbox{tr}(U_i^2)}{2j+1}  = \frac{1}{\kappa_{\rm c}^2}\Bigl( b^2 + \sum_{n=2}^{\mathcal{N}} a_n^2\Bigr)= \frac{b^2 +(1-a_1^2)}{\kappa_{\rm c}^2},
\end{align}
and the orthogonality to $V_i$ is ensured by
\begin{align}
 \label{mat19} 0 \stackrel{!}{=} ba_1 - \sum_{n=2}^{\mathcal{N}} a_n^2 = ba_1 -(1-a_1^2).
\end{align} 
These three equations determine the parameters $\kappa_{\rm c}, a_1, b$ to be
\begin{align}
 \label{mat20} \kappa_{\rm c} &= \sqrt{\frac{4}{3}j(j+1)-1},\\
 \label{mat21} a_1 &= \frac{1}{\sqrt{1+\kappa_{\rm c}^2}},\\
 \label{mat22} b &= \frac{\kappa_{\rm c}^2}{\sqrt{1+\kappa_{\rm c}^2}}.
\end{align}
The corresponding matrices $U_i=\frac{1}{\kappa_{\rm c}}(2J_i-V_i)$ for $V_i$ in Eqs. (\ref{mat28}) read
\begin{align}
 \nonumber j=\frac{3}{2}:\ U_i ={}& \frac{1}{2}\Bigl(\frac{13}{3}J_i-\frac{4}{3}J_i^3\Bigr),\\
 \nonumber j=\frac{5}{2}:\ U_i ={}& \frac{1}{4\sqrt{2/3}}\Bigl(-\frac{29}{60}J_i+2J_i^3-\frac{4}{15}J_i^5\Bigr),\\
 \nonumber j=\frac{7}{2}:\ U_i ={}&\frac{1}{2\sqrt{5}}\Bigl(\frac{3841}{840}J_i-\frac{217}{90}J_i^3+\frac{22}{45}J_i^5-\frac{8}{315} J_i^7\Bigr),\\
 \nonumber j=\frac{9}{2}:\ U_i ={}& \frac{1}{4\sqrt{2}}\Bigl(-\frac{12769}{20160}J_i+\frac{30571}{11340}J_i^3-\frac{179}{270}J_i^5\\
 \label{mat29}  &+\frac{52}{945}J_i^7-\frac{4}{2835}J_i^9\Bigr).
\end{align}

We now show relations (\ref{high7}), i.e. that $V_i$ commutes with $J_i$ and anticommutes with $J_{k\neq i}$. The first statement, $[V_i,J_i]=0$, immediately follows from the fact that $V_i$ is a linear combination of odd powers of $J_i$. To show the second statement, it is sufficient to work in the particular representation from Eqs. (\ref{high4})-(\ref{high6}) and (\ref{mat1})-(\ref{mat3}). We show that $V_3$ anticommutes with  $J_1$ and $J_2$, or, equivalently, that $V_3$ anticommutes with $J_+$ and $J_-$. For this compute the matrix elements 
\begin{align}
 \nonumber (V_3 J_+)_{mm'} &=(-1)^{m+j+1}(J_+)_{mm'}\\
 \nonumber &=(-1)^{m+j+1} \sqrt{j(j+1)-mm'} \delta_{m,m'+1},\\
 \nonumber (J_+ V_3)_{mm'} &= (-1)^{m'+1}(J_+)_{m m'} \\
 \nonumber &=(-1)^{m'+j+1} \sqrt{j(j+1)-mm'} \delta_{m,m'+1}\\
  \label{mat31} &=-(-1)^{m+j+1} \sqrt{j(j+1)-mm'} \delta_{m,m'+1},
\end{align}
and, consequently,
\begin{align}
 \label{mat32} (\{V_3,J_+\})_{mm'} = 0.
\end{align}
Analogously one shows $\{V_3,J_-\} =0$.

\subsection{Matrices $K_i^{(n)}$: General remarks}\label{AppKi}

In the following we construct the orthonormal matrices $\{K_i^{(1)},K_i^{(2)},\dots\}$ such that every $(2j+1) \times (2j+1)$ Hermitean matrix $\mathcal{V}_i$ that transforms as a vector under the cubic group can be written as a linear combination of the $K_i^{(n)}$ with real coefficients. We restrict the analysis to those $\mathcal{V}_i$ that are odd under time-reversal and thus satisfy $\{\mathcal{T},\mathcal{V}_i\}=0$. The general Hamiltonian can be written as
\begin{align}
 H(\textbf{p}) = \sum_{a=1}^\mathcal{N} u_a p_i K_i^{(n)}
\end{align}
with some velocity coefficients $(u_1,\dots,u_\mathcal{N})$. We can always rescale momentum such that one of these coefficients equals unity.

For general $j$, the number $\mathcal{N}$ is determined in the following way. Any single-particle Hamiltonian for $(2j+1)$-component fermions can be written as a linear combination (with real coefficients) of $(2j+1)^2$ Hermitean basis matrices. These basis elements may be constructed as the symmetric and traceless tensors that results from products $J_k J_l \cdots J_m$ of the spin matrices, see Ref. \cite{PhysRevB.95.075149}. Among these irreducible tensors, some will transform according to the desired $T_1$ representation of the cubic group. We refer to the number of such terms as "$\mathcal{N}$ without $\mathcal{T}$-symmetry", where $\mathcal{T}$ stands for time-reversal, see the definition below. Under $\mathcal{T}$, the spin matrices transform according to $J_i \to -J_i$. Therefore, only those $K_i^{(n)}$ that originate from a product of an odd number of spin matrices lead to a time-reversal invariant Hamiltonian. In this work, we restrict our attention to the time-reversal symmetric case, and refer to it as "$\mathcal{N}$ with $\mathcal{T}$-symmetry", since these are the terms that can be generated from the $\mathcal{T}$-symmetric Hamiltonian $p_iJ_i$ via self-energy corrections. The matrices $K_i^{(n)}$ are orthonormal according to
\begin{align}
 \label{mat11} &\mbox{tr}(K_k^{(n)}K_l^{(n')}) = (2j+1)\delta_{kl}\delta_{nn'}.
\end{align}

Some of the matrices $K_i^{(n)}$ can be constructed as odd powers of $J_i$. Note that the Cayley--Hamilton theorem implies $\mathcal{P}(J_i)=0$ for every $i=1,2,3$ with
\begin{align}
 \label{mat8} \mathcal{P}(X) = \prod_{m=-j}^{j} (X - m).
\end{align} 
Consequently, $J_i^{2j+1}$ is a linear combination of lesser powers of $J_i$. Using a Gram--Schmidt procedure to ensure Eq. (\ref{mat11}), one can generate the corresponding matrices $K_i^{(n)}$. The first such matrix is given by $J_i$ itself with an appropriate normalization,
\begin{align}
 \label{mat12} K_i^{(1)} = \sqrt{\frac{3}{j(j+1)}}J_i,
\end{align}
where we used 
\begin{align}
 \label{mat13} \mbox{tr}(J_z^2)=\sum_{m=-j}^{j} m^2=\frac{j(j+1)(2j+1)}{3}.
\end{align}
For $j\geq 3/2$, the second matrix can be determined from an ansatz $K_i^{(2)}=\bar{a} J_i^3 + \bar{b} J_i$ in Eq. (\ref{mat11}). To determine both parameters $\bar{a}$, $\bar{b}$, it is sufficient to consider the case $k=l=3$ only, where the matrices involved are diagonal. In this procedure, we can use that for even $n$ we have
\begin{align}
 \label{mat13b} \mbox{tr}(J_z^n) = \sum_{m=-j}^j m^n = \frac{2}{n+1}B_{n+1}(j+1)
\end{align}
with Bernoulli polynomials $B_n(x)$, while the trace vanishes for odd $n$. We then find
\begin{align}
 \nonumber K_i^{(2)} ={}& \frac{5\sqrt{7}}{\sqrt{j(j+1)(4j^4+8j^3-7j^2-11j+6)}}\\
 \label{mat13c} &\times\Bigl(J_i^3 -\frac{3j^2+3j-1}{5}J_i\Bigr).
\end{align}
Similarly, all higher orders are constructed. Clearly, the expressions are a little unwieldy for general $j$, but it is easy to determine them for any fixed value of $j$.

However, this construction utilizing linear combinations of odd powers of $J_i$ does not comprise all basis matrices for vectors under the cubic group when $j>3/2$. This is most easily seen by explicitly constructing the orthogonal basis (over $\mathbb{R}$) for \emph{any} Hermitean $(2j+1) \times (2j+1)$. Such a basis $\{\Sigma^A\}$ with $A=1,\dots,(2j+1)^2$ can be constructed by starting from products $J_k J_l J_m\cdots$ with at most $2j$ factors (again due to Cayley--Hamilton), making them symmetric and traceless with respect to all indices, and a successive Gram--Schmidt orthogonalization to ensure
\begin{align}
\mbox{tr}(\Sigma^A \Sigma^B) = (2j+1) \delta^{AB}.
\end{align}
The procedure is described in detail for $j=3/2$ in Ref. \cite{PhysRevB.95.075149}. A product of $j_{\rm tot}$ spin matrices constitutes a tensor of rank $j_{\rm tot}$ under $\text{SO}(3)$, and so each of the basis elements is a symmetric traceless tensor of rank $j_{\rm tot}$. By restricting the rotation group $\text{SO}(3)$ to transformations $J_i \to R_{ik}J_k$ with $R\in \text{O}$, these tensors transform according to irreducible representations of the cubic rotational group $\text{O}$.

Let us recall the irreducible representations of the cubic rotational group $\text{O}$. The group comprises 24 elements and permits five distinct irreducible  representations with dimensions $(d_1,d_2,d_3,d_4,d_5)=(1,1,2,3,3)$ satisfying
\begin{align}
 d_1^2+d_2^2+d_3^2+d_4^2+d_5^2 = 24,
\end{align}
which, in this order, are labeled $A_1$, $A_2$, $E$, $T_1$, and $T_2$. The one-dimensional $A_1$ is the trivial representation, whereas the three-dimensional $T_1$ is the "vector" representation we are after. The second-quantized Hamiltonian $\psi^\dagger p_i h_i \psi$ has $\psi$ transforming under rotations through the spin-$j$ representation, and so $h_i$ transforms under $j \otimes j = 0 \oplus 1 \oplus \dots \oplus 2j$. For each $j_{\rm tot}$ on the right-hand side of this equation, the corresponding $(2j_{\rm tot}+1)$ elements divide into multiplets that individually transform under $\text{O}$ according to the $A_1, A_2, E, T_1,$ or $T_2$ representations. Importantly, the number of such multiplets is fixed for every $j_{\rm tot}$ from group theory. The number of $T_1$ representations contained for $j_{\rm tot}\leq 2j$ then constitutes the number of matrices $K_i^{(n)}$ that span the space of vectors under the cubic group. We summarize the irreducible representations for $j_{\rm tot}\leq 11$ in Table \ref{TabIrrep}. Since we are interested in time-reversal symmetric Hamiltonians, only matrices that result from a product of an odd number of spin matrices are relevant, and so we discard vector representations for even $j_{\rm tot}$ from our analysis.

\renewcommand{\arraystretch}{1.8}
\begin{table}[t]
\begin{tabular}{|c|c|}
\hline
 \ $j_{\rm tot}$ \ & \ Irreducible representations of $\text{O}$ \ \\
\hline
 $0$ & $A_1$ \\
\hline
 $1$ & $T_1$ \\
\hline
 $2$ & $E\oplus T_2$ \\
\hline
 $3$ & $A_2 \oplus T_1 \oplus T_2$ \\
\hline
 $4$ & $A_1 \oplus E \oplus T_1 \oplus T_2$ \\
\hline
 $5$ & $E\oplus 2T_1 \oplus T_2$ \\
\hline
 $6$ & $A_1 \oplus A_2 \oplus E \oplus T_1 \oplus 2T_2$ \\
\hline
 $7$ & $A_2\oplus E \oplus 2T_1 \oplus 2T_2 $ \\
\hline
 $8$ & $A_1 \oplus 2E \oplus 2T_1 \oplus 2T_2$\\
\hline
 $9$ & $A_1 \oplus A_2 \oplus E \oplus 3T_1 \oplus 2T_2$ \\
\hline
 $10$ & $A_1 \oplus A_2 \oplus 2E \oplus 2T_1 \oplus 3T_2$\\
\hline
 $11$ & $A_2 \oplus 2E \oplus 3T_1 \oplus 3T_2$\\
\hline
\end{tabular}
\caption{Irreducible representations of the cubic rotational group $\text{O}$ for total spin $j_{\rm tot}$, see Table 5.6 in Ref. \cite{BookDresselhaus}. Vectors under the cubic group transform as $T_1$, and so the number of $T_1$-entries  for $j_{\rm tot} \leq 2 j$ counts the number $\mathcal{N}$ of admissible terms in Eq. (\ref{high3b}). In this work, we restrict to time-reversal symmetric Hamiltonians and, therefore, only include $T_1$-representations occurring for odd spin.}
\label{TabIrrep}
\end{table}
\renewcommand{\arraystretch}{1}

\subsection{Matrices $K_i^{(n)}$: Spin 5/2}\label{AppSpin52}

In this section, we present the complete orthonormal basis $\{\Sigma^A\}$ of $6\times 6$ Hermitean matrices starting from products of spin-5/2 matrices. Among these basis elements are five triplets that transform under the $T_1$ representation of the cubic rotational group. Only four of them, called $K_i^{(1)},\dots, K_i^{(4)}$, respect time-reversal symmetry according to $\{\mathcal{T},K_i^{(n)}\}=0$. These four matrices enter the Hamiltonian $H(\textbf{p})$ for $j=5/2$. Note that the matrices displayed here are linear combinations of the well-known Stevens operators \cite{StevensBook}. However, the conventional expressions for the Stevens operators give no hint on their transformation properties under the group $\text{O}$, and so we feel it is necessary to include a full list here. The complete set of matrices is presented in Table \ref{TabBasis}.

\renewcommand{\arraystretch}{2}
\begin{table*}[t]
\centering
\begin{tabular}{cc}
\begin{tabular}{|c|c|c|c|}
\hline
 \ $j_{\rm tot}$ \ & \ Rep \ & Elements & \\
\hline
 $0$ & $A_1$  & $\Sigma^1=\mathbb{1}$ & \\
\hline
 $1$ & $T_1$  & \begin{tabular}{c} $\Sigma^2 = 2\sqrt{\frac{3}{35}}J_1$ \\ $\Sigma^3 = 2\sqrt{\frac{3}{35}}J_2$  \\ $\Sigma^4 = 2\sqrt{\frac{3}{35}}J_3$  \end{tabular} & \ $K_i^{(1)}$ \ \\
\hline
 \multirow{2}{*}{$2$}  & $E$ & \begin{tabular}{c} $\Sigma^5 =  \frac{1}{2}\sqrt{\frac{3}{14}}(J_1^2-J_2^2)$ \\ $\Sigma^6 = \frac{1}{2\sqrt{14}}(3J_3^2-\frac{35}{4}\mathbb{1})$ \end{tabular} & \begin{tabular}{c} $\gamma_1$ \\ $\gamma_2$ \end{tabular} \\ \cline{2-4}  & $T_2$ & \begin{tabular}{c} $\Sigma^7= \frac{1}{2}\sqrt{\frac{3}{14}}\{J_2,J_3\}$ \\  $\Sigma^8 = \frac{1}{2}\sqrt{\frac{3}{14}}\{J_3,J_1\}$ \\   $\Sigma^9 = \ \frac{1}{2}\sqrt{\frac{3}{14}}\{J_1,J_2\}$ \end{tabular} & \begin{tabular}{c} $\gamma_3$ \\ $\gamma_4$ \\ $\gamma_5$ \end{tabular} \\ 
\hline
\multirow{3}{*}{$3$} & $A_2$ & \ $\Sigma^{10}=\frac{1}{\sqrt{18}}(J_1J_2J_3+J_3J_2J_1)$ & $\mathcal{W}$ \ \\  \cline{2-4}  & $T_1$ & \begin{tabular}{c} $\Sigma^{11} = \frac{1}{3}\sqrt{\frac{5}{6}}\Bigl(J_1^3-\frac{101}{20}J_1\Bigr) $ \\ $\Sigma^{12} = \frac{1}{3}\sqrt{\frac{5}{6}}\Bigl(J_2^3-\frac{101}{20}J_2\Bigr) $ \\ $\Sigma^{13} = \frac{1}{3}\sqrt{\frac{5}{6}}\Bigl(J_3^3-\frac{101}{20}J_3\Bigr)$ \end{tabular} & $K_i^{(2)}$ \\   \cline{2-4}  & $T_2$ & \begin{tabular}{c}  $\Sigma^{14}= \frac{1}{6\sqrt{2}}\{J_1,J_2^2-J_3^2\}$ \\  $\Sigma^{15} = \frac{1}{6\sqrt{2}}\{J_2,J_3^2-J_1^2\}$ \\ $\Sigma^{16} = \frac{1}{6\sqrt{2}}\{J_3,J_1^2-J_2^2\}$  \end{tabular} & \\ 
\hline
\multirow{1}{*}{$4$} & $A_1$ & \begin{tabular}{c} $\Sigma^{17}=\frac{1}{6\sqrt{2}}\Bigl(J_1^2J_2^2+J_1^2J_3^2+J_2^2J_3^2$\\ $+ J_2^2J_1^2+J_3^2J_1^2+J_3^2J_2^2 -\frac{259}{8}\mathbb{1}\Bigr)$ \\  $= \frac{-7\sqrt{2}}{9}\Bigl( \gamma_1^2+\gamma_2^2 - 2 \mathbb{1}\Bigr)$\end{tabular} & $\mathcal{I}$ \\ 
\hline
\end{tabular}
&
\begin{tabular}{|c|c|c|c|}
\hline
 \ $j_{\rm tot}$ \ & \ Rep \ & Elements & \\
\hline
\multirow{3}{*}{$4$}  & $E$ & \begin{tabular}{c} $\Sigma^{18} =  \frac{-7}{9}\sqrt{\frac{14}{5}}\Bigl(\gamma_1^2-\gamma_2^2 +\frac{10}{7}\sqrt{\frac{2}{7}}\gamma_2\Bigr)$ \\ $\Sigma^{19} = \frac{7}{9}\sqrt{\frac{14}{5}}\Bigl(\{\gamma_1,\gamma_2\} +\frac{10}{7}\sqrt{\frac{2}{7}}\gamma_1\Bigr)$ \end{tabular} &   \\ \cline{2-4} & $T_1$ & \begin{tabular}{c} $\Sigma^{20} = \frac{-7}{3\sqrt{30}}\{\gamma_3,(\gamma_1+\sqrt{3}\gamma_2)\}$ \\  $\Sigma^{21} = \frac{-7}{3\sqrt{30}}\{\gamma_4,(\gamma_1-\sqrt{3}\gamma_2)\}$ \\ $\Sigma^{22} = \frac{7}{3}\sqrt{\frac{2}{15}}\{\gamma_5,\gamma_1\}$ \end{tabular} &  \\ \cline{2-4} & $T_2$ & \begin{tabular}{c} $\Sigma^{23} = \frac{1}{4}\sqrt{\frac{21}{5}}\Bigl( \{J_1,\mathcal{W}\}-3\sqrt{\frac{3}{7}}\gamma_3\Bigr)$ \\ $\Sigma^{24} = \frac{1}{4}\sqrt{\frac{21}{5}}\Bigl( \{J_2,\mathcal{W}\}-3\sqrt{\frac{3}{7}}\gamma_4\Bigr)$ \\ $\Sigma^{25} = \frac{1}{4}\sqrt{\frac{21}{5}}\Bigl( \{J_3,\mathcal{W}\}-3\sqrt{\frac{3}{7}}\gamma_5\Bigr)$ \end{tabular} & \\
\hline
\multirow{5}{*}{$5$} & $E$ & \begin{tabular}{c} $\Sigma^{26}=\sqrt{\frac{14}{5}}\{\gamma_1,\mathcal{W}\}$ \\ $\Sigma^{27}=\sqrt{\frac{14}{5}}\{\gamma_2,\mathcal{W}\}$ \end{tabular} &  \\ \cline{2-4} & $T_1$ & \begin{tabular}{c} $\Sigma^{28}=\frac{1}{10}\sqrt{\frac{21}{2}}\Bigl(J_1^5 -\frac{145}{18}J_1^3 +\frac{11567}{1008}J_1\Bigr)$ \\ $\Sigma^{29}=\frac{1}{10}\sqrt{\frac{21}{2}}\Bigl(J_2^5 -\frac{145}{18}J_2^3 +\frac{11567}{1008}J_2\Bigr)$ \\ $\Sigma^{30}=\frac{1}{10}\sqrt{\frac{21}{2}}\Bigl(J_3^5 -\frac{145}{18}J_3^3 +\frac{11567}{1008}J_3\Bigr)$ \end{tabular} & \ $K_i^{(3)}$ \ \\ \cline{2-4} & $T_1$ & \begin{tabular}{c} $\Sigma^{31} =2\sqrt{\frac{3}{10}}\Bigl(\sqrt{2}\{J_1,\mathcal{I}\}+\frac{7}{12}J_1^5-\frac{95}{24}J_1^3+\frac{189}{64}J_1\Bigr)$ \\ $\Sigma^{32}=2\sqrt{\frac{3}{10}}\Bigl(\sqrt{2}\{J_2,\mathcal{I}\}+\frac{7}{12}J_2^5-\frac{95}{24}J_2^3+\frac{189}{64}J_2\Bigr)$ \\ $\Sigma^{33} =2\sqrt{\frac{3}{10}}\Bigl(\sqrt{2}\{J_3,\mathcal{I}\}+\frac{7}{12}J_3^5-\frac{95}{24}J_3^3+\frac{189}{64}J_3\Bigr)$ \end{tabular} & $K_i^{(4)}$ \\ \cline{2-4} & $T_2$ & \begin{tabular}{c} $\Sigma^{34}=\frac{-3}{5}\sqrt{\frac{7}{2}}\Bigl(\{K_1^{(2)},(\gamma_1+\sqrt{3}\gamma_2)\}-\frac{2}{3}\sqrt{\frac{2}{35}}\Sigma^{14}\Bigr)$ \\ $\Sigma^{35}=\frac{-3}{5}\sqrt{\frac{7}{2}}\Bigl(\{K_2^{(2)},(\gamma_1-\sqrt{3}\gamma_2)\}-\frac{2}{3}\sqrt{\frac{2}{35}}\Sigma^{15}\Bigr)$ \\ $\Sigma^{36}=\frac{3\sqrt{14}}{5}\Bigl(\{K_3^{(2)},\gamma_1\}+\frac{1}{3}\sqrt{\frac{2}{35}}\Sigma^{16}\Bigr)$ \end{tabular} & \\
\hline
\end{tabular}
\end{tabular}
\caption{Orthonormal basis (over $\mathbb{R}$) of Hermitean $6\times 6$ matrices constructed from products of spin-5/2 matrices. The four matrices $K_i^{(n)}$ that enter the Hamiltonian $H(\textbf{p})$ for $j=5/2$ transform under the $T_1$ representation of the cubic rotational group and are odd under time-reversal symmetry, i.e. belong to an odd $j_{\rm tot}$. We denote the entries $\Sigma^{10}$ and $\Sigma^{17}$ by $\mathcal{W}$ and $\mathcal{I}$, which are related, but not identical, to the symmetry operators $\hat{\mathcal{W}}$ and $\hat{\mathcal{I}}$ defined in the main text. We denote the five entries for $j_{\rm tot}=2$ by $\gamma_1,\dots,\gamma_5$, since their analogues (with different coefficients) for $j=3/2$ satisfy a Clifford algebra. This, however, is not true for $j=5/2$, implying that, for instance, $\mathcal{I}\neq 0$.}
\label{TabBasis}
\end{table*}
\renewcommand{\arraystretch}{1}

\subsection{Matrices $K_i^{(n)}$: Spin 7/2}\label{AppSpin72}

In this section, we display the matrices $K_i^{(n)}$ that enter the Hamiltonian $H(\textbf{p})$ for $j=7/2$. From Tables \ref{TabN} and \ref{TabIrrep} we deduce that, assuming time-reversal symmetry, we need to construct six orthonormal vectors. This is achieved easily by a Gram--Schmidt orthogonalization starting from the expressions $J_i$, $J_i^3$, $J_i^5$, $J_i^7$, and $\{\mathcal{I},J_i\}$, $\{\mathcal{I},J_i^3\}$ with the invariant tensor
\begin{align}
 \label{NewSpin1} \mathcal{I} = \mathcal{I}_{7/2} = \frac{1}{6\sqrt{33}}\Bigl(\sum_{k<l}(J_k^2J_l^2+J_l^2J_k^2) -\frac{819}{8}\mathbb{1}\Bigr).
\end{align}
We find
\begin{align}
  \label{NewSpin2} K_i^{(1)} ={}& \frac{2}{\sqrt{21}}J_i,\\
  \label{NewSpin3} K_i^{(2)} ={}& \frac{2}{3\sqrt{33}}\Bigl(J_i^3 - \frac{37}{4}J_i\Bigr),\\
  \label{NewSpin4} K_i^{(3)} ={}& \frac{\sqrt{7}}{10\sqrt{39}}\Bigl(J_i^5 -\frac{95}{6}J_i^3 +\frac{15709}{336}J_i\Bigr),\\
  \nonumber K_i^{(4)} ={}& 2\sqrt{\frac{11}{65}}\Bigl(\{J_i,\mathcal{I}\} +\frac{7}{12\sqrt{33}}J_i^5\\
  \label{NewSpin5} &-\frac{179}{24\sqrt{33}}J_i^3+\frac{21\sqrt{33}}{64}J_i\Bigr),
\end{align} 
and
\begin{align}
  \nonumber K_i^{(5)} ={}& \frac{\sqrt{143}}{210\sqrt{3}}\Bigl(J_i^7 -\frac{1043}{52} J_i^5 \\
  \label{NewSpin6} &+\frac{242837}{2288}J_i^3-\frac{1172307}{9152} J_i\Bigr),\\
  \nonumber K_i^{(6)} ={}& \frac{33}{5}{\sqrt{\frac{13}{7}}}\Bigl( \{K_i^{(2)},\mathcal{I}\}+\frac{16}{3\sqrt{21}}K_i^{(1)}\\
  \nonumber &-\frac{4}{11\sqrt{33}}K_i^{(2)}-\frac{100}{11\sqrt{273}}K_i^{(3)}\\
  \label{NewSpin7} &-\frac{20}{11}\sqrt{\frac{5}{39}}K_i^{(4)}+\frac{245}{33\sqrt{429}}K_i^{(5)}\Bigr).
\end{align}

\subsection{Basis change matrices}\label{AppBasis}
In this section, we list the basis change matrices employed in the discussion of $\text{O}(3)$ symmetric fixed points for $j=5/2$ in Sec. \ref{SecTopo}. We have
\begin{align}
 \label{band9} \mathcal{S}_1 = \begin{pmatrix} 1 & 0 & 0 & 0 & 0 & 0 \\ 0 & 0 & 0 & 0 & 0 & 1  \\ 0 & 1 & 0 & 0 & 0 & 0  \\ 0 & 0 & 0 & 0 & 1 & 0  \\ 0 & 0 & 1 & 0 & 0 & 0  \\ 0 & 0 & 0 & 1 & 0 & 0  \end{pmatrix}.
\end{align}
The matrix $\mathcal{S}_2$ is given by 
\begin{align}
 \label{band9b} \mathcal{S}_2 & =  \hat{\mathcal{S}}_2 \mathcal{S}_1,\\
  \hat{\mathcal{S}}_2&= \frac{1}{\sqrt{6}} \begin{pmatrix} \sqrt{5} & 0 & 0 & 1 & 0 & 0 \\ 0 & \sqrt{5} & 1 & 0 & 0 & 0 \\ 0 & -1 & \sqrt{5} & 0 & 0 & 0 \\ -1 & 0 & 0 & \sqrt{5} & 0 & 0 \\ 0 & 0 & 0 & 0 & \sqrt{6} & 0 \\ 0 & 0 & 0 & 0 & 0 & \sqrt{6} \end{pmatrix},
\end{align}
The matrix $\mathcal{S}_3$ reads
\begin{align}
 \nonumber \mathcal{S}_3 &= \begin{pmatrix} 0 & 0 & -1 & 0 & 0 & 0 \\ 0 & -b & 0 & 0 & 0 & \frac{2+\sqrt{10}}{6}  \\ \frac{1}{2-\sqrt{10}} & 0 & 0 & 0 & \frac{3}{\sqrt{22+4\sqrt{10}}} & 0  \\ 0 & 0 & 0 & 1 & 0 & 0  \\ 0 & \frac{2+\sqrt{10}}{6} & 0 & 0 & 0 & \frac{3}{\sqrt{22+4\sqrt{10}}}  \\ b & 0 & 0 & 0 & \frac{2+\sqrt{10}}{6} & 0  \end{pmatrix},\\ 
 \label{band9c}  b&=\frac{4\sqrt{2}+\sqrt{5}}{\sqrt{14-4\sqrt{10}}(7+2\sqrt{10})}.
\end{align}

\section{Renormalization group}

In this appendix, we present some more detailed equations that are useful for the computation of the fermion and photon self-energies in Fig. \ref{FigDiagrams}. In several cases we present multiple formulas for the same quantities, because they are more suitable in certain regimes, both analytically or numerically.

\subsection{Fermion self-energy: General remarks}\label{AppSelf}

\begin{figure}[t!]
\centering
\begin{minipage}{0.48\textwidth}
\includegraphics[width=\textwidth]{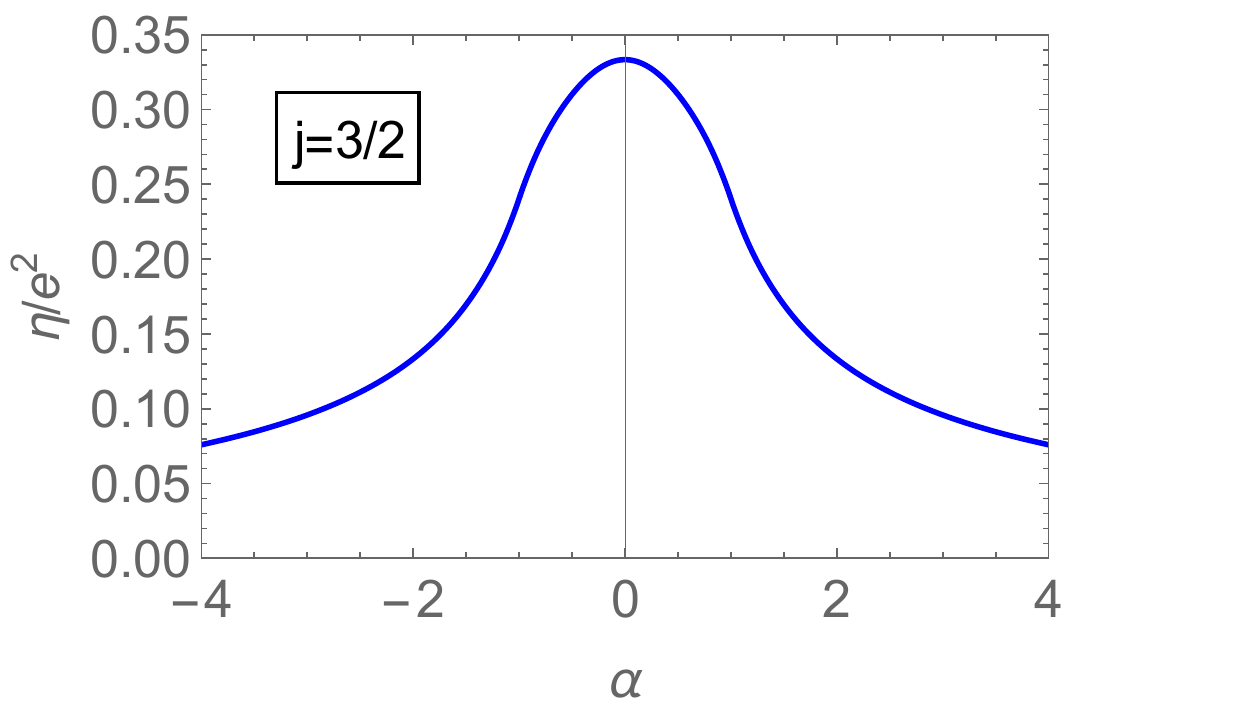}
\includegraphics[width=\textwidth]{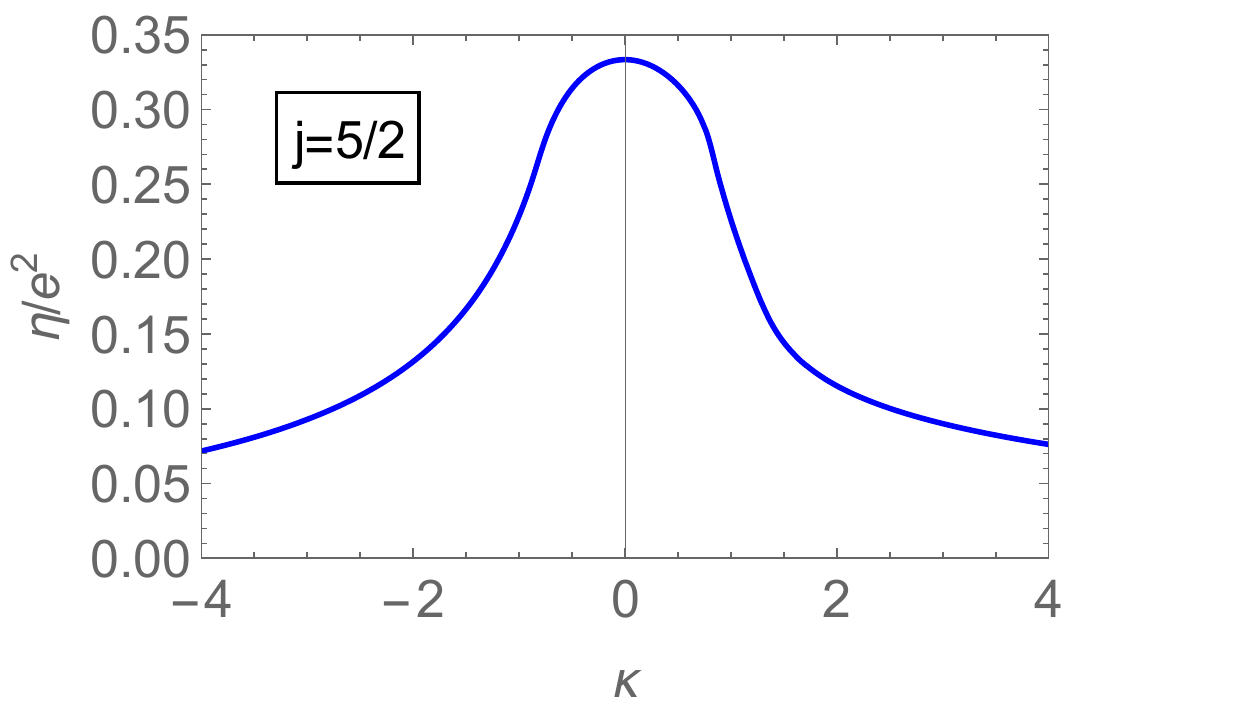}
\caption{Anomalous dimension for $j=3/2$ and $j=5/2$. We use the same scheme as in Figs. \ref{FigP} and \ref{FigFermions}, where we restrict the three-dimensional parameter space for spin 5/2 to the line $(\alpha,\beta,\gamma)=\frac{1}{3}(1,\sqrt{3},\sqrt{5})\kappa$ connecting the $\text{O}(3)$ and $\text{SO}(3)$ symmetric fixed points. The anomalous dimension for $\vec{\alpha}=0$ is $\eta=e^2/3$ for every $j$.}
\label{FigEta}
\end{minipage}
\end{figure}

In this section, we present the setup for computing the fermion self-energy and derive Eqs. (\ref{rg1})-(\ref{rg5}) involving the function $f$. For this purpose, we start from the one-loop correction to the fermion self-energy. To linear order in external momentum and up to an overall momentum-independent constant, it is given by
\begin{align}
 \label{rg16} \Sigma_\psi(\textbf{p}) = 2\bar{e}^2 \int_{q_0} \int_{\textbf{q}}^\prime \frac{\textbf{p}\cdot\textbf{q}}{q^4} G_\psi(Q).
\end{align}
We denote  $Q=(q_0,\textbf{q})$ with Euclidean frequency $q_0$ and
\begin{align}
 \label{rg17} \int_{q_0}(\dots) &= \int_{-\infty}^\infty \frac{\mbox{d}q_0}{2\pi}\ (\dots),\\
 \label{rg18} \int_{\textbf{q}}^\prime (\dots) &= \frac{1}{2\pi^2}\int_{\Lambda/b}^\Lambda \mbox{d}q\ q^2\int_{\Omega}(\dots).
\end{align}
The perturbative fermion propagator reads
\begin{align}
 \label{rg19} G_\psi(Q) &= \Bigl(\rmi q_0 \mathbb{1}+H(\textbf{q})\Bigr)^{-1}.
\end{align}
The $q_0$-integration in Eq. (\ref{rg16}) is understood as the principal value. The $q$-integration is trivial and we have
\begin{align}
 \label{rg19b} \Sigma_\psi(\textbf{p}) = 2e^2 \int_{q_0} \int_{\Omega} \frac{\textbf{p}\cdot\textbf{q}}{q} G_\psi(Q)\Bigr|_{q=1}.
\end{align}
We parametrize the self-energy correction according to
\begin{align}
 \label{rg20} \Sigma_\psi(\textbf{p}) = p_i\Bigl(\eta V_i + \sum_n \delta\alpha_n U_i^{(n)}\Bigr).
\end{align}
Utilizing the orthogonality from Eqs. (\ref{high3c}) and (\ref{high3d}), and choosing the external momentum $\textbf{p}=p\textbf{e}_3=(0,0,p)^T$ along the z-direction, we arrive at
\begin{align}
 \label{rg21} \eta & = \frac{1}{(2j+1)p} \mbox{tr}[V_3 \Sigma_\psi(p\textbf{e}_3)],\\
 \label{rg22} \delta\alpha_n & = \frac{1}{(2j+1)p} \mbox{tr}[U_3^{(n)} \Sigma_\psi(p\textbf{e}_3)].
\end{align}
After performing the trace in Eqs. (\ref{rg21}) and (\ref{rg22}), the frequency integration can be performed analytically. We are left with the angular integral $\int_\Omega(\dots)$, which, however, due to the cubic-only symmetry of the integrand can typically only be evaluated numerically. The anomalous dimension for $j=3/2$ and $j=5/2$ is shown in Fig. \ref{FigEta}.

There is a more elegant way to compute $\eta$ and $\delta \alpha_n$, which relies on the fact that the frequency dependence of the integrand in $\Sigma_\psi$ in Eq. (\ref{rg19b}) is solely due to the fermion propagator $G_\psi(Q)$. This results from to the perturbative photon propagator $G_a(\textbf{p})$ being frequency independent. Indeed, we have
\begin{align}
 \label{rg24} G_a(\textbf{p}) = \frac{\bar{e}^2}{p^2},
\end{align}
and any frequency dependence would be of the form $\propto (c^2p_0^2+p^2)^{-1}$, but the coupling $c$ is perturbatively irrelevant so that we can set $c=0$ when considering the infrared. Note, however, that the following argument does not rely on the specific form of the Hamiltonian $H(\textbf{p})$ and so applies to \emph{all} systems with a Lagrangian of the type (\ref{rg1}), where the electromagnetic field $A_\mu(\textbf{x},t)$ is approximated by the electrostatic component $a(\textbf{x})=A_0(0,\textbf{x})$. (This approximation, however, is common and typically fully sufficient in condensed matter systems.)

Denote the eigenvalues and eigenvectors of $\hat{H}(\phi,\theta)=H(\textbf{p})/p$ by $\hat{E}_\lambda(\phi,\theta)$ and $|\lambda(\phi,\theta)\rangle$. From contour integration we find the identity
\begin{align}
 \label{rg25} \int_{-\infty}^\infty \frac{\mbox{d}q_0}{2\pi} \frac{1}{\rmi q_0 \mathbb{1}+ \hat{H}} = \frac{1}{2}\sum_\lambda \sgn(\hat{E}_\lambda) |\lambda\rangle\langle \lambda|,
\end{align}
where the left hand side is defined through the principal value. Starting from Eqs. (\ref{rg21}) and (\ref{rg22}) with external momentum $\textbf{p}=p\textbf{e}_3$, we then have
\begin{align}
 \label{rg26} \eta &= \frac{e^2}{2j+1} \sum_\lambda \int_\Omega \hat{q}_3\ \sgn(\hat{E}_\lambda) \langle \lambda|V_3|\lambda\rangle,\\
 \label{rg27} \delta \alpha_{n} &=\frac{e^2}{2j+1} \sum_\lambda \int_\Omega \hat{q}_3\ \sgn(\hat{E}_\lambda) \langle \lambda|U_3^{(n))}|\lambda\rangle,
\end{align}
with $\hat{q}_i=q_i/q$. On the other hand, due to cubic symmetry, we may equally well project onto the 1- or 2-components, and sum each contribution with equal weight $1/3$ to obtain
\begin{align}
 \label{rg28} \eta &=\frac{e^2}{3(2j+1)} \sum_\lambda \int_\Omega \ \sgn(\hat{E}_\lambda) \langle \lambda|\hat{q}_iV_i|\lambda\rangle,
\end{align}
and similarly
\begin{align}
 \label{rg29}  \delta \alpha_n &=\frac{e^2}{3(2j+1)} \sum_\lambda \int_\Omega \ \sgn(\hat{E}_\lambda) \langle \lambda|\hat{q}_iU_i^{(n)}|\lambda\rangle.
\end{align}
Both equation together yield
\begin{align}
 \nonumber \eta + \vec{\alpha}\cdot\delta\vec{\alpha} &= \frac{e^2}{3(2j+1)} \sum_\lambda\int_\Omega \sgn(\hat{E}_\lambda) \langle \lambda| \hat{H}(\textbf{q})|\lambda\rangle\\
 \nonumber &= \frac{e^2}{3(2j+1)} \sum_\lambda\int_\Omega \sgn(\hat{E}_\lambda) \hat{E}_\lambda \\
 \label{rg30} &= \frac{e^2}{3(2j+1)} f(\vec{\alpha}).
\end{align}
This relation provides the first part towards proving Eqs. (\ref{rg4}) and (\ref{rg5}). In order to proceed, we need a second, linearly independent relation.

To obtain the second identity, we rewrite Eq. (\ref{rg29}) as
\begin{align}
 \label{rg32} \delta\alpha_n  &= \frac{e^2}{3(2j+1)} \sum_\lambda \int_\Omega \sgn(\hat{E}_\lambda) \Bigl\langle \lambda\Bigr|  \frac{\partial \hat{H}}{\partial \alpha_n} \Bigl|\lambda\Bigr\rangle.
\end{align}
Now employ the Feynman--Hellmann theorem
\begin{align}
 \label{rg33} \Bigl\langle \lambda\Bigr|  \frac{\partial \hat{H}}{\partial \alpha_n} \Bigl|\lambda\Bigr\rangle = \frac{\partial}{\partial\alpha_n} \hat{E}_\lambda
\end{align}
and (as we verified explicitly for many values of $j$) the fact that $\sgn(\hat{E}_\lambda)$ is independent of $\alpha_n$. We can then write
\begin{align}
 \label{rg34} \delta\alpha_n &= \frac{e^2}{3(2j+1)} \frac{\partial}{\partial \alpha_n} \sum_\lambda \int_\Omega \sgn(\hat{E}_\lambda)  \hat{E}_\lambda\\
 \label{rg35} &= \frac{e^2}{3(2j+1)} \frac{\partial f}{\partial \alpha_n}.
\end{align}
Together with Eq. (\ref{rg30}), this yields Eqs. (\ref{rg4}) and (\ref{rg5}).

\subsection{Fermion self-energy: Spin 5/2}

In this section, we explicitly compute the fermion self-energy for $j=5/2$ without relying on the function $f$ from the previous section. The Hamiltonian is given by Eq. (\ref{high16}) with
\begin{align}
 \vec{\alpha}^2 = \alpha^2+\beta^2+\gamma^2.
\end{align}
For $M=V,\ A,\ B,\ C$ define
\begin{align}
 \alpha_M = \begin{cases} 1 & (M=V)\\ \alpha & (M=A) \\ \beta & (M=B) \\ \gamma & (M=C) \end{cases}
\end{align}
Write the inverse propagator as 
\begin{align}
\mathcal{A}=q \hat{\mathcal{A}} = \rmi q_0\mathbb{1}+H(\textbf{q})
\end{align}
and $\hat{q}_i=q_i/q$, $\hat{q}_0=q_0/q$. We determine $G_\psi = \mathcal{A}^{-1}$ through the Cayley--Hamilton theorem, which implies that the inverse of any $6\times 6$ matrix $\mathcal{A}$ is given by 
\begin{align}
 \nonumber &\mathcal{A}^{-1}  =\frac{1}{\mbox{det}(\mathcal{A})}\Biggl( \frac{1}{120}\Bigl( [\mbox{tr}\mathcal{A}]^5-10[\mbox{tr}\mathcal{A}]^3\mbox{tr}(\mathcal{A}^2)\\
 \nonumber &+15\mbox{tr}\mathcal{A} [\mbox{tr}(\mathcal{A}^2)]^2+20[\mbox{tr} \mathcal{A}]^2\mbox{tr}(\mathcal{A}^3)-20\mbox{tr}(\mathcal{A}^2)\mbox{tr}(\mathcal{A}^3)\\
 \nonumber &-30\mbox{tr}\mathcal{A}\ \mbox{tr}(\mathcal{A}^4)+24\mbox{tr}(\mathcal{A}^5)\Bigr)\mathbb{1}_6-\frac{1}{24}\Bigl([\mbox{tr} \mathcal{A}]^4 \\
 \nonumber &-6[\mbox{tr} \mathcal{A}]^2\mbox{tr}(\mathcal{A}^2)+3[\mbox{tr}(\mathcal{A}^2)]^2+8\mbox{tr}\mathcal{A}\ \mbox{tr}(\mathcal{A}^3)\\
 \nonumber &-6\mbox{tr}(\mathcal{A}^4)\Bigr)\mathcal{A}+\frac{1}{6}\Bigl([\mbox{tr} \mathcal{A}]^3 -3\mbox{tr}\mathcal{A}\ \mbox{tr}(\mathcal{A}^2)+2\mbox{tr}(\mathcal{A}^3)\Bigr) A^2 \\
 &-\frac{1}{2}\Bigl([\mbox{tr} \mathcal{A}]^2-\mbox{tr}(\mathcal{A}^2)\Bigr)\mathcal{A}^3 +[\mbox{tr} \mathcal{A}]\mathcal{A}^4-\mathcal{A}^5\Biggr).
\end{align}
For this, determine the coefficients $e_1,\dots, e_5$ in the expansion
\begin{align}
 \nonumber \mbox{det}&(\mathcal{A}) = -q_0^6 - 3(1+\alpha^2+\beta^2+\gamma^2)q_0^4 q^2 +e_1 q_0^2q^4 \\
 &+ e_2 q^6 +e_3 q_0^2\sum_{k<l}q_k^2q_l^2 + e_4 q^2 \sum_{k<l}q_k^2q_l^2 + e_5 q_x^2q_y^2q_z^2,
\end{align}
and compute the functions $g_1^{(M)},\dots, g_6^{(M)}$ such that
\begin{align}
 \mbox{tr}(\mathcal{A}) &= 6\rmi q_0,\\
 \mbox{tr}(\mathcal{A}^2) &= 6[- q_0^2+(1+\alpha^2+\beta^2+\gamma^2)q^2],\\
 \mbox{tr}(\mathcal{A}^3) &= -6\rmi q_0[q_0^2-3(1+\alpha^2+\beta^2+\gamma^2)q^2],\\
 \nonumber \mbox{tr}(\mathcal{A}^4) &=6 q_0^4 -36(1+\alpha^2+\beta^2+\gamma^2)q_0^2q^2\\
 &+f_1(\alpha,\beta,\gamma)q^4+f_2(\alpha,\beta,\gamma) \sum_k q_k^4,
\end{align}
and
\begin{align}
 \mbox{tr}(M_i H) &= 6\alpha_M  q_i,\\
  \mbox{tr}(M_iH^3) &= 6\alpha_M\Bigl[g_1^{(M)}q_iq^2 + g_2^{(M)}q_i^3\Bigr],\\
 \nonumber \mbox{tr}(M_iH^5) &=6\alpha_M\Bigl[g_3^{(M)}q_iq^4+g_4^{(M)}q_i \sum_k q_k^4 \\
 &+g_5^{(M)}q_i^3q^2 +g_6^{(M)}q_i^5\Bigr].
\end{align}
Note that $\mbox{tr}(\mathcal{A}^5)$ is antisymmetric in $q_0$ and so vanishes from the frequency integration. We define
\begin{align}
 \nonumber f_1(\alpha,\beta,\gamma) &=6\Bigl(g_1^{(V)}+\alpha^2 g_1^{(A)}+\beta^2 g_1^{(B)}+\gamma^2 g_1^{(C)}\Bigr),\\
 f_2(\alpha,\beta,\gamma) &=6\Bigl(g_2^{(V)}+\alpha^2 g_2^{(A)}+\beta^2 g_2^{(B)}+\gamma^2 g_2^{(C)}\Bigr).
\end{align}
With these coefficient functions we find
\begin{widetext}
\begin{align}
 \nonumber \eta ={}& \frac{2}{3}e^2  \int_{\hat{q}_0}\int_\Omega \frac{1}{\mbox{det}(\hat{\mathcal{A}})}\Biggl(-\hat{q}_0^4+[-3(1+\vec{\alpha}^2)+g_1^{(V)}]\hat{q}_0^2+\frac{1}{4}f_1-\frac{9}{2}(1+\vec{\alpha}^2)^2+3g_1^{(V)}(1+\vec{\alpha}^2)\\
 \label{fix3b} &-g_3^{(V)}+\Bigl( \frac{1}{4}f_2 +g_2^{(V)}[\hat{q}_0^2+3(1+\vec{\alpha}^2)]-g_4^{(V)}-g_5^{(V)}\Bigr)\sum_k \hat{q}_k^4-g_6^{(V)}\sum_k \hat{q}_k^6\Biggr)
\end{align}
 and
\begin{align}
 \nonumber \delta \alpha ={}& \frac{2}{3}e^2 \alpha \int_{\hat{q}_0}\int_\Omega \frac{1}{\mbox{det}(\hat{\mathcal{A}})}\Biggl(-\hat{q}_0^4+[-3(1+\vec{\alpha}^2)+g_1^{(A)}]\hat{q}_0^2+\frac{1}{4}f_1-\frac{9}{2}(1+\vec{\alpha}^2)^2+3g_1^{(A)}(1+\vec{\alpha}^2)\\
 &-g_3^{(A)}+\Bigl( \frac{1}{4}f_2 +g_2^{(A)}[\hat{q}_0^2+3(1+\vec{\alpha}^2)]-g_4^{(A)}-g_5^{(A)}\Bigr)\sum_k \hat{q}_k^4-g_6^{(A)}\sum_k \hat{q}_k^6\Biggr),\\
 \nonumber \delta \beta ={}& \frac{2}{3}e^2 \beta \int_{\hat{q}_0}\int_\Omega \frac{1}{\mbox{det}(\hat{\mathcal{A}})}\Biggl(-\hat{q}_0^4+[-3(1+\vec{\alpha}^2)+g_1^{(B)}]\hat{q}_0^2+\frac{1}{4}f_1-\frac{9}{2}(1+\vec{\alpha}^2)^2+3g_1^{(B)}(1+\vec{\alpha}^2)\\
 &-g_3^{(B)}+\Bigl( \frac{1}{4}f_2 +g_2^{(B)}[\hat{q}_0^2+3(1+\vec{\alpha}^2)]-g_4^{(B)}-g_5^{(B)}\Bigr)\sum_k \hat{q}_k^4-g_6^{(B)}\sum_k \hat{q}_k^6\Biggr),\\
 \nonumber \delta \gamma ={}& \frac{2}{3}e^2 \gamma \int_{\hat{q}_0}\int_\Omega \frac{1}{\mbox{det}(\hat{\mathcal{A}})}\Biggl(-\hat{q}_0^4+[-3(1+\vec{\alpha}^2)+g_1^{(C)}]\hat{q}_0^2+\frac{1}{4}f_1-\frac{9}{2}(1+\vec{\alpha}^2)^2+3g_1^{(C)}(1+\vec{\alpha}^2)\\
  &-g_3^{(C)}+\Bigl( \frac{1}{4}f_2 +g_2^{(C)}[\hat{q}_0^2+3(1+\vec{\alpha}^2)]-g_4^{(C)}-g_5^{(C)}\Bigr)\sum_k \hat{q}_k^4-g_6^{(B)}\sum_k \hat{q}_k^6\Biggr).
\end{align}
The functions $h_{1,2,3}(\alpha,\beta,\gamma)$ read
\begin{align}
 \nonumber h_1(\alpha,\beta,\gamma) &= \frac{2}{3} \alpha \int_{\hat{q}_0}\int_\Omega \frac{1}{\mbox{det}(\hat{\mathcal{A}})}\Biggl((g_1^{(A)}-g_1^{(V)})[\hat{q}_0^2+3(1+\vec{\alpha}^2)]-(g_3^{(A)}-g_3^{(V)})\\
 \label{fix3c} &+\Bigl((g_2^{(A)}-g_2^{(V)})[\hat{q}_0^2+3(1+\vec{\alpha}^2)]-g_4^{(A)}-g_5^{(A)}+g_4^{(V)}+g_5^{(V)}\Bigr)\sum_k \hat{q}_k^4-(g_6^{(A)}-g_6^{(V)})\sum_k \hat{q}_k^6\Biggr),\\
 \nonumber h_2(\alpha,\beta,\gamma) &= \frac{2}{3} \beta \int_{\hat{q}_0}\int_\Omega \frac{1}{\mbox{det}(\hat{\mathcal{A}})}\Biggl((g_1^{(B)}-g_1^{(V)})[\hat{q}_0^2+3(1+\vec{\alpha}^2)]-(g_3^{(B)}-g_3^{(V)})\\
 \label{fix3d} &+\Bigl((g_2^{(B)}-g_2^{(V)})[\hat{q}_0^2+3(1+\vec{\alpha}^2)]-g_4^{(B)}-g_5^{(B)}+g_4^{(V)}+g_5^{(V)}\Bigr)\sum_k \hat{q}_k^4-(g_6^{(B)}-g_6^{(V)})\sum_k \hat{q}_k^6\Biggr).\\
 \nonumber h_3(\alpha,\beta,\gamma) &= \frac{2}{3} \gamma \int_{\hat{q}_0}\int_\Omega \frac{1}{\mbox{det}(\hat{\mathcal{A}})}\Biggl((g_1^{(C)}-g_1^{(V)})[\hat{q}_0^2+3(1+\vec{\alpha}^2)]-(g_3^{(C)}-g_3^{(V)})\\
 \label{fix3e} &+\Bigl((g_2^{(C)}-g_2^{(V)})[\hat{q}_0^2+3(1+\vec{\alpha}^2)]-g_4^{(C)}-g_5^{(C)}+g_4^{(V)}+g_5^{(V)}\Bigr)\sum_k \hat{q}_k^4-(g_6^{(C)}-g_6^{(V)})\sum_k \hat{q}_k^6\Biggr).
\end{align}
\end{widetext}

\subsection{Photon self-energy}\label{AppPhot}

In this section, we compute the photon self-energy and prove Eq. (\ref{rg11}). We write the one-loop correction to the photon self-energy as
\begin{align}
 \label{phot1} \Sigma_a(\textbf{p}) = \delta\frac{1}{e^2}\ p^2 =: P\ p^2 ,
\end{align}
which implies the one-loop correction
\begin{align}
 \label{phot2} \delta e^2 = -e^4 \delta \frac{1}{e^2} = -e^4 P.
\end{align}
(For convenience, we set $q=1$ for the loop momentum and we suppress the factor $2\pi^2$ from the momentum integration, thus $\bar{e}=e$ in this section.) The corresponding flow equation for the charge reads
\begin{align}
 \label{phot3} \frac{\mbox{d} e^2}{\mbox{d}\log b} &= -\eta e^2 -P(\vec{\alpha}) e^4.
\end{align}
We show that the function $P(\vec{\alpha})$ is positive.

The one-loop correction to the photon self-energy is given by
\begin{align}
 \label{phot4} \Sigma_a(\textbf{p}) = - \mbox{tr} \int_{q_0}\int_{\textbf{q}}^\prime G_\psi(Q+P) G_\psi(Q).
\end{align} 
We compute $P$ by means of
\begin{align}
 \label{phot5} P(\vec{\alpha})\ p^2 = \frac{1}{2} \frac{\partial^2 \Sigma_a(t\textbf{p})}{\partial t^2}\Bigr|_{t=0}.
\end{align}
Now recall $G_\psi(Q) = (\rmi q_0 \mathbb{1}+H(\textbf{q}))^{-1}$ and use the matrix formula
\begin{align}
 \label{phot6} \frac{\mbox{d}}{\mbox{d}t} M^{-1} = - M^{-1} \Bigl( \frac{\mbox{d}}{\mbox{d}t} M\Bigr) M^{-1},
\end{align}
to arrive at
\begin{align}
 \label{phot7} P(\vec{\alpha})\ p^2 &= -  \int_{q_0}\int_{\textbf{q}}^\prime \mbox{tr} \Bigl[ H(\textbf{p}) G_\psi(Q) H(\textbf{p}) G_\psi(Q)^3\Bigr].
\end{align}
We write $H(\textbf{p})=p_ih_i$ with
\begin{align}
 \label{phot8} h_i = V_i + \sum_{n=1}^{j-1/2} \alpha_n U_i^{(n)}
\end{align}
and find
\begin{align}
 \label{phot9} P(\vec{\alpha})\ p^2 &= - p_kp_l \int_{q_0}\int_\Omega \mbox{tr} \Bigl[ h_k G_\psi(Q) h_l G_\psi(Q)^3\Bigr].
\end{align}
The integral on the right-hand side must be proportional to $\delta_{kl}$, which is the only tensor with respect to cubic transformations with two indices.

To show positivity we start from the spectral decomposition of the fermion propagator
\begin{align}
 \label{phot9d} G_\psi(Q) = \sum_\lambda \frac{1}{\rmi q_0+\lambda}|\lambda\rangle\langle \lambda|
\end{align} 
and conclude that Eq. (\ref{phot9}), where we use cubic symmetry and set $k=l=3$, reads
\begin{align}
 \label{phot9e} P(\vec{\alpha}) &= - \int_{q_0} \int_\Omega \mbox{tr}(h_3 G_\psi h_3 G_\psi^3)\\
 \label{phot9f} &= -  \int_{q_0} \int_\Omega \sum_{\lambda} \sum_{\lambda'\neq \lambda} \frac{\langle \lambda|h_3|\lambda'\rangle \langle \lambda'|h_3|\lambda\rangle}{(\rmi q_0+E_\lambda)(\rmi q_0+E_{\lambda'})^3}.
\end{align}
We can assume $\lambda\neq \lambda'$ since only energies with opposite sign contribute to the frequency integration. Performing the latter we find
\begin{align}
 \nonumber P(\vec{\alpha})  &=  -\int_\Omega \sum_{\lambda} \sum_{\lambda'\neq \lambda} \Bigl[\theta(-E_{\lambda})-\theta(-E_{\lambda'})\Bigr] \frac{|\langle \lambda|h_3|\lambda'\rangle |^2}{(E_{\lambda}-E_{\lambda'})^3}\\
 \label{phot9g} & = 2 \sum_{E_\lambda>0}\sum_{E_{\lambda'}<0}  \int_\Omega \frac{|\langle \lambda|h_3|\lambda'\rangle |^2}{(E_{\lambda}-E_{\lambda'})^3}.
\end{align}
The expression on the right is manifestly positive, which proves our claim.

For $\vec{\alpha}=0$, using Clifford algebra, one verifies that
\begin{align}
 \label{phot9b} \int_{q_0}\int_\Omega \mbox{tr} \Bigl[ V_k G_\psi(Q) V_l G_\psi(Q)^3\Bigr] = - \frac{2j+1}{12}\delta_{kl}
\end{align}
so that
\begin{align}
P(\vec{0}) = \frac{2j+1}{12}.
\end{align}
The flow equation for the charge close to the relativistic fixed point becomes
\begin{align}
 \label{phot9c} \frac{\mbox{d} e^2}{\mbox{d}\log b}\Bigr|_{\vec{\alpha}=0} = - \eta_\star e^2 -\frac{2j+1}{12}e^4 = - \frac{2j+5}{12}e^4,
\end{align}
where we used $\eta_\star=e^2/3$ at the fixed point.

\subsection{Stability matrix}\label{AppStab}

In this section, we present two methods for computing the stability matrix. The first method employs the function $f$ defined in Eq. (\ref{rg1}). From Eqs. (\ref{rg4}) and (\ref{rg5}) we have
\begin{align}
 \label{stab2} \frac{\partial \eta}{\partial \alpha_{n'}} &= -\frac{e^2}{3(2j+1)}\vec{\alpha}\cdot \frac{\partial^2 f}{\partial \vec{\alpha}\partial \alpha_{n'}},\\
 \label{stab3} \frac{\partial \delta\alpha_n}{\partial \alpha_{n'}} &= \frac{e^2}{3(2j+1)}\frac{\partial^2 f}{\partial \alpha_n \partial \alpha_{n'}},
\end{align} 
and so, using $\dot{\alpha}_n = -\eta \alpha_n + \delta \alpha_n$, we arrive at
\begin{align}
 \label{stab4} M_{nn'} = -\eta \delta_{nn'} + \frac{e^2}{3(2j+1)}\Bigl[ \alpha_n\vec{\alpha}\cdot \frac{\partial^2 f}{\partial \vec{\alpha}\partial \alpha_{n'}}+\frac{\partial^2 f}{\partial \alpha_n \partial \alpha_{n'}}\Bigr],
\end{align}
as given in Eq. (\ref{rg9}). In particular, at the relativistic $\text{O}(3)$ symmetric fixed point with $\vec{\alpha}=0$ we have
\begin{align}
 \label{stab5} M_{nn'}\Bigr|_{\vec{\alpha}=0} &= -\frac{e^2}{3} \delta_{nn'} +\frac{e^2}{3(2j+1)} \frac{\partial^2 f}{\partial \alpha_n \partial \alpha_{n'}}
\end{align}

The second method to compute the stability matrix directly employs Eq. (\ref{rg19})-(\ref{rg22}). For this use
\begin{align}
 \label{stab7} \frac{\partial}{\partial \alpha_n} G_\psi = - G_\psi q_i U_i^{(n)}G_\psi
\end{align}
to arrive at
\begin{align}
 \nonumber \frac{\partial \eta}{\partial \alpha_{n'}} &= -\frac{2e^2}{(2j+1)} \int_{q_0}\int_\Omega\hat{q}_3\hat{q}_i\ \mbox{tr}\Bigl[ V_3 G_\psi U_i^{(n')}G_\psi\Bigr]_{q=1} \\
 \label{stab8} \frac{\partial \delta\alpha_n}{\partial \alpha_{n'}} &= -\frac{2e^2}{(2j+1)} \int_{q_0}\int_\Omega\hat{q}_3\hat{q}_i\ \mbox{tr}\Bigl[ U_3^{(n)} G_\psi U_i^{(n')}G_\psi\Bigr]_{q=1}.
\end{align}

In the case of the $\text{O}(3)$ and $\text{SO}(3)$ symmetric fixed points, the propagator $G_\psi$ has a fairly simple structure and the trace and integrals can be evaluated analytically. In particular, for the relativistic case with $\vec{\alpha}_\star=0$ we employ the perturbative propagator
\begin{align}
 G_\psi(Q) = \frac{-\rmi q_0 \mathbb{1}+q_i V_i}{q_0^2+q^2}
\end{align}
to find
\begin{align}
 \nonumber \frac{\partial \delta\alpha_n}{\partial \alpha_{n'}} ={}& \frac{e^2}{2(2j+1)} \int_\Omega\hat{q}_3\hat{q}_i\ \mbox{tr}[U_3^{(n)}  U_i^{(n')}] \\
 \nonumber &-\frac{e^2}{2(2j+1)} \int_\Omega\hat{q}_3\hat{q}_i\hat{q}_k \hat{q}_l\ \mbox{tr}[U_3^{(n)} V_k U_i^{(n')}V_l]\\
 \nonumber ={}&\frac{e^2}{6} \delta_{nm}-\frac{e^2}{30(2j+1)}  (\delta_{3i}\delta_{kl}+\delta_{3k}\delta_{il}+\delta_{3l}\delta_{ik})\\
 \label{stab9} &\times \mbox{tr}[U_3^{(n)} V_k U_i^{(n')}V_l].
\end{align}
In the last line we used
\begin{align}
 \label{stab10} &\int_\Omega  \hat{q}_i \hat{q}_j = \frac{1}{3}\delta_{ij},\\
 \label{stab10b}  \int_\Omega  &\hat{q}_i \hat{q}_j\hat{q}_k\hat{q}_l = \frac{1}{15} (\delta_{ij}\delta_{kl}+\delta_{ik}\delta_{jl}+\delta_{il}\delta_{jk}).
\end{align}
We can now further employ the fact that $V_i$ commutes with $U_i^{(n)}$ but anticommutes with $U_{k\neq i}^{(n)}$, $V_i^2=\mathbb{1}$, and the symmetry of the given trace with respect to $1\leftrightarrow 2$. This yields
\begin{align}
 \nonumber &(\delta_{3i}\delta_{kl}+\delta_{3k}\delta_{il}+\delta_{3l}\delta_{ik}) \mbox{tr}[U_3^{(n)} V_k U_i^{(n')}V_l] \\
 \label{stab11} &=(2j+1)\delta_{nm}+4 \mbox{tr}[U_3^{(n)} V_3 U_1^{(n')}V_1]
\end{align}
Together with $h_n = -\eta \alpha_n + \delta\alpha_n$ this yields Eq. (\ref{rel3}). (The term containing $\partial \eta/\partial \alpha_{n'}$ vanishes at the relativistic fixed point, because it is multiplied by $\alpha_{\star,n}=0$.)

\section{Band topology}\label{AppTopo}

We determine the band topology in the normal phase through the Chern numbers of the bands that cross at the origin. For this purpose we denote the eigenvalues and eigenvectors of $H(\textbf{q})$ by $E_\lambda$ and $|\lambda\rangle$, respectively, with the $\textbf{q}$-dependence being implicit. The Berry flux of the band $\lambda$ is
\begin{align}
 \label{top1} \mathcal{F}^{(\lambda)}_{ij} &=\sum_{\lambda' \neq \lambda} \frac{\langle \lambda|\partial_i H|\lambda'\rangle \langle \lambda' | \partial_j H|\lambda\rangle - \{i\leftrightarrow j\}}{(E_\lambda-E_{\lambda'})^2}\\
 \label{top2} &=2\rmi\ \text{Im} \sum_{\lambda'\neq \lambda} \frac{\langle \lambda|\partial_i H|\lambda'\rangle \langle \lambda' | \partial_j H|\lambda\rangle}{(E_\lambda-E_{\lambda'})^2}
\end{align}
with
\begin{align}
 \label{top3} \partial_i = \frac{\partial}{\partial q_i}.
\end{align}
The Chern number $\mathcal{C}_\lambda$ of the band $\lambda$ follows from the pseudomagnetic field
\begin{align}
 \label{top4} \mathcal{B}_i^{(\lambda)} = \frac{1}{2}\vare_{ijk} \mathcal{F}^{(\lambda)}_{jk}
\end{align}
by means of the surface integral surrounding the origin according to
\begin{align}
 \label{top5} \mathcal{C}_\lambda &= \frac{1}{2\pi} \oint_{|\textbf{q}|=q_0} \mbox{d}\vec{S}\cdot \vec{\mathcal{B}}^{(\lambda)}(\textbf{q})\\
 \label{top6} &= 2 \int_\Omega \Bigl( \hat{q}_1 \mathcal{B}_1^{(\lambda)}+\hat{q}_2 \mathcal{B}_2^{(\lambda)}+\hat{q}_3\mathcal{B}_3^{(\lambda)}\Bigr)\\
 \label{top7} &=-2\rmi \int_\Omega\Bigl( \hat{q}_1 \mathcal{F}_{23}^{(\lambda)}+\hat{q}_2\mathcal{F}_{31}^{(\lambda)}+\hat{q}_3 \mathcal{F}_{12}^{(\lambda)}\Bigr).
\end{align}
Due to cubic symmetry, the three contributions to the integral $\mathcal{C}_\lambda$ are identical and so we have
\begin{align}
 \label{top8} \mathcal{C}_\lambda=-6\rmi \int_\Omega \hat{q}_3 \mathcal{F}_{12}^{(\lambda)}.
\end{align}
The total monopole charge $\mathcal{Q}$ is the sum of the Chern numbers of the positive energy bands. For the practical implementation, it is useful to utilize the time-reversal symmetry of the Hamiltonian through $\{\mathcal{T},H(\textbf{q})\}=0$. Indeed, this equation implies that only the positive energy bands $E_\lambda(\textbf{q})>0$ and their eigenvectors $|\lambda\rangle$ need to be computed, because the negative energy bands $-E_\lambda(\textbf{q})$ have eigenvectors $\mathcal{T}|\lambda\rangle$.

\end{appendix}

\bibliographystyle{apsrev4-1}
\bibliography{refs_opt}

\begin{thebibliography}{65}%
\makeatletter
\providecommand \@ifxundefined [1]{%
 \@ifx{#1\undefined}
}%
\providecommand \@ifnum [1]{%
 \ifnum #1\expandafter \@firstoftwo
 \else \expandafter \@secondoftwo
 \fi
}%
\providecommand \@ifx [1]{%
 \ifx #1\expandafter \@firstoftwo
 \else \expandafter \@secondoftwo
 \fi
}%
\providecommand \natexlab [1]{#1}%
\providecommand \enquote  [1]{``#1''}%
\providecommand \bibnamefont  [1]{#1}%
\providecommand \bibfnamefont [1]{#1}%
\providecommand \citenamefont [1]{#1}%
\providecommand \href@noop [0]{\@secondoftwo}%
\providecommand \href [0]{\begingroup \@sanitize@url \@href}%
\providecommand \@href[1]{\@@startlink{#1}\@@href}%
\providecommand \@@href[1]{\endgroup#1\@@endlink}%
\providecommand \@sanitize@url [0]{\catcode `\\12\catcode `\$12\catcode
  `\&12\catcode `\#12\catcode `\^12\catcode `\_12\catcode `\%12\relax}%
\providecommand \@@startlink[1]{}%
\providecommand \@@endlink[0]{}%
\providecommand \url  [0]{\begingroup\@sanitize@url \@url }%
\providecommand \@url [1]{\endgroup\@href {#1}{\urlprefix }}%
\providecommand \urlprefix  [0]{URL }%
\providecommand \Eprint [0]{\href }%
\providecommand \doibase [0]{http://dx.doi.org/}%
\providecommand \selectlanguage [0]{\@gobble}%
\providecommand \bibinfo  [0]{\@secondoftwo}%
\providecommand \bibfield  [0]{\@secondoftwo}%
\providecommand \translation [1]{[#1]}%
\providecommand \BibitemOpen [0]{}%
\providecommand \bibitemStop [0]{}%
\providecommand \bibitemNoStop [0]{.\EOS\space}%
\providecommand \EOS [0]{\spacefactor3000\relax}%
\providecommand \BibitemShut  [1]{\csname bibitem#1\endcsname}%
\let\auto@bib@innerbib\@empty
\bibitem [{\citenamefont {Geim}\ and\ \citenamefont
  {Novoselov}(2007)}]{GeimReview}%
  \BibitemOpen
  \bibfield  {author} {\bibinfo {author} {\bibfnamefont {A.~K.}\ \bibnamefont
  {Geim}}\ and\ \bibinfo {author} {\bibfnamefont {K.~S.}\ \bibnamefont
  {Novoselov}},\ }\href {\doibase 10.1038/nmat1849} {\bibfield  {journal}
  {\bibinfo  {journal} {Nat. Mater.}\ }\textbf {\bibinfo {volume} {6}},\
  \bibinfo {pages} {183} (\bibinfo {year} {2007})}\BibitemShut {NoStop}%
\bibitem [{\citenamefont {Hasan}\ and\ \citenamefont
  {Kane}(2010)}]{RevModPhys.82.3045}%
  \BibitemOpen
  \bibfield  {author} {\bibinfo {author} {\bibfnamefont {M.~Z.}\ \bibnamefont
  {Hasan}}\ and\ \bibinfo {author} {\bibfnamefont {C.~L.}\ \bibnamefont
  {Kane}},\ }\href {\doibase 10.1103/RevModPhys.82.3045} {\bibfield  {journal}
  {\bibinfo  {journal} {Rev. Mod. Phys.}\ }\textbf {\bibinfo {volume} {82}},\
  \bibinfo {pages} {3045} (\bibinfo {year} {2010})}\BibitemShut {NoStop}%
\bibitem [{\citenamefont {Witczak-Krempa}\ \emph {et~al.}(2014)\citenamefont
  {Witczak-Krempa}, \citenamefont {Chen}, \citenamefont {Kim},\ and\
  \citenamefont {Balents}}]{WKrempa}%
  \BibitemOpen
  \bibfield  {author} {\bibinfo {author} {\bibfnamefont {W.}~\bibnamefont
  {Witczak-Krempa}}, \bibinfo {author} {\bibfnamefont {G.}~\bibnamefont
  {Chen}}, \bibinfo {author} {\bibfnamefont {Y.~B.}\ \bibnamefont {Kim}}, \
  and\ \bibinfo {author} {\bibfnamefont {L.}~\bibnamefont {Balents}},\ }\href
  {\doibase 10.1146/annurev-conmatphys-020911-125138} {\bibfield  {journal}
  {\bibinfo  {journal} {Annu. Rev. Condens. Matter Phys.}\ }\textbf {\bibinfo
  {volume} {5}},\ \bibinfo {pages} {57} (\bibinfo {year} {2014})}\BibitemShut
  {NoStop}%
\bibitem [{\citenamefont {Armitage}\ \emph {et~al.}(2018)\citenamefont
  {Armitage}, \citenamefont {Mele},\ and\ \citenamefont
  {Vishwanath}}]{RevModPhys.90.015001}%
  \BibitemOpen
  \bibfield  {author} {\bibinfo {author} {\bibfnamefont {N.~P.}\ \bibnamefont
  {Armitage}}, \bibinfo {author} {\bibfnamefont {E.~J.}\ \bibnamefont {Mele}},
  \ and\ \bibinfo {author} {\bibfnamefont {A.}~\bibnamefont {Vishwanath}},\
  }\href {\doibase 10.1103/RevModPhys.90.015001} {\bibfield  {journal}
  {\bibinfo  {journal} {Rev. Mod. Phys.}\ }\textbf {\bibinfo {volume} {90}},\
  \bibinfo {pages} {015001} (\bibinfo {year} {2018})}\BibitemShut {NoStop}%
\bibitem [{\citenamefont {Gao}\ \emph {et~al.}(2019)\citenamefont {Gao},
  \citenamefont {Venderbos}, \citenamefont {Kim},\ and\ \citenamefont
  {Rappe}}]{GaoReview}%
  \BibitemOpen
  \bibfield  {author} {\bibinfo {author} {\bibfnamefont {H.}~\bibnamefont
  {Gao}}, \bibinfo {author} {\bibfnamefont {J.~W.}\ \bibnamefont {Venderbos}},
  \bibinfo {author} {\bibfnamefont {Y.}~\bibnamefont {Kim}}, \ and\ \bibinfo
  {author} {\bibfnamefont {A.~M.}\ \bibnamefont {Rappe}},\ }\href {\doibase
  10.1146/annurev-matsci-070218-010049} {\bibfield  {journal} {\bibinfo
  {journal} {Annu. Rev. Mater. Res.}\ }\textbf {\bibinfo {volume} {49}},\
  \bibinfo {pages} {153} (\bibinfo {year} {2019})}\BibitemShut {NoStop}%
\bibitem [{\citenamefont {Weinberg}(2005)}]{BookWeinberg}%
  \BibitemOpen
  \bibfield  {author} {\bibinfo {author} {\bibfnamefont {S.}~\bibnamefont
  {Weinberg}},\ }\href@noop {} {\emph {\bibinfo {title} {{The Quantum Theory of
  Fields — Volume I: Foundations}}}}\ (\bibinfo  {publisher} {Cambridge
  University Press, Cambridge},\ \bibinfo {year} {2005})\BibitemShut {NoStop}%
\bibitem [{\citenamefont {Rarita}\ and\ \citenamefont
  {Schwinger}(1941)}]{PhysRev.60.61}%
  \BibitemOpen
  \bibfield  {author} {\bibinfo {author} {\bibfnamefont {W.}~\bibnamefont
  {Rarita}}\ and\ \bibinfo {author} {\bibfnamefont {J.}~\bibnamefont
  {Schwinger}},\ }\href {\doibase 10.1103/PhysRev.60.61} {\bibfield  {journal}
  {\bibinfo  {journal} {Phys. Rev.}\ }\textbf {\bibinfo {volume} {60}},\
  \bibinfo {pages} {61} (\bibinfo {year} {1941})}\BibitemShut {NoStop}%
\bibitem [{\citenamefont {Bradlyn}\ \emph {et~al.}(2016)\citenamefont
  {Bradlyn}, \citenamefont {Cano}, \citenamefont {Wang}, \citenamefont
  {Vergniory}, \citenamefont {Felser}, \citenamefont {Cava},\ and\
  \citenamefont {Bernevig}}]{Bradlynaaf5037}%
  \BibitemOpen
  \bibfield  {author} {\bibinfo {author} {\bibfnamefont {B.}~\bibnamefont
  {Bradlyn}}, \bibinfo {author} {\bibfnamefont {J.}~\bibnamefont {Cano}},
  \bibinfo {author} {\bibfnamefont {Z.}~\bibnamefont {Wang}}, \bibinfo {author}
  {\bibfnamefont {M.~G.}\ \bibnamefont {Vergniory}}, \bibinfo {author}
  {\bibfnamefont {C.}~\bibnamefont {Felser}}, \bibinfo {author} {\bibfnamefont
  {R.~J.}\ \bibnamefont {Cava}}, \ and\ \bibinfo {author} {\bibfnamefont
  {B.~A.}\ \bibnamefont {Bernevig}},\ }\href
  {https://science.sciencemag.org/content/353/6299/aaf5037} {\bibfield
  {journal} {\bibinfo  {journal} {Science}\ }\textbf {\bibinfo {volume} {353}}
  (\bibinfo {year} {2016})}\BibitemShut {NoStop}%
\bibitem [{\citenamefont {Tang}\ \emph {et~al.}(2017)\citenamefont {Tang},
  \citenamefont {Zhou},\ and\ \citenamefont {Zhang}}]{PhysRevLett.119.206402}%
  \BibitemOpen
  \bibfield  {author} {\bibinfo {author} {\bibfnamefont {P.}~\bibnamefont
  {Tang}}, \bibinfo {author} {\bibfnamefont {Q.}~\bibnamefont {Zhou}}, \ and\
  \bibinfo {author} {\bibfnamefont {S.-C.}\ \bibnamefont {Zhang}},\ }\href
  {\doibase 10.1103/PhysRevLett.119.206402} {\bibfield  {journal} {\bibinfo
  {journal} {Phys. Rev. Lett.}\ }\textbf {\bibinfo {volume} {119}},\ \bibinfo
  {pages} {206402} (\bibinfo {year} {2017})}\BibitemShut {NoStop}%
\bibitem [{\citenamefont {Chang}\ \emph {et~al.}(2018)\citenamefont {Chang},
  \citenamefont {Wieder}, \citenamefont {Schindler}, \citenamefont {Sanchez},
  \citenamefont {Belopolski}, \citenamefont {Huang}, \citenamefont {Singh},
  \citenamefont {Wu}, \citenamefont {Chang}, \citenamefont {Neupert},
  \citenamefont {Xu}, \citenamefont {Lin},\ and\ \citenamefont
  {Hasan}}]{chang2018topological}%
  \BibitemOpen
  \bibfield  {author} {\bibinfo {author} {\bibfnamefont {G.}~\bibnamefont
  {Chang}}, \bibinfo {author} {\bibfnamefont {B.~J.}\ \bibnamefont {Wieder}},
  \bibinfo {author} {\bibfnamefont {F.}~\bibnamefont {Schindler}}, \bibinfo
  {author} {\bibfnamefont {D.~S.}\ \bibnamefont {Sanchez}}, \bibinfo {author}
  {\bibfnamefont {I.}~\bibnamefont {Belopolski}}, \bibinfo {author}
  {\bibfnamefont {S.-M.}\ \bibnamefont {Huang}}, \bibinfo {author}
  {\bibfnamefont {B.}~\bibnamefont {Singh}}, \bibinfo {author} {\bibfnamefont
  {D.}~\bibnamefont {Wu}}, \bibinfo {author} {\bibfnamefont {T.-R.}\
  \bibnamefont {Chang}}, \bibinfo {author} {\bibfnamefont {T.}~\bibnamefont
  {Neupert}}, \bibinfo {author} {\bibfnamefont {S.-Y.}\ \bibnamefont {Xu}},
  \bibinfo {author} {\bibfnamefont {H.}~\bibnamefont {Lin}}, \ and\ \bibinfo
  {author} {\bibfnamefont {M.~Z.}\ \bibnamefont {Hasan}},\ }\href {\doibase
  10.1038/s41563-018-0169-3} {\bibfield  {journal} {\bibinfo  {journal} {Nat.
  Mater.}\ }\textbf {\bibinfo {volume} {17}},\ \bibinfo {pages} {978} (\bibinfo
  {year} {2018})}\BibitemShut {NoStop}%
\bibitem [{\citenamefont {Takane}\ \emph {et~al.}(2019)\citenamefont {Takane},
  \citenamefont {Wang}, \citenamefont {Souma}, \citenamefont {Nakayama},
  \citenamefont {Nakamura}, \citenamefont {Oinuma}, \citenamefont {Nakata},
  \citenamefont {Iwasawa}, \citenamefont {Cacho}, \citenamefont {Kim},
  \citenamefont {Horiba}, \citenamefont {Kumigashira}, \citenamefont
  {Takahashi}, \citenamefont {Ando},\ and\ \citenamefont
  {Sato}}]{PhysRevLett.122.076402}%
  \BibitemOpen
  \bibfield  {author} {\bibinfo {author} {\bibfnamefont {D.}~\bibnamefont
  {Takane}}, \bibinfo {author} {\bibfnamefont {Z.}~\bibnamefont {Wang}},
  \bibinfo {author} {\bibfnamefont {S.}~\bibnamefont {Souma}}, \bibinfo
  {author} {\bibfnamefont {K.}~\bibnamefont {Nakayama}}, \bibinfo {author}
  {\bibfnamefont {T.}~\bibnamefont {Nakamura}}, \bibinfo {author}
  {\bibfnamefont {H.}~\bibnamefont {Oinuma}}, \bibinfo {author} {\bibfnamefont
  {Y.}~\bibnamefont {Nakata}}, \bibinfo {author} {\bibfnamefont
  {H.}~\bibnamefont {Iwasawa}}, \bibinfo {author} {\bibfnamefont
  {C.}~\bibnamefont {Cacho}}, \bibinfo {author} {\bibfnamefont
  {T.}~\bibnamefont {Kim}}, \bibinfo {author} {\bibfnamefont {K.}~\bibnamefont
  {Horiba}}, \bibinfo {author} {\bibfnamefont {H.}~\bibnamefont {Kumigashira}},
  \bibinfo {author} {\bibfnamefont {T.}~\bibnamefont {Takahashi}}, \bibinfo
  {author} {\bibfnamefont {Y.}~\bibnamefont {Ando}}, \ and\ \bibinfo {author}
  {\bibfnamefont {T.}~\bibnamefont {Sato}},\ }\href {\doibase
  10.1103/PhysRevLett.122.076402} {\bibfield  {journal} {\bibinfo  {journal}
  {Phys. Rev. Lett.}\ }\textbf {\bibinfo {volume} {122}},\ \bibinfo {pages}
  {076402} (\bibinfo {year} {2019})}\BibitemShut {NoStop}%
\bibitem [{\citenamefont {Rao}\ \emph {et~al.}(2019)\citenamefont {Rao},
  \citenamefont {Li}, \citenamefont {Zhang}, \citenamefont {Tian},
  \citenamefont {Li}, \citenamefont {Fu}, \citenamefont {Tang}, \citenamefont
  {Wang}, \citenamefont {Li}, \citenamefont {Fan}, \citenamefont {Li},
  \citenamefont {Huang}, \citenamefont {Liu}, \citenamefont {Long},
  \citenamefont {Fang}, \citenamefont {Weng}, \citenamefont {Shi},
  \citenamefont {Lei}, \citenamefont {Sun}, \citenamefont {Qian},\ and\
  \citenamefont {Ding}}]{Rao2019ObservationOU}%
  \BibitemOpen
  \bibfield  {author} {\bibinfo {author} {\bibfnamefont {Z.}~\bibnamefont
  {Rao}}, \bibinfo {author} {\bibfnamefont {H.}~\bibnamefont {Li}}, \bibinfo
  {author} {\bibfnamefont {T.}~\bibnamefont {Zhang}}, \bibinfo {author}
  {\bibfnamefont {S.}~\bibnamefont {Tian}}, \bibinfo {author} {\bibfnamefont
  {C.}~\bibnamefont {Li}}, \bibinfo {author} {\bibfnamefont {B.}~\bibnamefont
  {Fu}}, \bibinfo {author} {\bibfnamefont {C.}~\bibnamefont {Tang}}, \bibinfo
  {author} {\bibfnamefont {L.}~\bibnamefont {Wang}}, \bibinfo {author}
  {\bibfnamefont {Z.~L.}\ \bibnamefont {Li}}, \bibinfo {author} {\bibfnamefont
  {W.}~\bibnamefont {Fan}}, \bibinfo {author} {\bibfnamefont {J.}~\bibnamefont
  {Li}}, \bibinfo {author} {\bibfnamefont {Y.}~\bibnamefont {Huang}}, \bibinfo
  {author} {\bibfnamefont {Z.}~\bibnamefont {Liu}}, \bibinfo {author}
  {\bibfnamefont {Y.}~\bibnamefont {Long}}, \bibinfo {author} {\bibfnamefont
  {C.}~\bibnamefont {Fang}}, \bibinfo {author} {\bibfnamefont {H.}~\bibnamefont
  {Weng}}, \bibinfo {author} {\bibfnamefont {Y.}~\bibnamefont {Shi}}, \bibinfo
  {author} {\bibfnamefont {H.}~\bibnamefont {Lei}}, \bibinfo {author}
  {\bibfnamefont {Y.}~\bibnamefont {Sun}}, \bibinfo {author} {\bibfnamefont
  {T.~W.}\ \bibnamefont {Qian}}, \ and\ \bibinfo {author} {\bibfnamefont
  {H.}~\bibnamefont {Ding}},\ }\href {\doibase 10.1038/s41586-019-1031-8}
  {\bibfield  {journal} {\bibinfo  {journal} {Nature}\ }\textbf {\bibinfo
  {volume} {567}},\ \bibinfo {pages} {496} (\bibinfo {year}
  {2019})}\BibitemShut {NoStop}%
\bibitem [{\citenamefont {S.~Sanchez}\ \emph {et~al.}(2019)\citenamefont
  {S.~Sanchez}, \citenamefont {Belopolski}, \citenamefont {A.~Cochran},
  \citenamefont {Xu}, \citenamefont {Yin}, \citenamefont {Chang}, \citenamefont
  {Xie}, \citenamefont {Manna}, \citenamefont {Süß}, \citenamefont {Huang},
  \citenamefont {Alidoust}, \citenamefont {Multer}, \citenamefont {S.~Zhang},
  \citenamefont {Shumiya}, \citenamefont {Wang}, \citenamefont {Wang},
  \citenamefont {Chang}, \citenamefont {Felser}, \citenamefont {Xu},\ and\
  \citenamefont {Hasan}}]{Sanchez}%
  \BibitemOpen
  \bibfield  {author} {\bibinfo {author} {\bibfnamefont {D.}~\bibnamefont
  {S.~Sanchez}}, \bibinfo {author} {\bibfnamefont {I.}~\bibnamefont
  {Belopolski}}, \bibinfo {author} {\bibfnamefont {T.}~\bibnamefont
  {A.~Cochran}}, \bibinfo {author} {\bibfnamefont {X.}~\bibnamefont {Xu}},
  \bibinfo {author} {\bibfnamefont {J.}~\bibnamefont {Yin}}, \bibinfo {author}
  {\bibfnamefont {G.}~\bibnamefont {Chang}}, \bibinfo {author} {\bibfnamefont
  {W.}~\bibnamefont {Xie}}, \bibinfo {author} {\bibfnamefont {K.}~\bibnamefont
  {Manna}}, \bibinfo {author} {\bibfnamefont {V.}~\bibnamefont {Süß}},
  \bibinfo {author} {\bibfnamefont {C.-Y.}\ \bibnamefont {Huang}}, \bibinfo
  {author} {\bibfnamefont {N.}~\bibnamefont {Alidoust}}, \bibinfo {author}
  {\bibfnamefont {D.}~\bibnamefont {Multer}}, \bibinfo {author} {\bibfnamefont
  {S.}~\bibnamefont {S.~Zhang}}, \bibinfo {author} {\bibfnamefont
  {N.}~\bibnamefont {Shumiya}}, \bibinfo {author} {\bibfnamefont
  {X.}~\bibnamefont {Wang}}, \bibinfo {author} {\bibfnamefont {G.-Q.}\
  \bibnamefont {Wang}}, \bibinfo {author} {\bibfnamefont {T.-R.}\ \bibnamefont
  {Chang}}, \bibinfo {author} {\bibfnamefont {C.}~\bibnamefont {Felser}},
  \bibinfo {author} {\bibfnamefont {S.-Y.}\ \bibnamefont {Xu}}, \ and\ \bibinfo
  {author} {\bibfnamefont {M.~Z.}\ \bibnamefont {Hasan}},\ }\href {\doibase
  10.1038/s41586-019-1037-2} {\bibfield  {journal} {\bibinfo  {journal}
  {Nature}\ }\textbf {\bibinfo {volume} {567}},\ \bibinfo {pages} {500}
  (\bibinfo {year} {2019})}\BibitemShut {NoStop}%
\bibitem [{\citenamefont {{{Schr{\"o}ter}, Niels B.~M. and {Pei}, Ding and
  {Vergniory}, Maia G. and {Sun}, Yan and {Manna}, Kaustuv and {de Juan},
  Fernando and {Krieger}, Jonas. A. and {S{\"u}ss}, Vicky and {Schmidt}, Marcus
  and {Dudin}, Pavel}}(2019)}]{SchroeterNature}%
  \BibitemOpen
  \bibfield  {author} {\bibinfo {author} {\bibnamefont {{{Schr{\"o}ter}, Niels
  B.~M. and {Pei}, Ding and {Vergniory}, Maia G. and {Sun}, Yan and {Manna},
  Kaustuv and {de Juan}, Fernando and {Krieger}, Jonas. A. and {S{\"u}ss},
  Vicky and {Schmidt}, Marcus and {Dudin}, Pavel}}},\ }\href {\doibase
  10.1038/s41567-019-0511-y} {\bibfield  {journal} {\bibinfo  {journal} {Nat.
  Phys.}\ } (\bibinfo {year} {2019}),\ 10.1038/s41567-019-0511-y}\BibitemShut
  {NoStop}%
\bibitem [{\citenamefont {Lv}\ \emph {et~al.}(2019)\citenamefont {Lv},
  \citenamefont {Feng}, \citenamefont {Zhao}, \citenamefont {Yuan},
  \citenamefont {Zong}, \citenamefont {Luo}, \citenamefont {Yu}, \citenamefont
  {Huang}, \citenamefont {Strocov}, \citenamefont {Chikina}, \citenamefont
  {Soluyanov}, \citenamefont {Gedik}, \citenamefont {Shi}, \citenamefont
  {Qian},\ and\ \citenamefont {Ding}}]{PhysRevB.99.241104}%
  \BibitemOpen
  \bibfield  {author} {\bibinfo {author} {\bibfnamefont {B.~Q.}\ \bibnamefont
  {Lv}}, \bibinfo {author} {\bibfnamefont {Z.-L.}\ \bibnamefont {Feng}},
  \bibinfo {author} {\bibfnamefont {J.-Z.}\ \bibnamefont {Zhao}}, \bibinfo
  {author} {\bibfnamefont {N.~F.~Q.}\ \bibnamefont {Yuan}}, \bibinfo {author}
  {\bibfnamefont {A.}~\bibnamefont {Zong}}, \bibinfo {author} {\bibfnamefont
  {K.~F.}\ \bibnamefont {Luo}}, \bibinfo {author} {\bibfnamefont
  {R.}~\bibnamefont {Yu}}, \bibinfo {author} {\bibfnamefont {Y.-B.}\
  \bibnamefont {Huang}}, \bibinfo {author} {\bibfnamefont {V.~N.}\ \bibnamefont
  {Strocov}}, \bibinfo {author} {\bibfnamefont {A.}~\bibnamefont {Chikina}},
  \bibinfo {author} {\bibfnamefont {A.~A.}\ \bibnamefont {Soluyanov}}, \bibinfo
  {author} {\bibfnamefont {N.}~\bibnamefont {Gedik}}, \bibinfo {author}
  {\bibfnamefont {Y.-G.}\ \bibnamefont {Shi}}, \bibinfo {author} {\bibfnamefont
  {T.}~\bibnamefont {Qian}}, \ and\ \bibinfo {author} {\bibfnamefont
  {H.}~\bibnamefont {Ding}},\ }\href {\doibase 10.1103/PhysRevB.99.241104}
  {\bibfield  {journal} {\bibinfo  {journal} {Phys. Rev. B}\ }\textbf {\bibinfo
  {volume} {99}},\ \bibinfo {pages} {241104} (\bibinfo {year}
  {2019})}\BibitemShut {NoStop}%
\bibitem [{\citenamefont {Cano}\ \emph {et~al.}(2019)\citenamefont {Cano},
  \citenamefont {Bradlyn},\ and\ \citenamefont
  {Vergniory}}]{cano2019multifold}%
  \BibitemOpen
  \bibfield  {author} {\bibinfo {author} {\bibfnamefont {J.}~\bibnamefont
  {Cano}}, \bibinfo {author} {\bibfnamefont {B.}~\bibnamefont {Bradlyn}}, \
  and\ \bibinfo {author} {\bibfnamefont {M.}~\bibnamefont {Vergniory}},\ }\href
  {\doibase 10.1063/1.5124314} {\bibfield  {journal} {\bibinfo  {journal} {APL
  Materials}\ }\textbf {\bibinfo {volume} {7}},\ \bibinfo {pages} {101125}
  (\bibinfo {year} {2019})}\BibitemShut {NoStop}%
\bibitem [{\citenamefont {Hsieh}\ \emph {et~al.}(2014)\citenamefont {Hsieh},
  \citenamefont {Liu},\ and\ \citenamefont {Fu}}]{PhysRevB.90.081112}%
  \BibitemOpen
  \bibfield  {author} {\bibinfo {author} {\bibfnamefont {T.~H.}\ \bibnamefont
  {Hsieh}}, \bibinfo {author} {\bibfnamefont {J.}~\bibnamefont {Liu}}, \ and\
  \bibinfo {author} {\bibfnamefont {L.}~\bibnamefont {Fu}},\ }\href {\doibase
  10.1103/PhysRevB.90.081112} {\bibfield  {journal} {\bibinfo  {journal} {Phys.
  Rev. B}\ }\textbf {\bibinfo {volume} {90}},\ \bibinfo {pages} {081112}
  (\bibinfo {year} {2014})}\BibitemShut {NoStop}%
\bibitem [{\citenamefont {Liang}\ and\ \citenamefont
  {Yu}(2016)}]{PhysRevB.93.045113}%
  \BibitemOpen
  \bibfield  {author} {\bibinfo {author} {\bibfnamefont {L.}~\bibnamefont
  {Liang}}\ and\ \bibinfo {author} {\bibfnamefont {Y.}~\bibnamefont {Yu}},\
  }\href {\doibase 10.1103/PhysRevB.93.045113} {\bibfield  {journal} {\bibinfo
  {journal} {Phys. Rev. B}\ }\textbf {\bibinfo {volume} {93}},\ \bibinfo
  {pages} {045113} (\bibinfo {year} {2016})}\BibitemShut {NoStop}%
\bibitem [{\citenamefont {Isobe}\ and\ \citenamefont
  {Fu}(2016)}]{PhysRevB.93.241113}%
  \BibitemOpen
  \bibfield  {author} {\bibinfo {author} {\bibfnamefont {H.}~\bibnamefont
  {Isobe}}\ and\ \bibinfo {author} {\bibfnamefont {L.}~\bibnamefont {Fu}},\
  }\href {\doibase 10.1103/PhysRevB.93.241113} {\bibfield  {journal} {\bibinfo
  {journal} {Phys. Rev. B}\ }\textbf {\bibinfo {volume} {93}},\ \bibinfo
  {pages} {241113} (\bibinfo {year} {2016})}\BibitemShut {NoStop}%
\bibitem [{\citenamefont {Ezawa}(2016)}]{PhysRevB.94.195205}%
  \BibitemOpen
  \bibfield  {author} {\bibinfo {author} {\bibfnamefont {M.}~\bibnamefont
  {Ezawa}},\ }\href {\doibase 10.1103/PhysRevB.94.195205} {\bibfield  {journal}
  {\bibinfo  {journal} {Phys. Rev. B}\ }\textbf {\bibinfo {volume} {94}},\
  \bibinfo {pages} {195205} (\bibinfo {year} {2016})}\BibitemShut {NoStop}%
\bibitem [{\citenamefont {Lin}\ and\ \citenamefont
  {Nandkishore}(2018)}]{PhysRevB.97.134521}%
  \BibitemOpen
  \bibfield  {author} {\bibinfo {author} {\bibfnamefont {Y.-P.}\ \bibnamefont
  {Lin}}\ and\ \bibinfo {author} {\bibfnamefont {R.~M.}\ \bibnamefont
  {Nandkishore}},\ }\href {\doibase 10.1103/PhysRevB.97.134521} {\bibfield
  {journal} {\bibinfo  {journal} {Phys. Rev. B}\ }\textbf {\bibinfo {volume}
  {97}},\ \bibinfo {pages} {134521} (\bibinfo {year} {2018})}\BibitemShut
  {NoStop}%
\bibitem [{\citenamefont {Kawakami}\ \emph {et~al.}(2018)\citenamefont
  {Kawakami}, \citenamefont {Okamura}, \citenamefont {Kobayashi},\ and\
  \citenamefont {Sato}}]{PhysRevX.8.041026}%
  \BibitemOpen
  \bibfield  {author} {\bibinfo {author} {\bibfnamefont {T.}~\bibnamefont
  {Kawakami}}, \bibinfo {author} {\bibfnamefont {T.}~\bibnamefont {Okamura}},
  \bibinfo {author} {\bibfnamefont {S.}~\bibnamefont {Kobayashi}}, \ and\
  \bibinfo {author} {\bibfnamefont {M.}~\bibnamefont {Sato}},\ }\href {\doibase
  10.1103/PhysRevX.8.041026} {\bibfield  {journal} {\bibinfo  {journal} {Phys.
  Rev. X}\ }\textbf {\bibinfo {volume} {8}},\ \bibinfo {pages} {041026}
  (\bibinfo {year} {2018})}\BibitemShut {NoStop}%
\bibitem [{\citenamefont {Roy}\ \emph {et~al.}(2018)\citenamefont {Roy},
  \citenamefont {Kennett}, \citenamefont {Yang},\ and\ \citenamefont
  {Juri\ifmmode \check{c}\else \v{c}\fi{}i\ifmmode~\acute{c}\else
  \'{c}\fi{}}}]{PhysRevLett.121.157602}%
  \BibitemOpen
  \bibfield  {author} {\bibinfo {author} {\bibfnamefont {B.}~\bibnamefont
  {Roy}}, \bibinfo {author} {\bibfnamefont {M.~P.}\ \bibnamefont {Kennett}},
  \bibinfo {author} {\bibfnamefont {K.}~\bibnamefont {Yang}}, \ and\ \bibinfo
  {author} {\bibfnamefont {V.}~\bibnamefont {Juri\ifmmode \check{c}\else
  \v{c}\fi{}i\ifmmode~\acute{c}\else \'{c}\fi{}}},\ }\href {\doibase
  10.1103/PhysRevLett.121.157602} {\bibfield  {journal} {\bibinfo  {journal}
  {Phys. Rev. Lett.}\ }\textbf {\bibinfo {volume} {121}},\ \bibinfo {pages}
  {157602} (\bibinfo {year} {2018})}\BibitemShut {NoStop}%
\bibitem [{\citenamefont {Boettcher}(2020)}]{PhysRevLett.124.127602}%
  \BibitemOpen
  \bibfield  {author} {\bibinfo {author} {\bibfnamefont {I.}~\bibnamefont
  {Boettcher}},\ }\href {\doibase 10.1103/PhysRevLett.124.127602} {\bibfield
  {journal} {\bibinfo  {journal} {Phys. Rev. Lett.}\ }\textbf {\bibinfo
  {volume} {124}},\ \bibinfo {pages} {127602} (\bibinfo {year}
  {2020})}\BibitemShut {NoStop}%
\bibitem [{\citenamefont {Link}\ \emph {et~al.}(2020)\citenamefont {Link},
  \citenamefont {Boettcher},\ and\ \citenamefont
  {Herbut}}]{PhysRevB.101.184503}%
  \BibitemOpen
  \bibfield  {author} {\bibinfo {author} {\bibfnamefont {J.~M.}\ \bibnamefont
  {Link}}, \bibinfo {author} {\bibfnamefont {I.}~\bibnamefont {Boettcher}}, \
  and\ \bibinfo {author} {\bibfnamefont {I.~F.}\ \bibnamefont {Herbut}},\
  }\href {\doibase 10.1103/PhysRevB.101.184503} {\bibfield  {journal} {\bibinfo
   {journal} {Phys. Rev. B}\ }\textbf {\bibinfo {volume} {101}},\ \bibinfo
  {pages} {184503} (\bibinfo {year} {2020})}\BibitemShut {NoStop}%
\bibitem [{\citenamefont {Lin}\ and\ \citenamefont
  {Hsiao}(2020)}]{lin2020dual}%
  \BibitemOpen
  \bibfield  {author} {\bibinfo {author} {\bibfnamefont {Y.-P.}\ \bibnamefont
  {Lin}}\ and\ \bibinfo {author} {\bibfnamefont {W.-H.}\ \bibnamefont
  {Hsiao}},\ }\href {https://arxiv.org/abs/2004.03609} {\bibfield  {journal}
  {\bibinfo  {journal} {arXiv:2004.03609}\ } (\bibinfo {year}
  {2020})}\BibitemShut {NoStop}%
\bibitem [{\citenamefont {Abrikosov}(1974)}]{abrikosov}%
  \BibitemOpen
  \bibfield  {author} {\bibinfo {author} {\bibfnamefont {A.~A.}\ \bibnamefont
  {Abrikosov}},\ }\href@noop {} {\bibfield  {journal} {\bibinfo  {journal}
  {Sov. Phys. JETP}\ }\textbf {\bibinfo {volume} {39}},\ \bibinfo {pages} {709}
  (\bibinfo {year} {1974})}\BibitemShut {NoStop}%
\bibitem [{\citenamefont {Abrikosov}\ and\ \citenamefont
  {Beneslavskii}(1971)}]{abrben}%
  \BibitemOpen
  \bibfield  {author} {\bibinfo {author} {\bibfnamefont {A.~A.}\ \bibnamefont
  {Abrikosov}}\ and\ \bibinfo {author} {\bibfnamefont {S.~D.}\ \bibnamefont
  {Beneslavskii}},\ }\href@noop {} {\bibfield  {journal} {\bibinfo  {journal}
  {Sov. Phys. JETP}\ }\textbf {\bibinfo {volume} {32}},\ \bibinfo {pages} {699}
  (\bibinfo {year} {1971})}\BibitemShut {NoStop}%
\bibitem [{\citenamefont {Luttinger}(1956)}]{luttinger}%
  \BibitemOpen
  \bibfield  {author} {\bibinfo {author} {\bibfnamefont {J.~M.}\ \bibnamefont
  {Luttinger}},\ }\href {\doibase 10.1103/PhysRev.102.1030} {\bibfield
  {journal} {\bibinfo  {journal} {Phys. Rev.}\ }\textbf {\bibinfo {volume}
  {102}},\ \bibinfo {pages} {1030} (\bibinfo {year} {1956})}\BibitemShut
  {NoStop}%
\bibitem [{\citenamefont {Moon}\ \emph {et~al.}(2013)\citenamefont {Moon},
  \citenamefont {Xu}, \citenamefont {Kim},\ and\ \citenamefont
  {Balents}}]{moon}%
  \BibitemOpen
  \bibfield  {author} {\bibinfo {author} {\bibfnamefont {E.-G.}\ \bibnamefont
  {Moon}}, \bibinfo {author} {\bibfnamefont {C.}~\bibnamefont {Xu}}, \bibinfo
  {author} {\bibfnamefont {Y.~B.}\ \bibnamefont {Kim}}, \ and\ \bibinfo
  {author} {\bibfnamefont {L.}~\bibnamefont {Balents}},\ }\href {\doibase
  10.1103/PhysRevLett.111.206401} {\bibfield  {journal} {\bibinfo  {journal}
  {Phys. Rev. Lett.}\ }\textbf {\bibinfo {volume} {111}},\ \bibinfo {pages}
  {206401} (\bibinfo {year} {2013})}\BibitemShut {NoStop}%
\bibitem [{\citenamefont {Herbut}\ and\ \citenamefont
  {Janssen}(2014)}]{PhysRevLett.113.106401}%
  \BibitemOpen
  \bibfield  {author} {\bibinfo {author} {\bibfnamefont {I.~F.}\ \bibnamefont
  {Herbut}}\ and\ \bibinfo {author} {\bibfnamefont {L.}~\bibnamefont
  {Janssen}},\ }\href {\doibase 10.1103/PhysRevLett.113.106401} {\bibfield
  {journal} {\bibinfo  {journal} {Phys. Rev. Lett.}\ }\textbf {\bibinfo
  {volume} {113}},\ \bibinfo {pages} {106401} (\bibinfo {year}
  {2014})}\BibitemShut {NoStop}%
\bibitem [{\citenamefont {Savary}\ \emph {et~al.}(2014)\citenamefont {Savary},
  \citenamefont {Moon},\ and\ \citenamefont {Balents}}]{PhysRevX.4.041027}%
  \BibitemOpen
  \bibfield  {author} {\bibinfo {author} {\bibfnamefont {L.}~\bibnamefont
  {Savary}}, \bibinfo {author} {\bibfnamefont {E.-G.}\ \bibnamefont {Moon}}, \
  and\ \bibinfo {author} {\bibfnamefont {L.}~\bibnamefont {Balents}},\ }\href
  {\doibase 10.1103/PhysRevX.4.041027} {\bibfield  {journal} {\bibinfo
  {journal} {Phys. Rev. X}\ }\textbf {\bibinfo {volume} {4}},\ \bibinfo {pages}
  {041027} (\bibinfo {year} {2014})}\BibitemShut {NoStop}%
\bibitem [{\citenamefont {Janssen}\ and\ \citenamefont
  {Herbut}(2015)}]{PhysRevB.92.045117}%
  \BibitemOpen
  \bibfield  {author} {\bibinfo {author} {\bibfnamefont {L.}~\bibnamefont
  {Janssen}}\ and\ \bibinfo {author} {\bibfnamefont {I.~F.}\ \bibnamefont
  {Herbut}},\ }\href {\doibase 10.1103/PhysRevB.92.045117} {\bibfield
  {journal} {\bibinfo  {journal} {Phys. Rev. B}\ }\textbf {\bibinfo {volume}
  {92}},\ \bibinfo {pages} {045117} (\bibinfo {year} {2015})}\BibitemShut
  {NoStop}%
\bibitem [{\citenamefont {Murray}\ \emph {et~al.}(2015)\citenamefont {Murray},
  \citenamefont {Vafek},\ and\ \citenamefont {Balents}}]{PhysRevB.92.035137}%
  \BibitemOpen
  \bibfield  {author} {\bibinfo {author} {\bibfnamefont {J.~M.}\ \bibnamefont
  {Murray}}, \bibinfo {author} {\bibfnamefont {O.}~\bibnamefont {Vafek}}, \
  and\ \bibinfo {author} {\bibfnamefont {L.}~\bibnamefont {Balents}},\ }\href
  {\doibase 10.1103/PhysRevB.92.035137} {\bibfield  {journal} {\bibinfo
  {journal} {Phys. Rev. B}\ }\textbf {\bibinfo {volume} {92}},\ \bibinfo
  {pages} {035137} (\bibinfo {year} {2015})}\BibitemShut {NoStop}%
\bibitem [{\citenamefont {Boettcher}\ and\ \citenamefont
  {Herbut}(2016)}]{PhysRevB.93.205138}%
  \BibitemOpen
  \bibfield  {author} {\bibinfo {author} {\bibfnamefont {I.}~\bibnamefont
  {Boettcher}}\ and\ \bibinfo {author} {\bibfnamefont {I.~F.}\ \bibnamefont
  {Herbut}},\ }\href {\doibase 10.1103/PhysRevB.93.205138} {\bibfield
  {journal} {\bibinfo  {journal} {Phys. Rev. B}\ }\textbf {\bibinfo {volume}
  {93}},\ \bibinfo {pages} {205138} (\bibinfo {year} {2016})}\BibitemShut
  {NoStop}%
\bibitem [{\citenamefont {Janssen}\ and\ \citenamefont
  {Herbut}(2016)}]{PhysRevB.93.165109}%
  \BibitemOpen
  \bibfield  {author} {\bibinfo {author} {\bibfnamefont {L.}~\bibnamefont
  {Janssen}}\ and\ \bibinfo {author} {\bibfnamefont {I.~F.}\ \bibnamefont
  {Herbut}},\ }\href {\doibase 10.1103/PhysRevB.93.165109} {\bibfield
  {journal} {\bibinfo  {journal} {Phys. Rev. B}\ }\textbf {\bibinfo {volume}
  {93}},\ \bibinfo {pages} {165109} (\bibinfo {year} {2016})}\BibitemShut
  {NoStop}%
\bibitem [{\citenamefont {Meinert}(2016)}]{PhysRevLett.116.137001}%
  \BibitemOpen
  \bibfield  {author} {\bibinfo {author} {\bibfnamefont {M.}~\bibnamefont
  {Meinert}},\ }\href {\doibase 10.1103/PhysRevLett.116.137001} {\bibfield
  {journal} {\bibinfo  {journal} {Phys. Rev. Lett.}\ }\textbf {\bibinfo
  {volume} {116}},\ \bibinfo {pages} {137001} (\bibinfo {year}
  {2016})}\BibitemShut {NoStop}%
\bibitem [{\citenamefont {Brydon}\ \emph {et~al.}(2016)\citenamefont {Brydon},
  \citenamefont {Wang}, \citenamefont {Weinert},\ and\ \citenamefont
  {Agterberg}}]{PhysRevLett.116.177001}%
  \BibitemOpen
  \bibfield  {author} {\bibinfo {author} {\bibfnamefont {P.~M.~R.}\
  \bibnamefont {Brydon}}, \bibinfo {author} {\bibfnamefont {L.}~\bibnamefont
  {Wang}}, \bibinfo {author} {\bibfnamefont {M.}~\bibnamefont {Weinert}}, \
  and\ \bibinfo {author} {\bibfnamefont {D.~F.}\ \bibnamefont {Agterberg}},\
  }\href {\doibase 10.1103/PhysRevLett.116.177001} {\bibfield  {journal}
  {\bibinfo  {journal} {Phys. Rev. Lett.}\ }\textbf {\bibinfo {volume} {116}},\
  \bibinfo {pages} {177001} (\bibinfo {year} {2016})}\BibitemShut {NoStop}%
\bibitem [{\citenamefont {Goswami}\ \emph {et~al.}(2017)\citenamefont
  {Goswami}, \citenamefont {Roy},\ and\ \citenamefont
  {Das~Sarma}}]{PhysRevB.95.085120}%
  \BibitemOpen
  \bibfield  {author} {\bibinfo {author} {\bibfnamefont {P.}~\bibnamefont
  {Goswami}}, \bibinfo {author} {\bibfnamefont {B.}~\bibnamefont {Roy}}, \ and\
  \bibinfo {author} {\bibfnamefont {S.}~\bibnamefont {Das~Sarma}},\ }\href
  {\doibase 10.1103/PhysRevB.95.085120} {\bibfield  {journal} {\bibinfo
  {journal} {Phys. Rev. B}\ }\textbf {\bibinfo {volume} {95}},\ \bibinfo
  {pages} {085120} (\bibinfo {year} {2017})}\BibitemShut {NoStop}%
\bibitem [{\citenamefont {Boettcher}\ and\ \citenamefont
  {Herbut}(2017)}]{PhysRevB.95.075149}%
  \BibitemOpen
  \bibfield  {author} {\bibinfo {author} {\bibfnamefont {I.}~\bibnamefont
  {Boettcher}}\ and\ \bibinfo {author} {\bibfnamefont {I.~F.}\ \bibnamefont
  {Herbut}},\ }\href {\doibase 10.1103/PhysRevB.95.075149} {\bibfield
  {journal} {\bibinfo  {journal} {Phys. Rev. B}\ }\textbf {\bibinfo {volume}
  {95}},\ \bibinfo {pages} {075149} (\bibinfo {year} {2017})}\BibitemShut
  {NoStop}%
\bibitem [{\citenamefont {Agterberg}\ \emph {et~al.}(2017)\citenamefont
  {Agterberg}, \citenamefont {Brydon},\ and\ \citenamefont
  {Timm}}]{PhysRevLett.118.127001}%
  \BibitemOpen
  \bibfield  {author} {\bibinfo {author} {\bibfnamefont {D.~F.}\ \bibnamefont
  {Agterberg}}, \bibinfo {author} {\bibfnamefont {P.~M.~R.}\ \bibnamefont
  {Brydon}}, \ and\ \bibinfo {author} {\bibfnamefont {C.}~\bibnamefont
  {Timm}},\ }\href {\doibase 10.1103/PhysRevLett.118.127001} {\bibfield
  {journal} {\bibinfo  {journal} {Phys. Rev. Lett.}\ }\textbf {\bibinfo
  {volume} {118}},\ \bibinfo {pages} {127001} (\bibinfo {year}
  {2017})}\BibitemShut {NoStop}%
\bibitem [{\citenamefont {Ghorashi}\ \emph {et~al.}(2017)\citenamefont
  {Ghorashi}, \citenamefont {Davis},\ and\ \citenamefont
  {Foster}}]{PhysRevB.95.144503}%
  \BibitemOpen
  \bibfield  {author} {\bibinfo {author} {\bibfnamefont {S.~A.~A.}\
  \bibnamefont {Ghorashi}}, \bibinfo {author} {\bibfnamefont {S.}~\bibnamefont
  {Davis}}, \ and\ \bibinfo {author} {\bibfnamefont {M.~S.}\ \bibnamefont
  {Foster}},\ }\href {\doibase 10.1103/PhysRevB.95.144503} {\bibfield
  {journal} {\bibinfo  {journal} {Phys. Rev. B}\ }\textbf {\bibinfo {volume}
  {95}},\ \bibinfo {pages} {144503} (\bibinfo {year} {2017})}\BibitemShut
  {NoStop}%
\bibitem [{\citenamefont {Timm}\ \emph {et~al.}(2017)\citenamefont {Timm},
  \citenamefont {Schnyder}, \citenamefont {Agterberg},\ and\ \citenamefont
  {Brydon}}]{PhysRevB.96.094526}%
  \BibitemOpen
  \bibfield  {author} {\bibinfo {author} {\bibfnamefont {C.}~\bibnamefont
  {Timm}}, \bibinfo {author} {\bibfnamefont {A.~P.}\ \bibnamefont {Schnyder}},
  \bibinfo {author} {\bibfnamefont {D.~F.}\ \bibnamefont {Agterberg}}, \ and\
  \bibinfo {author} {\bibfnamefont {P.~M.~R.}\ \bibnamefont {Brydon}},\ }\href
  {\doibase 10.1103/PhysRevB.96.094526} {\bibfield  {journal} {\bibinfo
  {journal} {Phys. Rev. B}\ }\textbf {\bibinfo {volume} {96}},\ \bibinfo
  {pages} {094526} (\bibinfo {year} {2017})}\BibitemShut {NoStop}%
\bibitem [{\citenamefont {Yang}\ \emph {et~al.}(2017)\citenamefont {Yang},
  \citenamefont {Xiang},\ and\ \citenamefont {Wu}}]{PhysRevB.96.144514}%
  \BibitemOpen
  \bibfield  {author} {\bibinfo {author} {\bibfnamefont {W.}~\bibnamefont
  {Yang}}, \bibinfo {author} {\bibfnamefont {T.}~\bibnamefont {Xiang}}, \ and\
  \bibinfo {author} {\bibfnamefont {C.}~\bibnamefont {Wu}},\ }\href {\doibase
  10.1103/PhysRevB.96.144514} {\bibfield  {journal} {\bibinfo  {journal} {Phys.
  Rev. B}\ }\textbf {\bibinfo {volume} {96}},\ \bibinfo {pages} {144514}
  (\bibinfo {year} {2017})}\BibitemShut {NoStop}%
\bibitem [{\citenamefont {{Cheng}}\ \emph {et~al.}(2017)\citenamefont
  {{Cheng}}, \citenamefont {{Ohtsuki}}, \citenamefont {{Chaudhuri}},
  \citenamefont {{Nakatsuji}}, \citenamefont {{Lippmaa}},\ and\ \citenamefont
  {{Armitage}}}]{BingNature}%
  \BibitemOpen
  \bibfield  {author} {\bibinfo {author} {\bibfnamefont {B.}~\bibnamefont
  {{Cheng}}}, \bibinfo {author} {\bibfnamefont {T.}~\bibnamefont {{Ohtsuki}}},
  \bibinfo {author} {\bibfnamefont {D.}~\bibnamefont {{Chaudhuri}}}, \bibinfo
  {author} {\bibfnamefont {S.}~\bibnamefont {{Nakatsuji}}}, \bibinfo {author}
  {\bibfnamefont {M.}~\bibnamefont {{Lippmaa}}}, \ and\ \bibinfo {author}
  {\bibfnamefont {N.~P.}\ \bibnamefont {{Armitage}}},\ }\href {\doibase
  10.1038/s41467-017-02121-y} {\bibfield  {journal} {\bibinfo  {journal} {Nat.
  Commun.}\ }\textbf {\bibinfo {volume} {8}},\ \bibinfo {eid} {2097} (\bibinfo
  {year} {2017})}\BibitemShut {NoStop}%
\bibitem [{\citenamefont {Savary}\ \emph {et~al.}(2017)\citenamefont {Savary},
  \citenamefont {Ruhman}, \citenamefont {Venderbos}, \citenamefont {Fu},\ and\
  \citenamefont {Lee}}]{PhysRevB.96.214514}%
  \BibitemOpen
  \bibfield  {author} {\bibinfo {author} {\bibfnamefont {L.}~\bibnamefont
  {Savary}}, \bibinfo {author} {\bibfnamefont {J.}~\bibnamefont {Ruhman}},
  \bibinfo {author} {\bibfnamefont {J.~W.~F.}\ \bibnamefont {Venderbos}},
  \bibinfo {author} {\bibfnamefont {L.}~\bibnamefont {Fu}}, \ and\ \bibinfo
  {author} {\bibfnamefont {P.~A.}\ \bibnamefont {Lee}},\ }\href {\doibase
  10.1103/PhysRevB.96.214514} {\bibfield  {journal} {\bibinfo  {journal} {Phys.
  Rev. B}\ }\textbf {\bibinfo {volume} {96}},\ \bibinfo {pages} {214514}
  (\bibinfo {year} {2017})}\BibitemShut {NoStop}%
\bibitem [{\citenamefont {Boettcher}\ and\ \citenamefont
  {Herbut}(2018)}]{PhysRevLett.120.057002}%
  \BibitemOpen
  \bibfield  {author} {\bibinfo {author} {\bibfnamefont {I.}~\bibnamefont
  {Boettcher}}\ and\ \bibinfo {author} {\bibfnamefont {I.~F.}\ \bibnamefont
  {Herbut}},\ }\href {\doibase 10.1103/PhysRevLett.120.057002} {\bibfield
  {journal} {\bibinfo  {journal} {Phys. Rev. Lett.}\ }\textbf {\bibinfo
  {volume} {120}},\ \bibinfo {pages} {057002} (\bibinfo {year}
  {2018})}\BibitemShut {NoStop}%
\bibitem [{\citenamefont {Venderbos}\ \emph {et~al.}(2018)\citenamefont
  {Venderbos}, \citenamefont {Savary}, \citenamefont {Ruhman}, \citenamefont
  {Lee},\ and\ \citenamefont {Fu}}]{PhysRevX.8.011029}%
  \BibitemOpen
  \bibfield  {author} {\bibinfo {author} {\bibfnamefont {J.~W.~F.}\
  \bibnamefont {Venderbos}}, \bibinfo {author} {\bibfnamefont {L.}~\bibnamefont
  {Savary}}, \bibinfo {author} {\bibfnamefont {J.}~\bibnamefont {Ruhman}},
  \bibinfo {author} {\bibfnamefont {P.~A.}\ \bibnamefont {Lee}}, \ and\
  \bibinfo {author} {\bibfnamefont {L.}~\bibnamefont {Fu}},\ }\href {\doibase
  10.1103/PhysRevX.8.011029} {\bibfield  {journal} {\bibinfo  {journal} {Phys.
  Rev. X}\ }\textbf {\bibinfo {volume} {8}},\ \bibinfo {pages} {011029}
  (\bibinfo {year} {2018})}\BibitemShut {NoStop}%
\bibitem [{\citenamefont {Kim}\ \emph {et~al.}(2018)\citenamefont {Kim},
  \citenamefont {Wang}, \citenamefont {Nakajima}, \citenamefont {Hu},
  \citenamefont {Ziemak}, \citenamefont {Syers}, \citenamefont {Wang},
  \citenamefont {Hodovanets}, \citenamefont {Denlinger}, \citenamefont
  {Brydon}, \citenamefont {Agterberg}, \citenamefont {Tanatar}, \citenamefont
  {Prozorov},\ and\ \citenamefont {Paglione}}]{Kimeaao4513}%
  \BibitemOpen
  \bibfield  {author} {\bibinfo {author} {\bibfnamefont {H.}~\bibnamefont
  {Kim}}, \bibinfo {author} {\bibfnamefont {K.}~\bibnamefont {Wang}}, \bibinfo
  {author} {\bibfnamefont {Y.}~\bibnamefont {Nakajima}}, \bibinfo {author}
  {\bibfnamefont {R.}~\bibnamefont {Hu}}, \bibinfo {author} {\bibfnamefont
  {S.}~\bibnamefont {Ziemak}}, \bibinfo {author} {\bibfnamefont
  {P.}~\bibnamefont {Syers}}, \bibinfo {author} {\bibfnamefont
  {L.}~\bibnamefont {Wang}}, \bibinfo {author} {\bibfnamefont {H.}~\bibnamefont
  {Hodovanets}}, \bibinfo {author} {\bibfnamefont {J.~D.}\ \bibnamefont
  {Denlinger}}, \bibinfo {author} {\bibfnamefont {P.~M.~R.}\ \bibnamefont
  {Brydon}}, \bibinfo {author} {\bibfnamefont {D.~F.}\ \bibnamefont
  {Agterberg}}, \bibinfo {author} {\bibfnamefont {M.~A.}\ \bibnamefont
  {Tanatar}}, \bibinfo {author} {\bibfnamefont {R.}~\bibnamefont {Prozorov}}, \
  and\ \bibinfo {author} {\bibfnamefont {J.}~\bibnamefont {Paglione}},\ }\href
  {\doibase 10.1126/sciadv.aao4513} {\bibfield  {journal} {\bibinfo  {journal}
  {Science Advances}\ }\textbf {\bibinfo {volume} {4}} (\bibinfo {year}
  {2018}),\ 10.1126/sciadv.aao4513}\BibitemShut {NoStop}%
\bibitem [{\citenamefont {Ghorashi}\ \emph {et~al.}(2018)\citenamefont
  {Ghorashi}, \citenamefont {Hosur},\ and\ \citenamefont
  {Ting}}]{PhysRevB.97.205402}%
  \BibitemOpen
  \bibfield  {author} {\bibinfo {author} {\bibfnamefont {S.~A.~A.}\
  \bibnamefont {Ghorashi}}, \bibinfo {author} {\bibfnamefont {P.}~\bibnamefont
  {Hosur}}, \ and\ \bibinfo {author} {\bibfnamefont {C.-S.}\ \bibnamefont
  {Ting}},\ }\href {\doibase 10.1103/PhysRevB.97.205402} {\bibfield  {journal}
  {\bibinfo  {journal} {Phys. Rev. B}\ }\textbf {\bibinfo {volume} {97}},\
  \bibinfo {pages} {205402} (\bibinfo {year} {2018})}\BibitemShut {NoStop}%
\bibitem [{\citenamefont {Mandal}\ and\ \citenamefont
  {Nandkishore}(2018)}]{PhysRevB.97.125121}%
  \BibitemOpen
  \bibfield  {author} {\bibinfo {author} {\bibfnamefont {I.}~\bibnamefont
  {Mandal}}\ and\ \bibinfo {author} {\bibfnamefont {R.~M.}\ \bibnamefont
  {Nandkishore}},\ }\href {\doibase 10.1103/PhysRevB.97.125121} {\bibfield
  {journal} {\bibinfo  {journal} {Phys. Rev. B}\ }\textbf {\bibinfo {volume}
  {97}},\ \bibinfo {pages} {125121} (\bibinfo {year} {2018})}\BibitemShut
  {NoStop}%
\bibitem [{\citenamefont {Yu}\ and\ \citenamefont
  {Liu}(2018)}]{PhysRevB.98.104514}%
  \BibitemOpen
  \bibfield  {author} {\bibinfo {author} {\bibfnamefont {J.}~\bibnamefont
  {Yu}}\ and\ \bibinfo {author} {\bibfnamefont {C.-X.}\ \bibnamefont {Liu}},\
  }\href {\doibase 10.1103/PhysRevB.98.104514} {\bibfield  {journal} {\bibinfo
  {journal} {Phys. Rev. B}\ }\textbf {\bibinfo {volume} {98}},\ \bibinfo
  {pages} {104514} (\bibinfo {year} {2018})}\BibitemShut {NoStop}%
\bibitem [{\citenamefont {Yao}\ and\ \citenamefont
  {Chen}(2018)}]{PhysRevX.8.041039}%
  \BibitemOpen
  \bibfield  {author} {\bibinfo {author} {\bibfnamefont {X.-P.}\ \bibnamefont
  {Yao}}\ and\ \bibinfo {author} {\bibfnamefont {G.}~\bibnamefont {Chen}},\
  }\href {\doibase 10.1103/PhysRevX.8.041039} {\bibfield  {journal} {\bibinfo
  {journal} {Phys. Rev. X}\ }\textbf {\bibinfo {volume} {8}},\ \bibinfo {pages}
  {041039} (\bibinfo {year} {2018})}\BibitemShut {NoStop}%
\bibitem [{\citenamefont {Roy}\ \emph {et~al.}(2019)\citenamefont {Roy},
  \citenamefont {Ghorashi}, \citenamefont {Foster},\ and\ \citenamefont
  {Nevidomskyy}}]{PhysRevB.99.054505}%
  \BibitemOpen
  \bibfield  {author} {\bibinfo {author} {\bibfnamefont {B.}~\bibnamefont
  {Roy}}, \bibinfo {author} {\bibfnamefont {S.~A.~A.}\ \bibnamefont
  {Ghorashi}}, \bibinfo {author} {\bibfnamefont {M.~S.}\ \bibnamefont
  {Foster}}, \ and\ \bibinfo {author} {\bibfnamefont {A.~H.}\ \bibnamefont
  {Nevidomskyy}},\ }\href {\doibase 10.1103/PhysRevB.99.054505} {\bibfield
  {journal} {\bibinfo  {journal} {Phys. Rev. B}\ }\textbf {\bibinfo {volume}
  {99}},\ \bibinfo {pages} {054505} (\bibinfo {year} {2019})}\BibitemShut
  {NoStop}%
\bibitem [{\citenamefont {Boettcher}(2019)}]{PhysRevB.99.125146}%
  \BibitemOpen
  \bibfield  {author} {\bibinfo {author} {\bibfnamefont {I.}~\bibnamefont
  {Boettcher}},\ }\href {\doibase 10.1103/PhysRevB.99.125146} {\bibfield
  {journal} {\bibinfo  {journal} {Phys. Rev. B}\ }\textbf {\bibinfo {volume}
  {99}},\ \bibinfo {pages} {125146} (\bibinfo {year} {2019})}\BibitemShut
  {NoStop}%
\bibitem [{\citenamefont {Sim}\ \emph {et~al.}(2019)\citenamefont {Sim},
  \citenamefont {Mishra}, \citenamefont {Park}, \citenamefont {Kim},
  \citenamefont {Cho},\ and\ \citenamefont {Lee}}]{2018arXiv181104046S}%
  \BibitemOpen
  \bibfield  {author} {\bibinfo {author} {\bibfnamefont {G.}~\bibnamefont
  {Sim}}, \bibinfo {author} {\bibfnamefont {A.}~\bibnamefont {Mishra}},
  \bibinfo {author} {\bibfnamefont {M.~J.}\ \bibnamefont {Park}}, \bibinfo
  {author} {\bibfnamefont {Y.~B.}\ \bibnamefont {Kim}}, \bibinfo {author}
  {\bibfnamefont {G.~Y.}\ \bibnamefont {Cho}}, \ and\ \bibinfo {author}
  {\bibfnamefont {S.}~\bibnamefont {Lee}},\ }\href {\doibase
  10.1103/PhysRevB.100.064509} {\bibfield  {journal} {\bibinfo  {journal}
  {Phys. Rev. B}\ }\textbf {\bibinfo {volume} {100}},\ \bibinfo {pages}
  {064509} (\bibinfo {year} {2019})}\BibitemShut {NoStop}%
\bibitem [{\citenamefont {Tchoumakov}\ and\ \citenamefont
  {Witczak-Krempa}(2019)}]{PhysRevB.100.075104}%
  \BibitemOpen
  \bibfield  {author} {\bibinfo {author} {\bibfnamefont {S.}~\bibnamefont
  {Tchoumakov}}\ and\ \bibinfo {author} {\bibfnamefont {W.}~\bibnamefont
  {Witczak-Krempa}},\ }\href {\doibase 10.1103/PhysRevB.100.075104} {\bibfield
  {journal} {\bibinfo  {journal} {Phys. Rev. B}\ }\textbf {\bibinfo {volume}
  {100}},\ \bibinfo {pages} {075104} (\bibinfo {year} {2019})}\BibitemShut
  {NoStop}%
\bibitem [{\citenamefont {Mauri}\ and\ \citenamefont
  {Polini}(2019)}]{PhysRevB.100.165115}%
  \BibitemOpen
  \bibfield  {author} {\bibinfo {author} {\bibfnamefont {A.}~\bibnamefont
  {Mauri}}\ and\ \bibinfo {author} {\bibfnamefont {M.}~\bibnamefont {Polini}},\
  }\href {\doibase 10.1103/PhysRevB.100.165115} {\bibfield  {journal} {\bibinfo
   {journal} {Phys. Rev. B}\ }\textbf {\bibinfo {volume} {100}},\ \bibinfo
  {pages} {165115} (\bibinfo {year} {2019})}\BibitemShut {NoStop}%
\bibitem [{\citenamefont {Tchoumakov}\ \emph {et~al.}(2020)\citenamefont
  {Tchoumakov}, \citenamefont {Godbout},\ and\ \citenamefont
  {Witczak-Krempa}}]{PhysRevResearch.2.013230}%
  \BibitemOpen
  \bibfield  {author} {\bibinfo {author} {\bibfnamefont {S.}~\bibnamefont
  {Tchoumakov}}, \bibinfo {author} {\bibfnamefont {L.~J.}\ \bibnamefont
  {Godbout}}, \ and\ \bibinfo {author} {\bibfnamefont {W.}~\bibnamefont
  {Witczak-Krempa}},\ }\href {\doibase 10.1103/PhysRevResearch.2.013230}
  {\bibfield  {journal} {\bibinfo  {journal} {Phys. Rev. Research}\ }\textbf
  {\bibinfo {volume} {2}},\ \bibinfo {pages} {013230} (\bibinfo {year}
  {2020})}\BibitemShut {NoStop}%
\bibitem [{\citenamefont {Herbut}(2007)}]{herbutbook}%
  \BibitemOpen
  \bibfield  {author} {\bibinfo {author} {\bibfnamefont {I.}~\bibnamefont
  {Herbut}},\ }\href@noop {} {\emph {\bibinfo {title} {{A Modern Approach to
  Critical Phenomena}}}}\ (\bibinfo  {publisher} {Cambridge University Press,
  Cambridge, England},\ \bibinfo {year} {2007})\BibitemShut {NoStop}%
\bibitem [{\citenamefont {Lan}\ \emph {et~al.}(2011)\citenamefont {Lan},
  \citenamefont {Goldman}, \citenamefont {Bermudez}, \citenamefont {Lu},\ and\
  \citenamefont {\"Ohberg}}]{PhysRevB.84.165115}%
  \BibitemOpen
  \bibfield  {author} {\bibinfo {author} {\bibfnamefont {Z.}~\bibnamefont
  {Lan}}, \bibinfo {author} {\bibfnamefont {N.}~\bibnamefont {Goldman}},
  \bibinfo {author} {\bibfnamefont {A.}~\bibnamefont {Bermudez}}, \bibinfo
  {author} {\bibfnamefont {W.}~\bibnamefont {Lu}}, \ and\ \bibinfo {author}
  {\bibfnamefont {P.}~\bibnamefont {\"Ohberg}},\ }\href {\doibase
  10.1103/PhysRevB.84.165115} {\bibfield  {journal} {\bibinfo  {journal} {Phys.
  Rev. B}\ }\textbf {\bibinfo {volume} {84}},\ \bibinfo {pages} {165115}
  (\bibinfo {year} {2011})}\BibitemShut {NoStop}%
\bibitem [{\citenamefont {Liu}\ \emph {et~al.}(2014)\citenamefont {Liu},
  \citenamefont {Law},\ and\ \citenamefont {Ng}}]{PhysRevLett.112.086401}%
  \BibitemOpen
  \bibfield  {author} {\bibinfo {author} {\bibfnamefont {X.-J.}\ \bibnamefont
  {Liu}}, \bibinfo {author} {\bibfnamefont {K.~T.}\ \bibnamefont {Law}}, \ and\
  \bibinfo {author} {\bibfnamefont {T.~K.}\ \bibnamefont {Ng}},\ }\href
  {\doibase 10.1103/PhysRevLett.112.086401} {\bibfield  {journal} {\bibinfo
  {journal} {Phys. Rev. Lett.}\ }\textbf {\bibinfo {volume} {112}},\ \bibinfo
  {pages} {086401} (\bibinfo {year} {2014})}\BibitemShut {NoStop}%
\bibitem [{\citenamefont {Wang}\ \emph {et~al.}(2014)\citenamefont {Wang},
  \citenamefont {Deng},\ and\ \citenamefont {Duan}}]{PhysRevLett.113.033002}%
  \BibitemOpen
  \bibfield  {author} {\bibinfo {author} {\bibfnamefont {S.-T.}\ \bibnamefont
  {Wang}}, \bibinfo {author} {\bibfnamefont {D.-L.}\ \bibnamefont {Deng}}, \
  and\ \bibinfo {author} {\bibfnamefont {L.-M.}\ \bibnamefont {Duan}},\ }\href
  {\doibase 10.1103/PhysRevLett.113.033002} {\bibfield  {journal} {\bibinfo
  {journal} {Phys. Rev. Lett.}\ }\textbf {\bibinfo {volume} {113}},\ \bibinfo
  {pages} {033002} (\bibinfo {year} {2014})}\BibitemShut {NoStop}%
\bibitem [{\citenamefont {Dresselhaus}\ \emph {et~al.}(2008)\citenamefont
  {Dresselhaus}, \citenamefont {Dresselhaus},\ and\ \citenamefont
  {Jorio}}]{BookDresselhaus}%
  \BibitemOpen
  \bibfield  {author} {\bibinfo {author} {\bibfnamefont {M.~S.}\ \bibnamefont
  {Dresselhaus}}, \bibinfo {author} {\bibfnamefont {G.}~\bibnamefont
  {Dresselhaus}}, \ and\ \bibinfo {author} {\bibfnamefont {A.}~\bibnamefont
  {Jorio}},\ }\href@noop {} {\emph {\bibinfo {title} {{Applications of Group
  Theory to the Physics of Solids}}}}\ (\bibinfo  {publisher} {Springer,
  Berlin, Heidelberg},\ \bibinfo {year} {2008})\BibitemShut {NoStop}%
\bibitem [{\citenamefont {Stevens}(1997)}]{StevensBook}%
  \BibitemOpen
  \bibfield  {author} {\bibinfo {author} {\bibfnamefont {K.~W.~H.}\
  \bibnamefont {Stevens}},\ }\href@noop {} {\emph {\bibinfo {title} {{Magnetic
  Ions in Crystals}}}}\ (\bibinfo  {publisher} {Princeton University
  Press,Princeton, NJ},\ \bibinfo {year} {1997})\BibitemShut {NoStop}%
\end{thebibliography}%

\end{document}